\documentclass[useAMS,usenatbib]{mn2e}
\usepackage{subfigure}
\usepackage{natbib}
\usepackage{lscape}
\usepackage[dvips]{color}
\usepackage{amssymb}
\usepackage{amsmath}
\usepackage{graphicx}
\usepackage{url} 
\usepackage{ulem} 
\usepackage{setspace}
\usepackage{threeparttable}
\usepackage{multirow}

\newcommand{\apjl}{ApJ Let.}

\newcommand{\apjs}{ApJS}
\newcommand{\aap}{A\&A}

\usepackage[displaymath, mathlines]{lineno}
\usepackage{blindtext}
\title[DES: Evolution of GLF \& GSMF since $z=1$]{Evolution of Galaxy Luminosity and Stellar-Mass Functions since $z=1$ with the Dark Energy Survey Science Verification Data}
\author[Diego Capozzi et al.] {D. Capozzi$^{1}$\thanks{E-mail: diego.capozzi80@gmail.com}, J.~Etherington$^{1}$, D.~Thomas$^{1,2}$, C.~Maraston$^{1, 2}$, E.~S.~Rykoff$^{3, 4}$, \and I.~Sevilla-Noarbe$^{5}$, K.~Bechtol$^{6, 7}$, M.~Carrasco Kind$^{8, 9}$, A.~Drlica-Wagner$^{10}$, J.~Pforr$^{11, 12}$, \and J.~Gschwend$^{13, 14}$, A.~Carnero Rosell$^{13, 14}$, P.~Pellegrini$^{13, 14}$, M.~A.~G.~Maia$^{13, 14}$, \and L.~N.~da Costa$^{13, 14}$, A.~Benoit-L\'{e}vy$^{15, 16, 17}$, M.~E.~C.~Swanson$^{9}$, R.~H.~Wechsler$^{3, 4, 18}$, \and M,~Banerji$^{19, 20}$, C.~Papovich$^{21}$, X.~Morice-Atkinson$^{1}$, F.~Abdalla$^{16, 22}$, D.~Brooks$^{16}$, \and J.~Carretero$^{23}$, C.~Cunha$^{3}$, C.~D'Andrea$^{24}$, S.~Desai$^{25}$, T.~H.~Diehl$^{10}$, A.~Evrards$^{26, 27}$, \and B.~Flaugher$^{10}$, P.~Fosalba$^{28}$, J.~Frieman$^{10, 29}$, J.~Garc\'ia-Bellido$^{30}$, E.~Gaztanaga$^{28}$, \and D.~W.~Gerdes$^{26, 27}$, D.~Gruen$^{3, 4}$, R.~A.~Gruendl$^{8, 9}$, G.~Gutierrez$^{10}$, W.~G.~Hartley$^{16, 31}$, \and D.~James$^{32, 33}$, T.~Jeltema$^{34}$, K.~Kuehn$^{35}$, S.~Kuhlmann$^{36}$, N.~Kuropatkin$^{10}$, O.~Lahav$^{16}$, \and M.~Lima$^{14, 37}$, J.~L.~Marshall$^{21}$, P.~Martini$^{38, 39}$, F.~Menanteau$^{8, 9}$, R.~Miquel$^{23, 40}$, B.~Nord$^{10}$, \and R.~L.~C.~Ogando$^{13, 14}$, A.~A.~Plazas~Malag\`{o}n$^{41}$, A.~K.~Romer$^{42}$, E.~Sanchez$^{5}$, V.~Scarpine$^{10}$, \and R.~Schindler$^{4}$, M.~Schubnell$^{27}$, M.~Smith$^{43}$, M.~Soares-Santos$^{10}$, F.~Sobreira$^{14, 44}$, \and E.~Suchyta$^{45}$, G.~Tarle$^{27}$\\} 
\date{Accepted ;
  Received ; in original form }
\pagerange{\pageref{firstpage}--\pageref{lastpage}} \pubyear{2017}

\begin{document}

\maketitle

\label{firstpage}

\begin{abstract}
We present the first study of the evolution of the galaxy luminosity and stellar-mass functions (GLF and GSMF) carried out by the Dark Energy Survey (DES). We describe the COMMODORE galaxy catalogue selected from Science Verification images. This catalogue is made of $\sim 4\times 10^{6}$ galaxies at $0<z\lesssim1.3$ over a sky area of $\sim155\ {\rm sq. \ deg}$ with {\it i}-band limiting magnitude ${\it i}=23\ {\rm mag}$. Such characteristics are unprecedented for galaxy catalogues and they enable us to study the evolution of GLF and GSMF at $0<z<1$ homogeneously with the same statistically-rich data-set and free of cosmic variance effects. 
The aim of this study is twofold: i) we want to test our method based on the use of photometric-redshift probability density functions against literature results obtained with spectroscopic redshifts; ii) we want to shed light on the way galaxies build up their masses over cosmic time. 
We find that both the {\it i}-band galaxy luminosity and stellar mass functions are characterised by a double-Schechter shape at $z<0.2$. Both functions agree well with those based on spectroscopic redshifts. The DES GSMF agrees especially with those measured for the GAlaxy Mass Assembly and the PRism MUlti-object Survey out to $z\sim1$. 
At $0.2<z<1$, we find the {\it i}-band luminosity and stellar-mass densities respectively to be constant ($\rho_{\rm L}\propto (1+z)^{-0.12\pm0.11}$) and decreasing ($\rho_{\rm Mstar}\propto (1+z)^{-0.5\pm0.1}$) with $z$. This indicates that, while at higher $z$ galaxies have less stellar mass, their luminosities do not change substantially because of their younger and brighter stellar populations. Finally, we also find evidence for a top-down mass-dependent evolution of the GSMF. 
\end{abstract}


\section{Introduction}
\label{sec:sec1}
The most widely accepted structure-formation paradigm predicts that structures are generated from primordial density perturbations in the power spectrum \citep{Blumenthal-1984, Davis-1985} and form via gravitational collapse following dark matter clustering \citep{White-1978}. Within this picture, galaxies are thought to assemble their mass (including its baryonic component, e.g. gas and stars) over cosmic time following this hierarchical pattern. However, galaxy baryonic mass growth is the result of the interplay of several processes \citep{Somerville-2015}, for example, to mention only some of them, star formation from accreted or in-situ gas (e.g., \citealp{Thomas-2005,Thomas-2010}), radiative cooling (e.g. \citealp{White-1978}), supernova and Active Galactic Nuclei (AGN) feedbacks (e.g. \citealp{Benson-2003, Croton-2006}), mergers and galaxy interactions (e.g., \citealp{Lotz-2011}). 
Hence, the hierarchical picture portrayed for dark matter might not be fully applicable to baryonic matter, because the latter is not subject only to gravity. In fact, despite the general belief that galaxies form hierarchically [the majority of semi-analytic models are built on this premise (e.g., \citealp{De-Lucia-2006})], recent observational studies suggest a less significant role for the halo-scale environment in influencing galaxy formation and evolution, particularly for high-mass galaxies (e.g., see \citealp[and references therein]{Cowie-1996, Thomas-2005, Cimatti-2006, Capozzi-2010, Pozzetti-2010, Stott-2010, Peng-2010,Thomas-2010, Maraston-2013}). 

At the same time, the influence that the physics driving galaxy formation has on halo properties is still a matter of debate (e.g., \cite{Peacock-2000, Berlind-2002, Lin-2004, Capozzi-2012, Behroozi-2013}). As a consequence of this contrast, a number of fundamental issues remain unsolved. For instance, the latest galaxy formation models regard the time when galaxy stellar mass forms and assemble onto the main galaxy halo as two different stages in the life of a galaxy. These two stages might not coincide and they are both mass dependent (e.g. \citealp{De-Lucia-2006, Moster-2013}). In general, the more massive the galaxy the earlier in time its stellar mass is formed, in the sense that at a given redshift, more massive galaxies will have formed a higher fraction of their final stellar mass than less massive galaxies. This is generally found both in observation- (e.g., \citealp{Thomas-2005, Thomas-2010}) and simulations-based studies (e.g., \citealp{De-Lucia-2006, Moster-2013}). When focusing instead on the time when stellar mass is assembled, the current picture is blurred by the discrepancies between the findings of studies based on theoretical models applied to simulations and those based on observed data. In fact, while the former generally find that the stellar mass assembly happens in a hierarchical (bottom-up) fashion (e.g., \citealp{De-Lucia-2006, Somerville-2008, Monaco-2007, Moster-2013}), the majority of the latter finds that the stellar mass assembly follows an anti-hierarchical (top-down) pattern (e.g., \citealp{Cimatti-2006,Perez-Gonzalez-2008, Cirasuolo-2010, Pozzetti-2010, Goncalves-2012, Ilbert-2013, Moustakas-2013}). However there are also observation-based studies which show agreement with simulations, i.e. identify a hierarchical pattern in the galaxy mass assembly (see for instance \citealp{Muzzin-2013}). As a result, galaxy stellar-mass assembly still remains a burning question for the scientific community.

Galaxy luminosity and stellar mass functions (respectively GLF and GSMF) constitute some of the most useful observables for gaining insight into the mechanisms actually governing galaxy formation and evolution, hence are fundamental for replying to this burning question. In particular, studying the evolution over cosmic time of galaxy number, luminosity and stellar mass densities allows us to directly probe the galaxy mass growth process. 
Several studies in the literature focused on the investigation of these functions in relation to galaxy formation and evolution (e. g. \citealp{Schechter-1976, Bell-2003, Blanton-2005, Baldry-2008, Montero-Dorta-2009, Bernardi-2010, Ramos-2011, Baldry-2012, Bernardi-2013, Pozzetti-2010, Loveday-2012, Marchesini-2009, Ilbert-2013, Muzzin-2013, Moustakas-2013, Maraston-2013, Mortlock-2015}). The majority of such studies were either performed out to high redshift ($z\sim2$ or higher) but in deep pencil-beam surveys (usually $\lesssim 2 \ {\rm deg^{2}}$ wide, e.g. \citealp{Marchesini-2009, Ramos-2011, Ilbert-2013, Muzzin-2013, Mortlock-2015}) or limited to low redshifts ($z\lesssim0.5$) in relatively shallow but large-area surveys [e.g., SDSS and GAlaxy Mass Assembly (GAMA), \citealp{Blanton-2005, Baldry-2012, Loveday-2012}].

Many of these studies made use of Spectral-Energy-Distribution (SED) fitting for deriving galaxy properties (such as stellar mass). The filter sets used generally listed more than 5 filters encompassing the spectral region from the UV-optical to the infrared. Variation of the filter sets and of the spectral coverage of galaxy SEDs is linked to variation in the precision of galaxy properties and hence of the GLF and GSMFs presented in these studies. In addition, especially when studying the GLF and GSMF out to $z>1$, the use of photometric redshifts was often adopted, leading to additional uncertainties on galaxy properties and on GLF and GSMF (e.g. \citealp{Marchesini-2009}). Additional sources of uncertainties and systematics, like those due to evolutionary stellar population synthesis models, SED-fitting related assumptions (e. g.,  model templates, initial mass function, metallicity, dust content and modeling) and others, were also treated in some of the studies of the GLF/GSMF in the literature (e.g., see \citealp{Marchesini-2009, Pozzetti-2010, Behroozi-2010, Moustakas-2013}). All these sources of uncertainty must be properly taken into account in order to reliably measure spatial number densities and their evolution with cosmic time.

In this paper we study the evolution of the GLF and GSMF out to $z\sim1$ with the Dark Energy Survey, using data taken during its Science Verification (SV) phase over 5 broad-band filters ({\it g, r, i, z, Y}). Such data are characterised by a relatively high depth ($\sim 23\ {\rm mag}$ in {\it i} band) and a large area ($\sim 155\ {\rm sq.~deg}$), characteristics which allow us to better study the GLF and GSMF with respect to previous surveys. Our work is carried out by using photometric redshifts, which allow us to analyse a galaxy sample made of about $4\times 10^{6}$ galaxies out to redshift $z\sim1$. The aim of our study is twofold. On one hand, we want to test if over a wide-field 5-band survey we can get consistent GLFs/GSMFs from photometric and spectroscopic redshifts (see \citealp{Ramos-2011} for a similar test but over a $\lesssim 1 \ {\rm deg^{2}}$ wide sky area). On the other hand, we want to shed light on the way galaxies build up their masses over cosmic time.

Despite assessing the effects on our analysis due to uncertainties on photometry, photometric redshifts and to galaxy completeness, in our study we do not take into account sources of uncertainties due to evolutionary population synthesis models and SED-fitting assumptions [such as model templates, initial mass function (IMF), metallicity, dust content and modeling]. We present here the current study of GLF and GSMF on DES SV data as a proof of concept and build up on our current work to subsequently produce more precise estimates based on the survey full data set and taking into account all systematics and sources of uncertainty.

The layout of the paper is the following: in Section \ref{sec:sec2} we provide a description of the DES, the SV data-set, photometric redshifts and the selection of the galaxy catalogue (including galaxy completeness assessment). Section \ref{sec:sec3} is dedicated to the determination of galaxy properties (both physical and detectability-related), while in Section \ref{sec:sec4} the analysis carried out to measure the GLF and GSMF is presented. Section \ref{sec:sec5} is dedicated to the description of our results in comparison with observational measurements in the literature based on spectroscopic redshifts. Finally, in Sections  \ref{sec:sec6} \& \ref{sec:sec7}  we respectively discuss our results and draw our conclusions.

Throughout the paper, we assume a flat $\Lambda$ cold dark matter ($\Lambda CDM$) cosmology with $H_{0}=70\ {\rm km~s^{-1}~Mpc^{-1}}$ and $\Omega_{m}=0.286$, use magnitudes in the $AB$ system and, unless indicated otherwise, utilise a Salpeter IMF \citep{Salpeter-1955}. In addition, $M^{*}$ stands for the cut-off value (either absolute magnitude or stellar mass) of the Schechter function, while $M_{*}$ is used to refer to stellar mass.
 
\begin{figure*}
\centering
\includegraphics[scale=0.5]{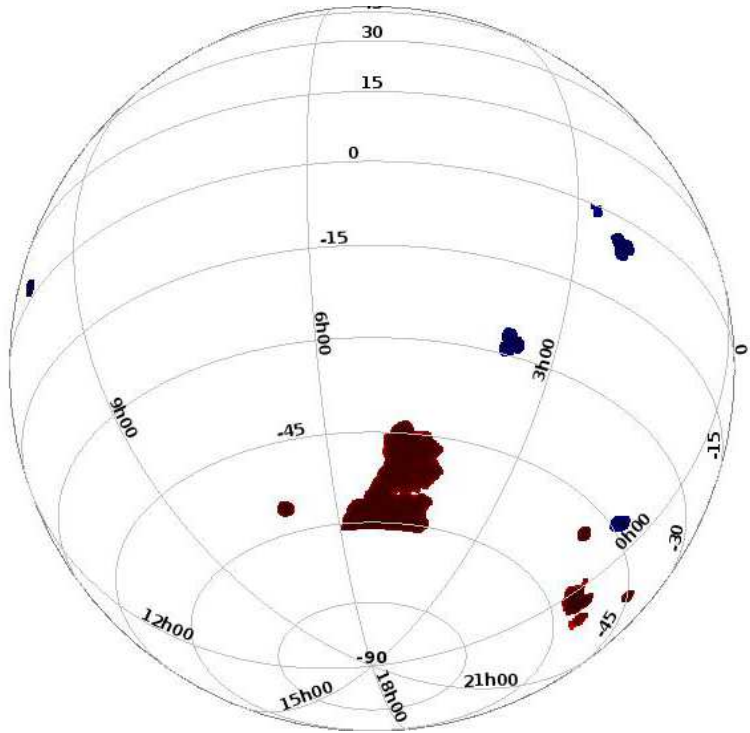}
\caption{DES SV footprint. Regions at standard depth (Wide survey: SPT-E, SPT-W, El Gordo, Bullet-cluster and RXJ2248 fields) are reported in red, while deeper regions (Deep survey: SN fields) are plotted in blue.}
\label{fig:Fig1}
\end{figure*}

\section{DES Science Verification Data}
\label{sec:sec2}
The Dark Energy Survey  is a photometric survey carried out in 5 bands ({\it g, r, i, z,Y}) with the Dark Energy Camera (DECam, \citealp{Flaugher-2015}) mounted on the 4-meter Blanco Telescope at Cerro Tololo Inter-American Observatory (CTIO). The survey started in August 2013 but from November 2012 to March 2013, DES carried out a Science Verification (SV) survey. These observations provide science-quality data for more than 250 {\rm sq.~ deg} at close to the main survey's nominal depth (standard depth), the latter being achieved by the coaddition of 10 single-epoch images (see further down).

The sky footprint over which these data were taken was chosen in order to contain a combination of large contiguous regions at standard depth ({\it i}-band 2-arcsec aperture magnitude$=24$), covering parts of the eastern (over $\sim 160\ {\rm sq. \ deg.}$) and western (over $\sim 35\ {\rm sq. \ deg.}$) areas of the South-Pole-Telescope (SPT, \citealp{Lueker-2010}) survey (referred to respectively as SPT-E and SPT-W fields), and smaller regions either at standard depth or deeper  (limiting {\it i}-band 2-arcsec aperture magnitude$=26$). The former extend individually over $<1\ {\rm sq. \ deg}$ and are respectively centred on the well-known El Gordo, Bullet and RXJ2248 galaxy clusters (we will refer to these fields respectively as El Gordo, Bullet-cluster and RXJ2248 fields). The latter extend over a total area of $\sim 30\ {\rm sq. \ deg}$ and we shall refer to them as Supernova (SN) fields.
From now on we will generally refer to SN fields as Deep survey and to the remaining fields at the survey standard depth as Wide survey. 

In Figure \ref{fig:Fig1}, the footprint of the SV area is shown.

\subsection{The Gold SVA1 galaxy catalogue}
\label{subsec:subsec2.1}
The data stored in the catalogue of the SV coadded imaging were created by the DES Data Management (DESDM) pipeline (\citealp{Sevilla-2011,Mohr-2012} and Morganson et al. in prep), which used \texttt{SExtractor} \citep{Bertin-1996} as the main source extraction software. In order to make these data science ready, a team of DES scientists thoroughly tested and analysed them. In particular, the following steps were carried out:

\begin{itemize}
\item Incorporating satellite trail and other artifact information to mask out specific areas.
\item Removing areas with colours severely affected by stray light in the images and areas with a small exposure count (at the borders of the footprint).
\item Applying an additional zero point correction via stellar locus regression (SLR), to tighten the calibration even further according to the distribution of star colours with respect to those of reference stars. 
\item Removing the area ($\sim 40\ {\rm sq.\ deg.}$) below declination of $-61^{\circ}$, largely occupied by the Large Magellanic Cloud (LMC). This was done because our SLR tests showed that this area could not be accurately calibrated to the same scale as the rest of the survey. This cut has also the advantage of removing $\sim 5\ {\rm sq.\ deg}$ contaminated by stray light from R Doradus, the second brightest star in the infrared sky.
\item Identifying a star/galaxy classifier (the `modest classifier') to perform star/galaxy separation (see Section \ref{subsubsec:subsubsec2.1.3}). 
\end{itemize}

These additional steps led to the identification of a new photometric catalogue, called the \textit{SV Annual 1} (\textit{SVA1}) \textit{Gold} catalogue, containing 25,227,559 objects extending over an area $\sim 250$ {\rm sq.~deg}.  This catalogue is now publicly available\footnote{The SVA1 Gold catalogue, all related data products and documentation are publicly available at: \url{http://des.ncsa.illinois.edu/releases/SVA1}.}. 

\subsubsection{Survey depth and systematics: \texttt{MANGLE} mask and depth maps}
\label{subsubsec:subsubsec2.1.1}
Despite the survey strategy being decided in such a way to minimize depth differences over the sky, variable observing conditions on different observing nights makes this difficult. In addition to variable depth over the sky, observations are also influenced by artificial effects like cosmic rays, airplane and satellite trails and by stray light from very bright stars. In order to take all these sky-region dependent effects into account, the creation of a mask via \texttt{MANGLE} \citep{Swanson-2008} software was implemented within the DESDM pipeline. \texttt{MANGLE} takes into account properties of DECam CCDs and the sky during each night and gives an estimation of the $10\sigma$-level depth for different regions (called Molygons). This depth is calculated for a $2\ {\rm arcsec}$ aperture magnitude (\texttt{MAG\_APER4}). Despite the resulting mask having very high resolution (at a pixel size of $6.44\ {\rm arcsec}$), allowing to eliminate regions severely affected by the aforementioned effects and to take variable depth with the sky into account, aperture magnitudes such as \texttt{MAG\_APER4} do not contain the total light of extended sources. Hence they are not typically used for the majority of galaxy studies, which mainly rely on integrated magnitudes. In addition, the depth value characterising a galaxy sample should be given in the same magnitude system used for galaxy selection. Hence, in our case, the \texttt{MAG\_APER4} depth value should be converted into the total magnitude system used for galaxy selection. Because of this, a new mask was created for each total magnitude estimation algorithm [i.e., \texttt{MAG\_AUTO}, \texttt{MAG\_DETMODEL} (similar to SDSS' {\it modelMag}) and \texttt{MAG\_MODEL} (similar to SDSS' {\it cmodelMag})] used during the derivation of the photometric catalogues. This was done by following the process described in \citet{Rykoff-2015} and resulted in new masks at a resolution of $0.18\ {\rm arcmin^{2}}$. These masks are used here for selecting a galaxy sample at $10\sigma$ level.

\subsubsection{Photometric Redshifts}
\label{subsubsec:subsubsec2.1.2}
Reliably measuring photometric redshifts is a key step for carrying out cosmological measurements, for a proper determination of galaxy properties and for a correct study of GLF and GSMF. Keeping this in mind, for our study we make use of photometric redshifts measured with the ``Trees for PHOTOZ''  (TPZ) algorithm \citep{Carrasco-2013,Carrasco-2014}, which provides individual photo-$z$ probability density functions (pdfs) which fold in photometric uncertainty. We can then use such pdfs in our study to determine the effects on GLF and GSMF due to photometry and photo-$z$ uncertainties. In particular, TPZ is a machine learning parallel algorithm which utilises prediction trees and random forest techniques to produce not only redshift pdfs but also ancillary information for a given galaxy sample. We refer the reader to \citet{Carrasco-2013, Carrasco-2014} for details on this algorithm
%, including a probabilistic star/galaxy classifier [\texttt{TPZ\_SG\_CLASS} (TPC), with values between 0 (for galaxy) and 1 (for star)], which we will use in the next section. We refer the % reader to \citet{Carrasco-2013, Carrasco-2014} for details on this algorithm and to \citet{Kim-2015} for details on the TPC star/galaxy separation method. 

Our choice of using TPZ photo-$z$'s was made also because the TPZ algorithm is one of the best performing methods in the study by \citet{Sanchez-2014}, who found that empirical methods using, for instance, artificial neural networks or random forests (as TPZ) yielded the best performance, achieving core photo-$z$ resolutions $\sigma_{\rm 68}\sim 0.08-0.1$, defined as the 68 percentile width of $\Delta z$ around the median.

We take into account photometry and photo-$z$ uncertainties on galaxy properties (e.g. absolute magnitude and stellar mass) and on GLF and GSMF by performing Monte Carlo simulations. In particular we draw 100 photo-$z$'s values from each galaxy pdf via inverse-cumulative-distribution-function resampling method, in order to construct 100 additional galaxy catalogues for our analysis. We carry out our analysis on each of these catalogues, from galaxy property estimation to GLF and GSMF measurements, so to derive uncertainties on galaxy properties and on number densities (see Sections \ref{sec:sec3} and \ref{sec:sec4}).
We find (see \citealp{Etherington-2017}) that 100 draws are sufficient for producing unbiased resampled photo-$z$ pdfs (both mean and standard deviation of the resampled photo-$z$ pdfs are unbiased on average). Such resampling is also characterised by RMSE of pdf mean and standard deviation (respectively 0.0076 and 0.0082) larger than the typical width of the median photo-$z$ pdf. However the precision quoted is sufficient for the study carried out here. We refer the reader to \citet{Etherington-2017} for more details on the photo-$z$ pdf resampling and to Sections \ref{sec:sec3} and \ref{sec:sec4} for the results of this Monte Carlo simulations on galaxy properties and on the study of GLF and GSMF.

Note that in this paper we do not study the effect of using different photometric redshifts estimated via different methods, which are also available for the SVA1 Gold catalogue\footnote{\url{https://des.ncsa.illinois.edu/releases/sva1/doc/photoz}.} and are described in \citet{Bonnett-2016}.

\subsubsection{Star/galaxy separation}
\label{subsubsec:subsubsec2.1.3}
The selection of galaxies from the Gold catalogue is carried out by using the so-called `modest classifier' (1 for galaxies and 2 for stars), which combines several SExtractor outputs in the {\it i} band: i) star/galaxy separation parameters, i.e. \texttt{CLASS\_STAR} \citep{Bertin-1996}, \texttt{SPREAD\_MODEL} and its uncertainty \texttt{SPREADERR\_MODEL} \citep{Desai-2012, Bouy-2013}; ii) magnitude measurements, i.e. \texttt{MAG\_AUTO} and \texttt{MAG\_PSF}; iii) SExtractor internal flags (\texttt{FLAGS}). The combination used to select galaxies is the following (see also \citealp{Jarvis-2016}):

\begin{itemize}
\item[] \noindent $\texttt{bright\_test}=\texttt{CLASS\_STAR\_I} > 0.3 \ \texttt{AND MAG\_AUTO\_I}<18.0$% < 18.0}$
\item[] \noindent $\texttt{locus\_test}=\texttt{SPREAD\_MODEL\_I} +3*\texttt{SPREADERR\_MODEL\_I} < 0.003$
\item[] \noindent $\texttt{faint\_psf\_test}=\texttt{MAG\_PSF\_I} > 30.0\ \texttt{AND MAG\_AUTO\_I} < 21.0$
\item[] \noindent $\texttt{flag\_test}=\texttt{FLAGS\_I} > 3$
\end{itemize}

\begin{itemize}
\item[] \noindent  $\texttt{galaxies}= \texttt{NOT bright\_test AND NOT locus\_test} \\ \texttt{AND NOT faint\_psf\_test AND NOT flag\_test}$
\end{itemize}

The above combination of \texttt{SExtractor} parameters takes into account the relative magnitude regions where \texttt{CLASS\_STAR} and \texttt{SPREAD\_MODEL} perform better and where star or galaxies are expected to dominate (hence the use of different magnitude types). In addition, \texttt{SExtractor} internal flags serve for artifacts removal. The modest classifier is found to simultaneously achieve $\geq 90$ per cent efficiency and purity for objects with {\it i}-band magnitude fainter that 19 {\rm mag}. %We refer the reader to the documentation available online\footnote{\url{http://des.ncsa.illinois.edu/releases/SVA1}.} for a detailed description of the `modest classifier'. 

In addition to the `modest classifier', a probabilistic star/galaxy classifier is also provided by the TPZ algorithm  [\texttt{TPZ\_SG\_CLASS} (TPC), with values between 0 (for galaxy) and 1 (for star)]. We refer the reader to \citet{Kim-2015} for details on the TPC star/galaxy separation method, which will be also used in the next section.

\begin{figure}
\centering
\includegraphics[width=0.4\textwidth]{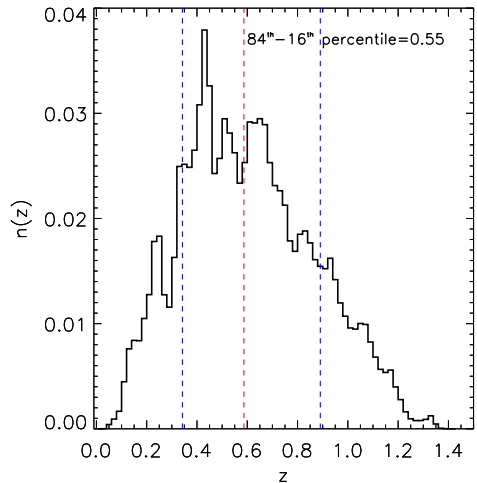}
\caption{Distribution of the median photometric redshifts by TPZ.} 
\label{fig:Fig2}
\end{figure}

\subsection{The COMMODORE galaxy catalogue}
\label{subsec:subsec2.2}
In this section we describe the identification of a galaxy catalogue taken from Gold SVA1, with the purpose of a proper study of galaxy formation and evolution, in general, and of GLF and GSMF in particular. 
One of the difficulties with the SVA1 Gold catalogue is the sky variation of galaxy depth, which not only can affect the results of galaxy evolution studies but also makes galaxy completeness characterisation challenging. For these reasons, we selected the COMMODORE (COnstant \texttt{Mag\_MOdel} Depth Originated REgion) galaxy catalogue, whose selection is described below.

\subsubsection{Sample selection}
\label{subsubsec:subsubsec2.2.1}
The main requirement we apply for the identification of this catalogue is to have a galaxy catalogue selected over a region of constant limiting magnitude (depth). Such a requirement can be met by using the depth maps available for the SVA1 Gold catalogue. By using the latter in the {\it i} band, we can identify sky pixels with measured depth higher than a minimum value ($\texttt{MAG\_MODEL}_{\rm i}=23 \ {\rm mag}$) and select only them for the galaxy catalogue construction. These pixels do not have to be contiguous, an ideal characteristic for minimising Large-Scale-Structure effects and ensuring statistical independence over the sky for studies like ours for which contiguity is not a necessary requirement. Once these pixels are identified, only galaxies with magnitudes brighter than $\texttt{MAG\_MODEL}_{\rm i}=23 \ {\rm mag}$ are kept. This enables us to identify a catalogue over a sky region with approximately homogeneous depth. We exclude SN fields for ensuring homogeneity of observing conditions and use \texttt{MAG\_MODEL} because it performs better than \texttt{MAG\_DETMODEL} in estimating galaxy total fluxes, while the latter is more indicated for colour estimation.
 
Hence, aiming at combining the most reliable information about galaxy SED shape and about galaxy total flux, we rescale galaxy \texttt{MAG\_DETMODEL} magnitudes by the flux ratio (i.e., magnitude difference) between the latter and the {\it i}-band \texttt{MAG\_MODEL} values (see \citealp{Maraston-2013}). These magnitudes (hereafter \texttt{RESCALED DETMODEL}) are then used for galaxy properties estimation, as described in Section \ref{sec:sec3}. The galaxy apparent magnitude depth of this catalogue in the various bands is then assessed via number counts drop. The values estimated, respectively for {\it g,r,i,z} and {\it Y} bands, are: 23.40, 22.98, 23.00 (by construction), 22.26 and 22.06.

In addition to the described selection, we also apply the following cuts:
\begin{itemize}
\item[i)] $\texttt{MAGERR\_MODEL}_{\rm i}<0.11$ 
\item[ii)] $1\leq \texttt{DETMODEL}\_({\it g-r}) \leq4$ \&  $1\leq \texttt{DETMODEL}\_({\it i-z}) \leq4$
\item[iii)]  $16\leq \texttt{MAG\_MODEL}_{\it i} \leq 23$ \& $16\leq \texttt{MU\_MAX}_{\it i} \leq 27$
\item[iv)] $\texttt{TPZ\_SG\_CLASS}<0.00023$.
\end{itemize}

Criterion i) is used to select a galaxy sample at $10\sigma$ level since the SVA1 Gold depth maps provide limiting magnitude values at this detection level, while criterion ii) is used to make sure to exclude objects with strong colours characteristics of diffraction artifacts. Criterion iii) is used to make sure the final catalogue contains only galaxies with reliable values of apparent magnitude and surface brightness, the latter referred to the galaxy brightest pixel as measured by \texttt{SExtractor}'s parameter \texttt{MU\_MAX}. Finally, criterion iv) is used to identify a galaxy catalogue at $\sim 99$ per cent galaxy purity but still at $>90$ per cent completeness level, a condition not ensured by the use of the `modest classifier' only in the apparent magnitude range selected with criterion iii). The conservativeness of the chosen galaxy purity level aims at making sure that contamination by stars does not affect our analysis, especially in low-number-count regimes as in the bright/massive ends of high-$z$ GLFs/GSMFs. 

The newly-obtained catalogue counts 3,711,833 galaxies over a sky area of $\sim 155\ {\rm sq.\  deg}$. We point out that using our Monte Carlo simulations (see Section \ref{subsubsec:subsubsec2.1.2}) we obtain 100 more such catalogues on which we repeat the entire analysis (from galaxy properties estimation to measuring GLF and GSMF). Figure \ref{fig:Fig2} shows the distribution of the median photo-$z$ derived by the Monte Carlo simulations. 

\subsubsection{Galaxy completeness as function of observed quantities}
\label{subsubsec:subsubsec2.2.2}
A fundamental property for a galaxy sample is represented by its characteristic completeness as a function of apparent integrated magnitude and surface brightness. This is indispensable for studies of galaxy evolution with cosmic time like the one presented here. We study galaxy completeness of the COMMODORE catalogue by investigating a few fields located within the deeper SN fields. 
The reason why this study focuses only in regions within the SN fields is that in SVA1 data there is no overlap between the Wide and Deep surveys. As a consequence we have to rely on our SN fields for a study of galaxy completeness. In order to do so, single-epoch images in the tiles listed in Table \ref{tab:Table1} were used to obtain new coadded images. 
The selected tiles used for this new set of coadded images were selected so to have limiting magnitude $\sim$ twice as deep as the Wide survey standard depth and to be far from footprint edges.

\begin{table}
%\begin{tiny}
\begin{center}
\caption{DES tiles used for constructing shallow and deep coadds for studying galaxy completeness.}
\begin{tabular}{l}
 \hline
  \multicolumn{1}{l}{\bf DES Tiles} \\
\hline
DES0223-0416 \\
DES0224-0458  \\
DES0226-0416  \\
DES0227-0458 \\
DES0328-2749 \\
DES0329-2832  \\
DES0332-2749  \\
DES0332-2832 \\
DES0957+0209 \\
DES0957+0252  \\
DES0959+0126 \\
DES1002+0126  \\
DES1000+0252  \\
DES1000+0209 \\
DES1003+0209  \\
DES1003+0252 \\
\hline
\end{tabular}
\label{tab:Table1}							    
\end{center}
%  \end{tiny}
\end{table}

\begin{figure*}
\centering
\includegraphics[scale=0.6]{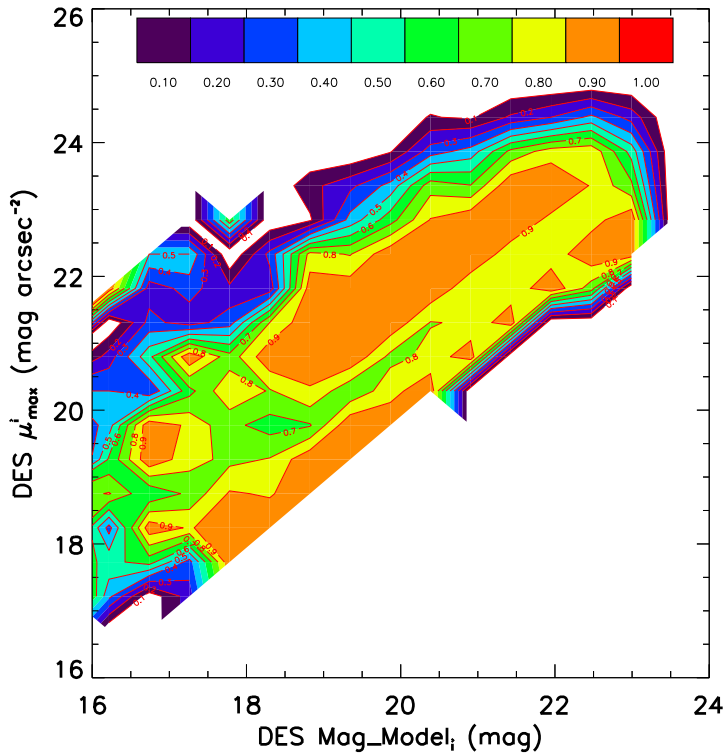}
\caption{Galaxy completeness map used to assign completeness factors to each galaxy in the COMMODORE catalogue. Red contours identify region of constant completeness.}
\label{fig:Fig3}
\end{figure*}

Two new sets of coadded images are obtained for our completeness study. The first set is created such that to obtain coadded images equivalent to those of the Wide survey, hence matching the total exposure time and the standard limiting magnitude characterising them. We will refer to this set of coadded images as `shallow coadds'. The second set of coadded images is instead obtained by coadding all the single-epoch images available in the selected tiles, resulting in coadded images significantly deeper than ($\sim$ twice as deep as) those obtained with the shallow coadds. This second set of coadded images will be hereafter referred to as `deep coadds'. Galaxy catalogues for both the shallow and deep coadds are then built up as done with SVA1 catalogues and selected following the same approach used for the SVA1 Gold catalogue.  

In order to study galaxy completeness as a function of total apparent magnitude and surface brightness, the ratio of the number of detected galaxies in both the deep and shallow coadds to those detected in the deep ones are studied as a function of these two quantities within $0.5\ {\rm mag}$ x $0.5\ {\rm mag}$ 2-dimensional bins and considering the deep coadds as truth. The identification of those galaxies detected in both the coadds is carried out by sky cross-matching the two galaxy catalogues with an angular radius of $0.5 \ {\rm arcsec}$.  Before doing this, the $10-\sigma$ limiting {\it i}-band magnitude of the shallow-coadds galaxy catalogue was determined to be $\texttt{MAG\_MODEL}_{\rm i}=23$ and the shallow- and deep-coadds galaxy catalogues were respectively deprived of galaxies with $\texttt{MAG\_MODEL}_{\rm i}>23.0$ and $\texttt{MAG\_MODEL}_{\rm i}>23.5$. The magnitude cut carried out on the deep-coadds catalogue was made in order to avoid mismatch of brighter galaxies in the shallow-coadds catalogue with fainter ones in the deep-coadds one.

\begin{figure*}
\centering
\includegraphics[scale=0.6]{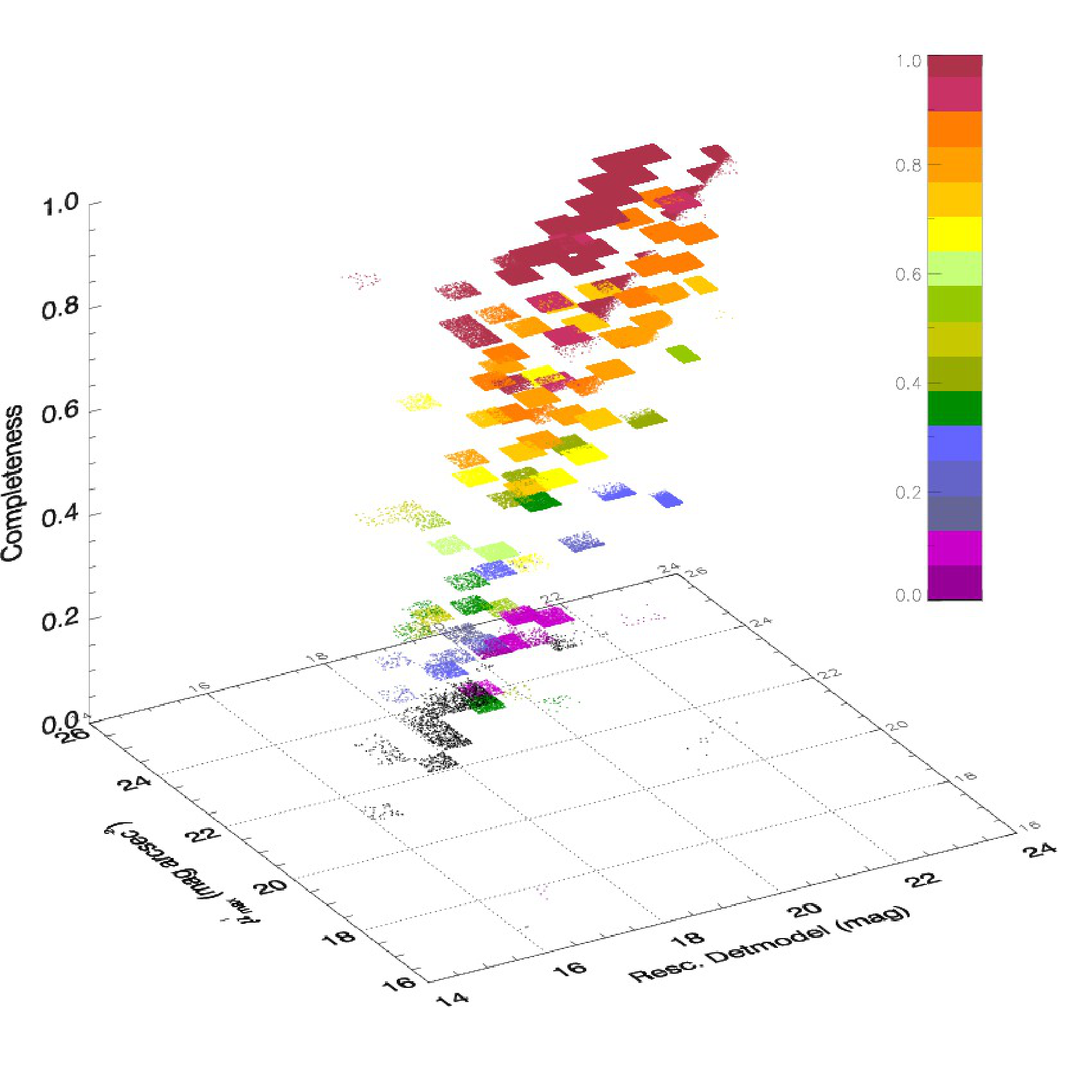}
\caption{3D visualisation of galaxy completeness (z axis) as function of apparent magnitude (x axis, showing the rescaled {\it i}-band Detmodel magnitude) and surface brightness (y axis, showing the {\it i}-band $\mu_{\rm max}$ in units of ${\rm mag\ arcsec^{-2}}$).}
\label{fig:Fig4}
\end{figure*}

The results of our study of galaxy completeness are visualised in Figure \ref{fig:Fig3}, showing the completeness map as a function of {\it i}-band apparent magnitude and surface brightness, obtained by using the galaxy catalogues just described. 
This map is then used to assign a completeness factor to each galaxy contained in the COMMODORE galaxy catalogue according to their  {\it i}-band values of apparent magnitude and surface brightness (in our case of \texttt{MAG\_MODEL} and \texttt{MU\_MAX}). The results of this process are visualised in Figure \ref{fig:Fig4}, where the completeness factor is plotted against {\it i}-band \texttt{MAG\_MODEL} (in this band equivalent to the \texttt{RESCALED DETMODEL} magnitude) and \texttt{MU\_MAX} for the entire COMMODORE galaxy catalogue. These completeness factors will be then used in our study of the GLF and GSMF, as explained in Section \ref{sec:sec4} and are also used in the study of the environmental dependence of the GSMF by \citet{Etherington-2017}.

\section{Galaxy properties determination}
\label{sec:sec3}
In order to measure galaxy properties such as absolute magnitudes, stellar ages, stellar masses and k-correctionsste, we carry out a two-step analysis. The two steps consist of: i) fitting observed spectral energy distributions (SEDs) with theoretical ones; ii) using the resulting best fit models to calculate star-formation-history dependent properties.

\subsection{SED fitting}
\label{subsec:subsec3.1}
For carrying out step i), we use an adapted version of HYPERZ \citep{Bolzonella-2000} (hereafter referred to as HYPERZSPEC), which allows to keep redshifts fixed at the measured values (in our case at the one measured photometrically, see Section \ref{subsubsec:subsubsec2.1.2}). For each galaxy, the fitting procedure is performed by fitting different theoretical SEDs to observed ones and evaluating for each of them the reduced $\chi^{2}$ ($\chi^{2}_{\rm r}$), which is used as figure of merit to identify the best-fitting one. Once the latter is found, all the characteristics of the stellar populations generating the chosen model (i.e., SFH, age and metallicity) are assigned to the fitted galaxy SED. The SED fitting code is implemented with theoretical templates based on the evolutionary population synthesis models by \citet{Maraston-2005}. The template setup used here (the same as the one utilised in \citealp{Maraston-2006,Pforr-2012} \& \citealp{Capozzi-2016} and referred to as wide template setup) is made of 32 theoretical spectra covering a broad range of SFHs: i) SSP (simple stellar population, corresponding to a single star burst); ii) exponentially declining star formation rate (SFR, $\tau$-model with $\tau=$0.1, 0.3, and 1 {\rm Gyr}); iii) truncated SFR (step-like star formation, i.e. constant for a time interval $t=$0.1, 0.3 and 1 {\rm Gyr} since galaxy formation, null afterwards); iv) constant SFR. Metallicity varies among four values (1/5, 1/2, 1 and 2 $Z_{\rm \odot}$), while 221 values of age out to $15\ {\rm Gyr}$ are investigated. The templates are calculated for a Salpeter IMF \citep{Salpeter-1955}. To avoid unrealistic solutions, galaxies are also constrained to have stellar ages younger than the age of the Universe at their redshifts, within the used cosmological model. In addition, we apply a low-age cutoff at $age<0.1\ {\rm Gyr}$, so to avoid fitting unrealistically young ages \citep{Maraston-2010}. In order to avoid age-dust degeneracy effects (for instance see \citealp{Renzini-2006} for a review of this problem), we do not use reddening in the SED fitting procedure. This is because this degeneracy increases when reddening is included as a free parameter, favouring dusty solutions with unlikely too young ages, as shown by \citet{Pforr-2012} in their study of simulated galaxies. In practice, the inclusion of reddening was shown to produce well-recovered SEDs but significantly underestimated stellar masses, especially for old galaxies which have experienced a recent, small star-formation burst. For each galaxy, the best-fitting model is then used to calculate the remaining properties (see Section \ref{subsec:subsec3.2}).

\subsection{K correction, physical and detectability-related galaxy properties}
\label{subsec:subsec3.2}
The second step for determining the remaining galaxy properties consists of measuring k-correction, absolute magnitudes, stellar mass and the maximum ($z_{\rm max}$) and minimum ($z_{\rm min}$) detectability redshifts (i.e., the maximum and minimum redshifts at which each galaxy is detectable given the characteristics of the survey). These quantities are all estimated using the galaxy best-fit SEDs. 
Note that we do not apply evolutionary corrections to the obtained absolute magnitudes or to the calculation of $z_{\rm max}$ and $z_{\rm min}$ (see further below). Stellar masses are calculated by re-normalising the best-fitting model template SED to the observed one and allowing for mass loss (see \citealp{Maraston-1998} \& \citealp{Maraston-2005}), according the prescriptions of \citet{Renzini-1993}. 
The actual calculation is carried out using a routine developed by E. Daddi and C. Maraston and already used in used \citet{Daddi-2005,Maraston-2006} and \citet{Capozzi-2016}.

The calculation of $z_{\rm max}$ and $z_{\rm min}$ values depends on the depth ($m_{\rm u}$) of the survey, its lower apparent magnitude limit ($m_{\rm l}$), the lower ($z_{\rm l}$) and upper ($z_{u}$) redshift limits of the survey 
and k-correction. Once the absolute magnitude $M^{j}$ ($j=${\it g,r,i,z,Y}) and stellar age of a given galaxy are determined, we calculate what would be the redshift values at which the apparent magnitude of such a galaxy, with the same properties, will equal  $m_{\rm u}$ and  $m_{\rm l}$ (the apparent magnitude limits of the survey).

We always make sure that the value of $z_{\rm max}$ is such that the age of the galaxy considered is lower than the age of the Universe at this redshift within the cosmological model used. If this is not the case, we then correct the value of $z_{\rm max}$ downward by the minimum redshift difference needed for the galaxy age to equal that of the Universe.  

\begin{figure*}
\centering
\includegraphics[width=0.4\textwidth]{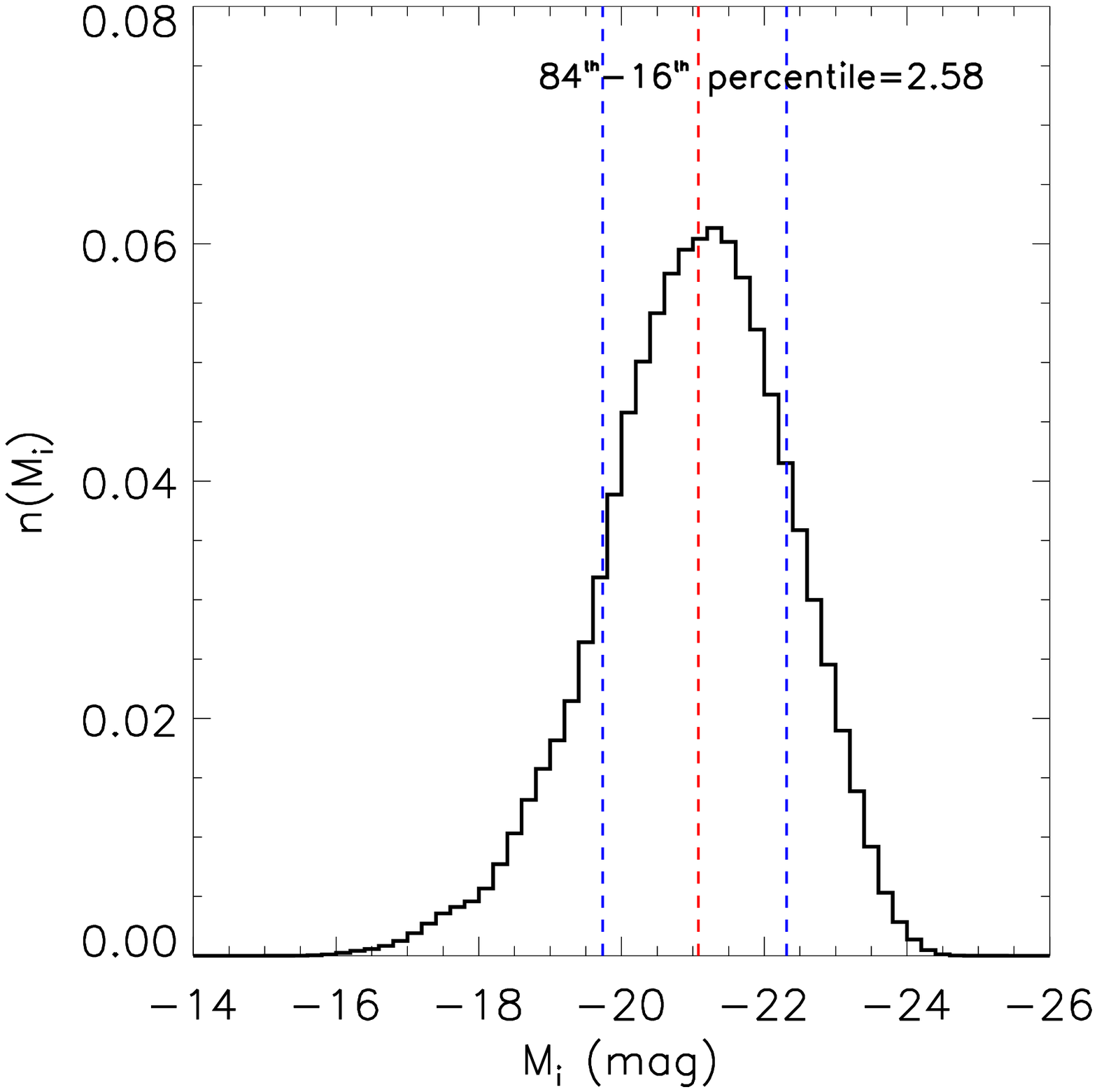}
\includegraphics[width=0.4\textwidth]{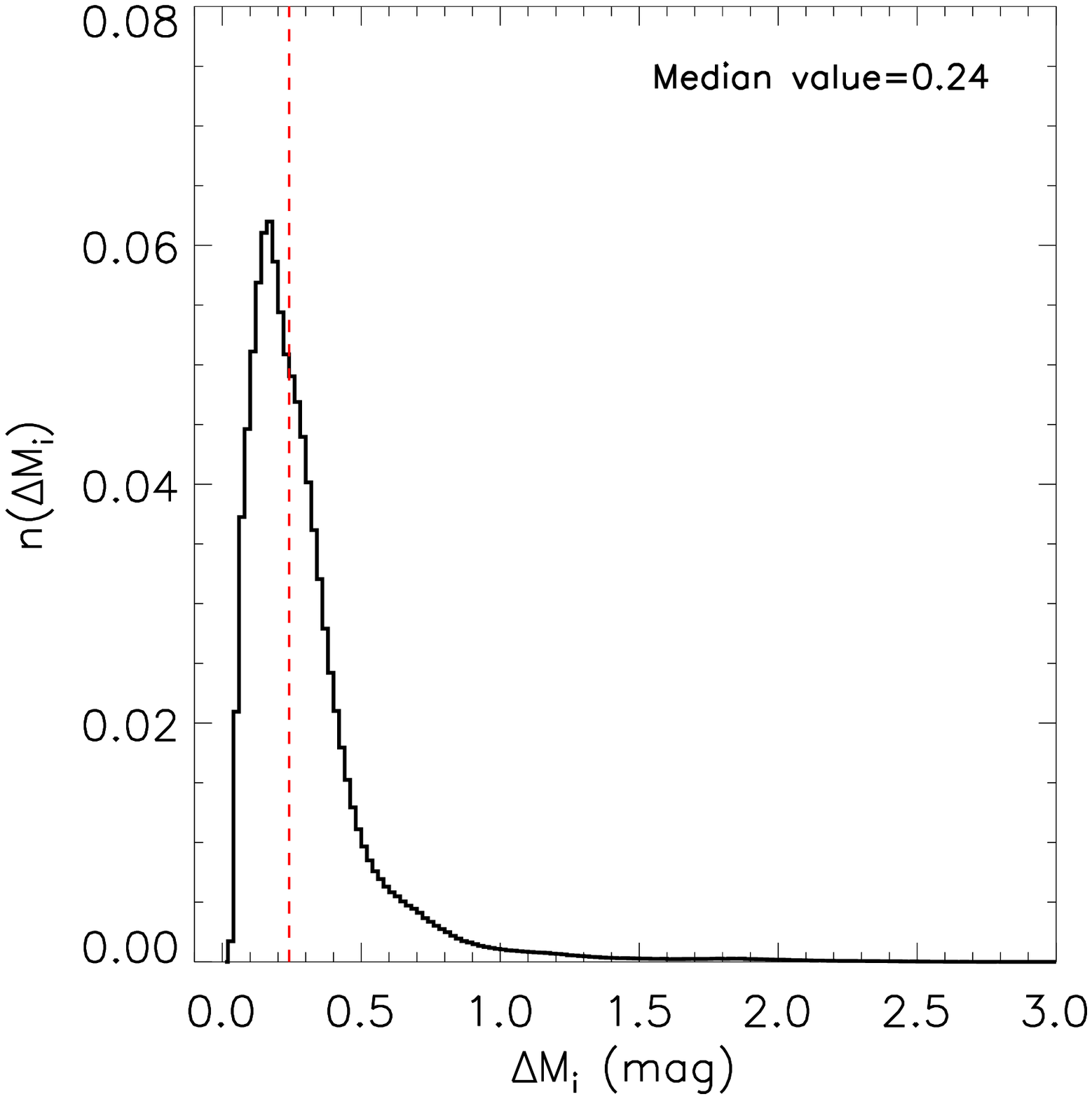}
\includegraphics[width=0.4\textwidth]{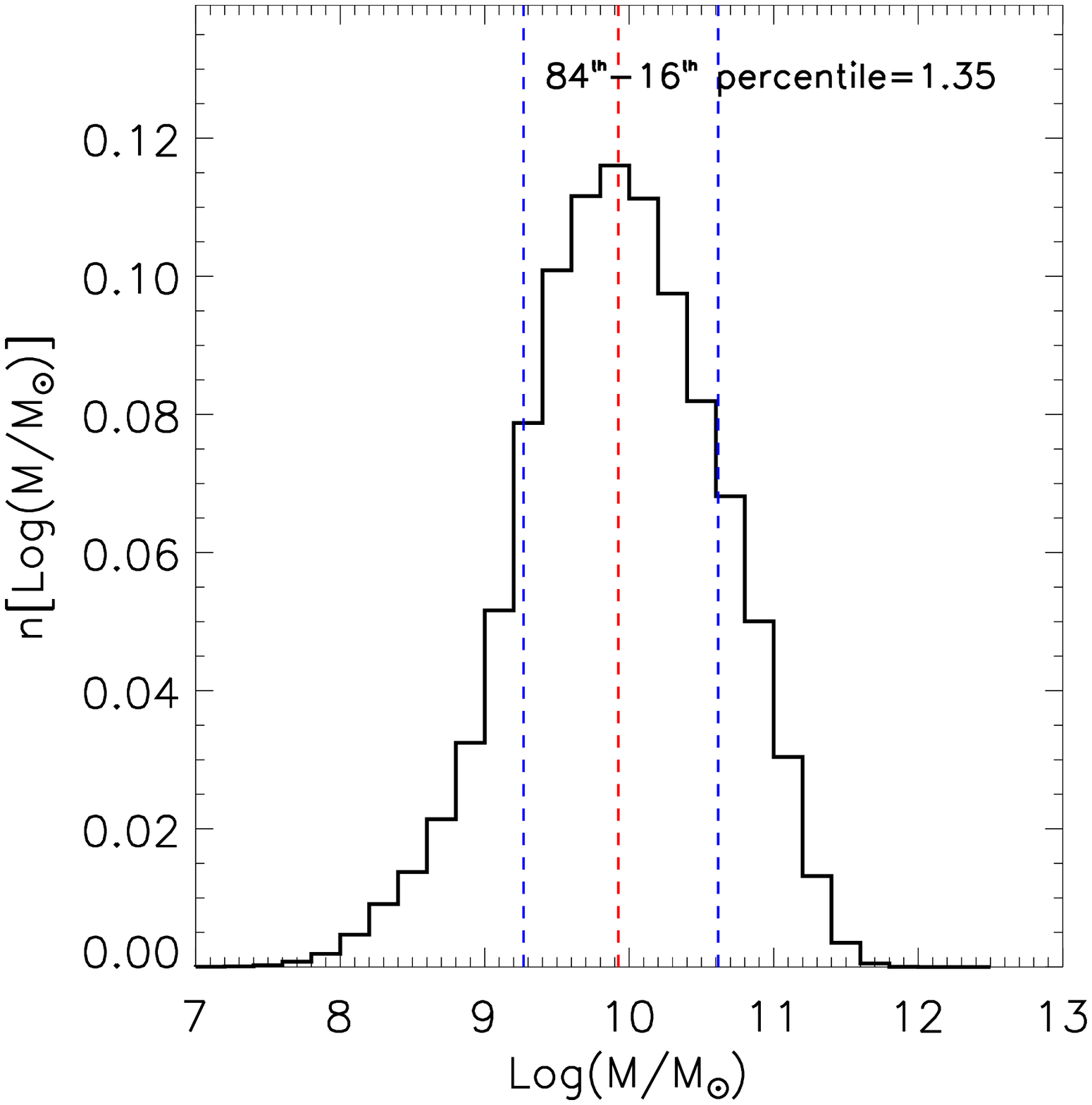}
\includegraphics[width=0.4\textwidth]{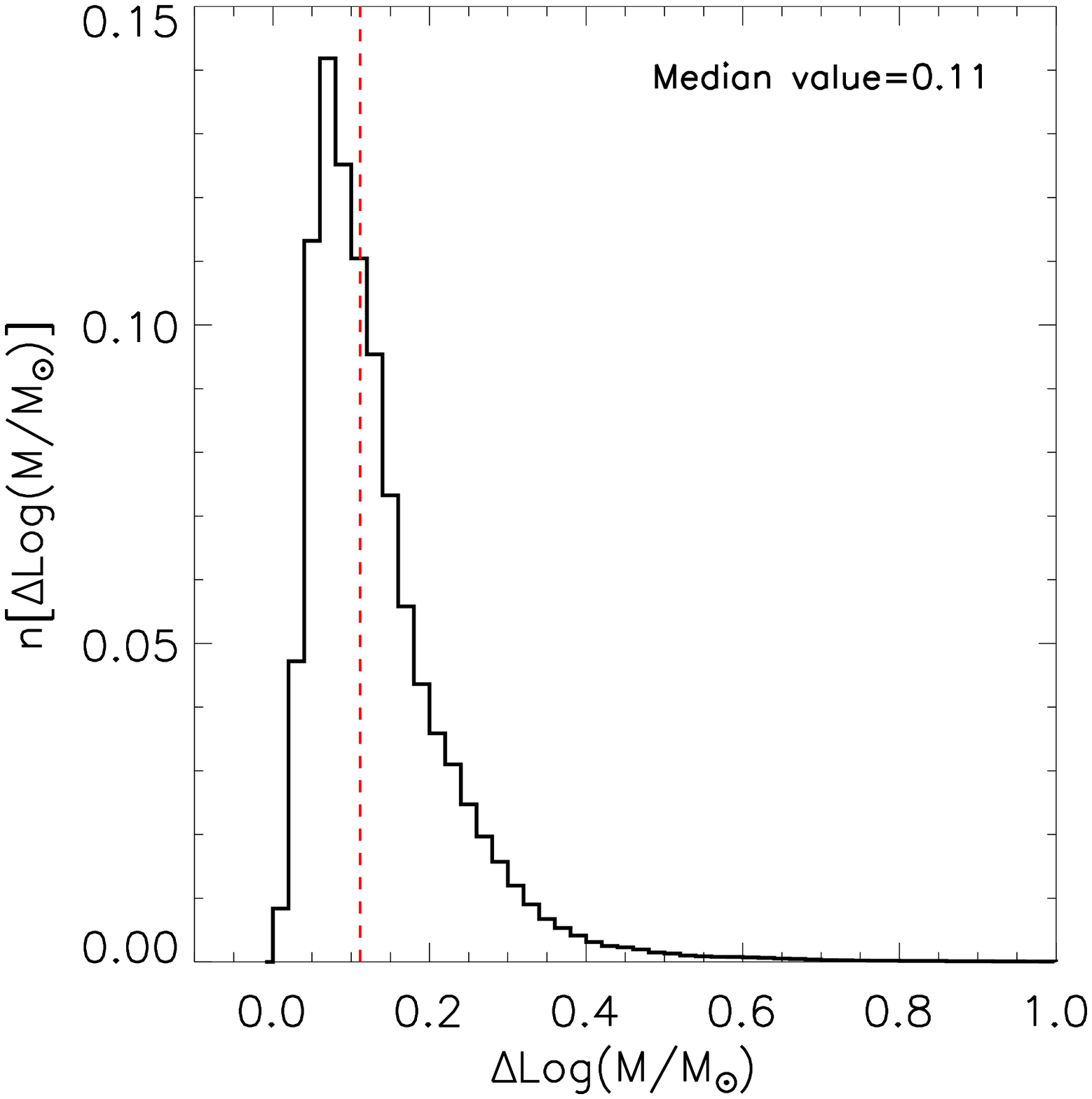}
\includegraphics[width=0.4\textwidth]{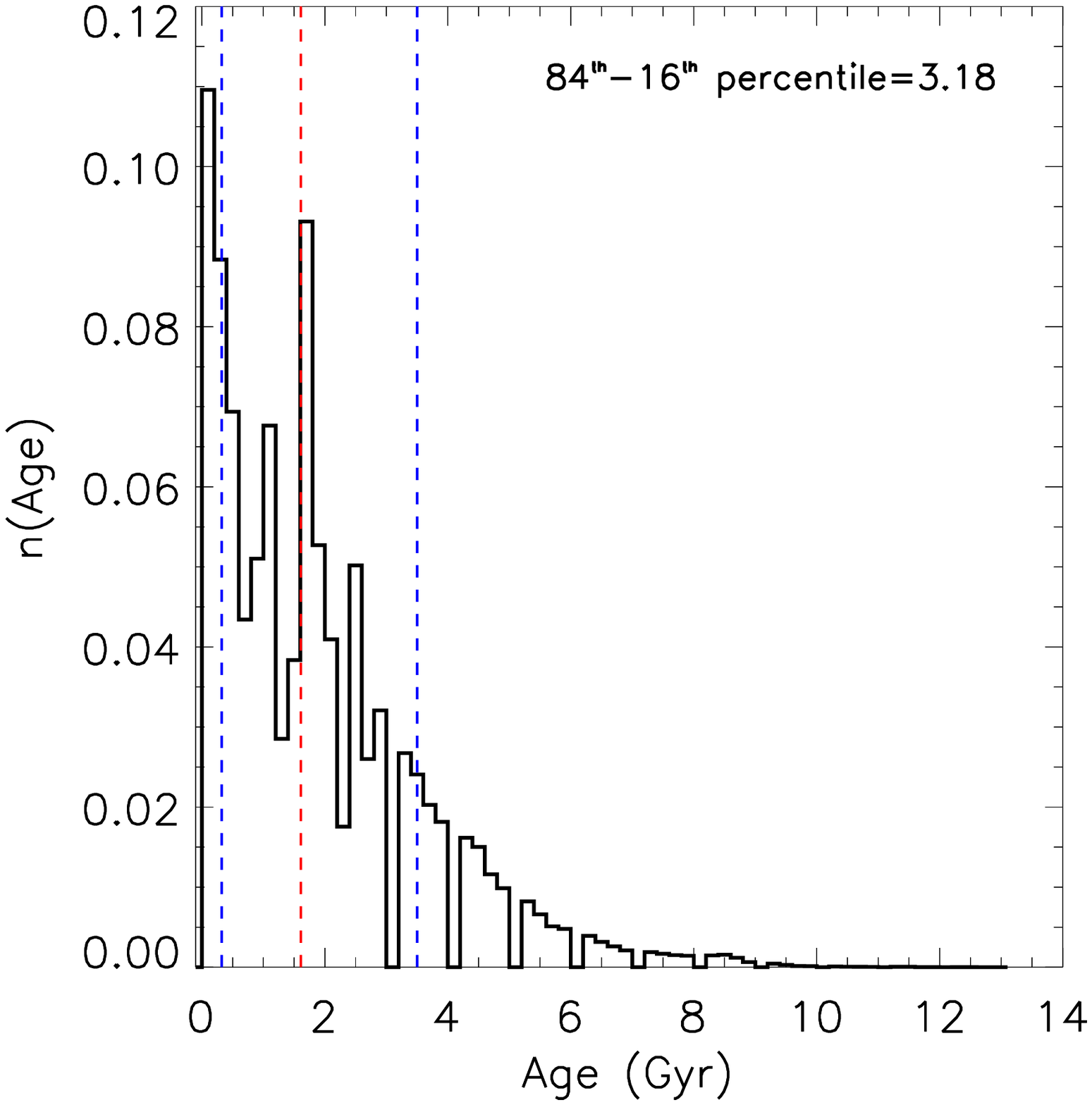}
\includegraphics[width=0.4\textwidth]{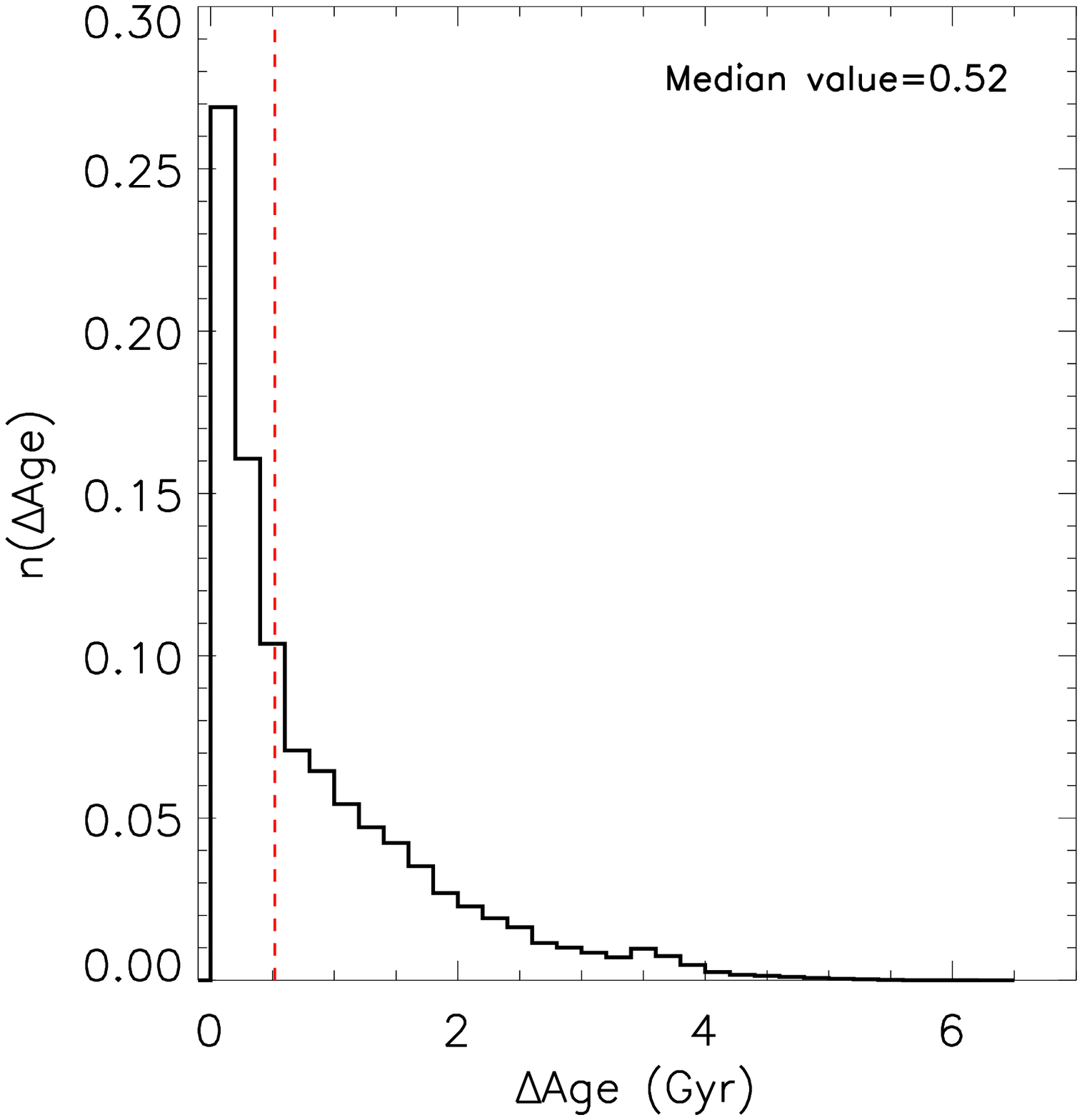}
\caption{Left-hand panels: median distributions of galaxy {\it i}-band absolute magnitudes (top) stellar masses (centre), stellar age (bottom). Right-hand panels: distributions of uncertainties on these quantities.}
\label{fig:Fig5}
\end{figure*}

As all galaxy properties will depend on uncertainties on photometric redshifts and photometry, we use our Monte Carlo simulations described in Sections \ref{subsubsec:subsubsec2.1.2} and \ref{subsec:subsec2.2} to estimate their uncertainties. Figure \ref{fig:Fig5} shows the median distributions of {\it i}-band absolute magnitude, logarithm of stellar mass and stellar age values (left-hand panels) and their uncertainties (right-hand panels). The latter are defined as half the difference between the $84^{\rm th}$ and $16^{\rm th}$ percentiles of each galaxy property (absolute magnitude, stellar mass and age) distributions. The range and median uncertainties are respectively 2.6 and 0.2 {\rm mag} for absolute magnitudes, 1.3 and 0.1 {\rm dex} for stellar masses and 3.2 and 0.5 {\rm Gyr} for stellar age. As the median uncertainty due to photometric redshifts uncertainty and photometry is an order of magnitude smaller than the property range for both absolute magnitudes and stellar masses, these properties can be safely studied. Note that we find the median uncertainty on the logarithm of stellar mass to be lower than that on absolute magnitude. This is in agreement with the finding of \citet{Taylor-2009}, who found error bars due to photometric redshift uncertainties on absolute magnitude and the logarithm of stellar mass to be respectively $\sim 0.3\ {\rm mag}$ and $\sim 0.1\ {\rm dex}$, showing that the effect of photometric redshift uncertainties is significantly larger for absolute magnitudes than for stellar masses.

More details on the study of the dependence of property uncertainties with redshifts, also in connection with environment, can be found in the companion paper by \citet{Etherington-2017}, which also uses the COMMODORE catalogue.

%\addtocounter{figure}{-1}

\subsection{Sample physical-property completeness as a function of redshift}
\label{subsec:subsec3.3}
When studying the luminosity and the stellar mass functions of a galaxy sample, it is important to estimate the completeness in absolute magnitude and stellar mass characterising the sample as a function of redshift, in addition to that in observed quantities like flux and surface brightness (see Section \ref{subsubsec:subsubsec2.2.2}). 
In order to do so, we estimate the higher and lower $\gtrsim 90$ per cent completeness limits in absolute magnitude (respectively $M_{\rm faint}^{\rm j}$ and $M_{\rm bright}^{\rm j}$) and stellar mass (respectively $M_{\rm *, high}$ and $M_{\rm *, low}$) of our sample as a function of redshift.

\begin{figure*}
\centering
\includegraphics[width=0.49\textwidth]{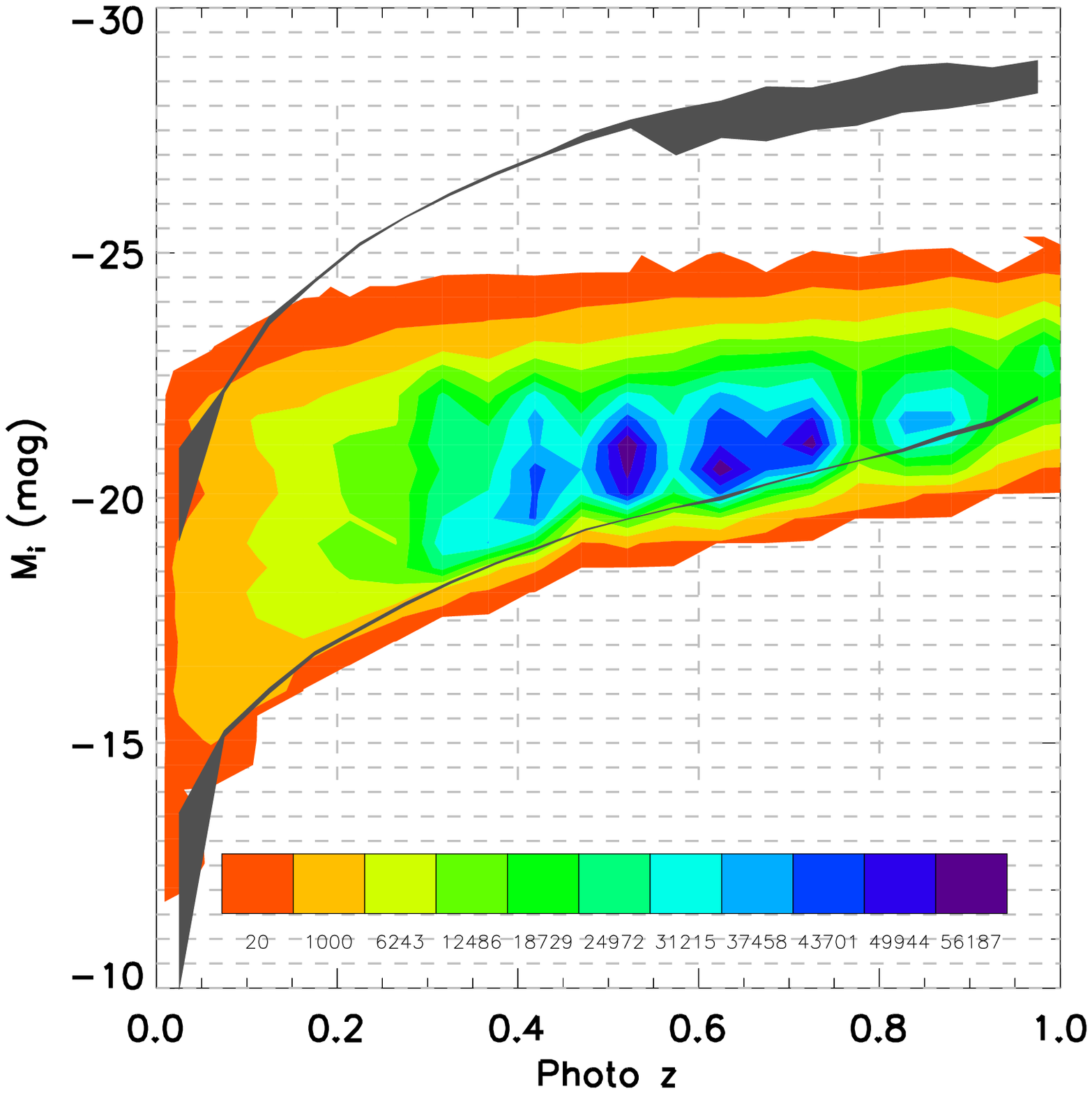}
\includegraphics[width=0.49\textwidth]{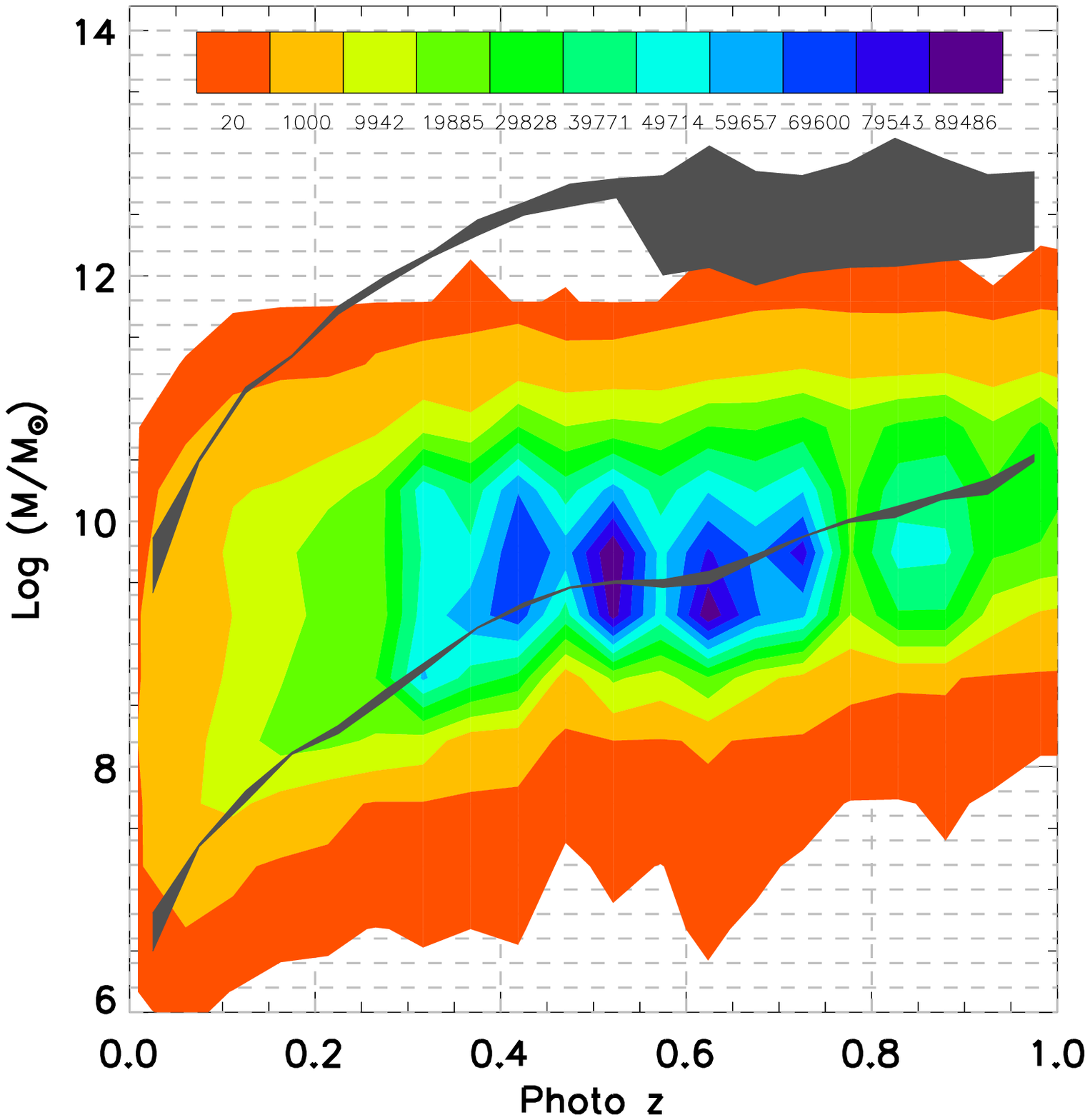}
\caption{Left-hand panel: TPZ photometric redshift vs.  {\it i} absolute magnitude. Right-hand panel: TPZ photometric redshift vs. stellar mass. In both density plots, the plotted values refer to one Monte Carlo realisation. Grey shaded regions identify the variation of the $\gtrsim 90$ per cent luminosity/stellar mass completeness with redshift at the bright/high-mass end and at the faint/low-mass end of our 100 Monte Carlo simulations (see Section \ref{subsec:subsec3.3}).} 
\label{fig:Fig6}
\end{figure*}

Such limits are shown in Figure \ref{fig:Fig6} (left panel for {\it i}-band absolute magnitudes, right panel for stellar masses). We will use these limits to identify absolute-magnitude and stellar-mass intervals complete at $\gtrsim 90$ per cent level to be used for evolutionary studies (see Sections \ref{subsubsec:subsubsec5.1.2} and  \ref{subsubsec:subsubsec5.2.2}) and to exclude incomplete $z-M_{\rm i}$ and  $z-M_{\rm *}$ 2-dimensional bins when calculating GLFs and GSMFs in our redshift bins.

Obviously, the completeness limits identified depend on the individual galaxy photometric redshifts. As with other properties, the variation of these limits due to photometric redshifts' uncertainties is estimated via our Monte Carlo simulations. %In this way we are able to investigate the range spanned by the completeness limits just described and to take this variation into account. 
In Figure \ref{fig:Fig6} we show one realisation of the estimate of galaxy properties for the COMMODORE sample together with absolute magnitude (left-hand panel) and stellar mass (right-hand panel) completenesses as function of redshift. The latter are both identified as shaded regions, representing their total variation when taking into account all the 100 Monte Carlo realisations of the COMMODORE sample, as we calculate such completenesses for each of them in order to be able to properly carry out our 100 measurements of GLF and GSMF.

\section{Galaxy luminosity and stellar mass functions}
\label{sec:sec4}

\subsection{The method}
\label{subsec:subsec4.1}
To estimate DES GLF (GSMF), we use the classic Schmidt-Eales ($1/V_{\rm max}$, \citealp{Schmidt-1968}) method, which consists of determining the largest volume accessible to each galaxy \citep{Schmidt-1968,Takeuchi-2000}, given the galaxy absolute magnitude, the apparent magnitude depth of the survey ($m_{\rm u}$), its lower apparent magnitude limit ($m_{\rm l}$) and the solid angle $\Omega$ corresponding to the analysed sky area. We now briefly explain the method used only for the case of the GLF. However, the GSMF can be derived in the same way, as the majority of the quantities needed for such a calculation are only flux dependent (e.g., $V_{\rm max}$). 

 The number density of galaxies within a considered absolute magnitude range ($M_{\rm l}<M<M_{\rm u}$) can be written as:

\begin{eqnarray}
\label{eq:eq6}
%\int_{M_{l}}^{M_{u}} \! \phi(M) \, dM = \sum\limits_{s=1}^{N_{obs}} \frac{1}{V_{max}(s)}\, ,\\
\int_{M_{l}}^{M_{u}} \! \phi(M) \, dM = \sum\limits_{s=1}^{N_{obs}} \frac{1}{C_{\rm s}V_{max}(s)}\, ,\\
V_{max}(s)\equiv \int_{\Omega}\int_{z_{min,s}}^{z_{max,s}} \! \frac{d^{2} V}{d\Omega dz} \, dz d\Omega \, ,
\end{eqnarray}

\noindent where $z_{\rm max,s}$ and $z_{\rm min,s}$ are the upper and lower redshift limits within which a galaxy with absolute magnitude $M_{\rm s}$ can be detected in the survey (see Section \ref{subsec:subsec3.2}), $V_{\rm max}(s)$ being the maximum co-moving volume accessible to it and $C_{\rm s}$ is the galaxy completeness factor (see Section \ref{subsubsec:subsubsec2.2.2}), which takes into account incompleteness due to missed detection as a function of surface brightness and apparent magnitude. Note also that one has to take also into account that $z_{\rm l}\leq z_{\rm min,s} < z_{\rm max,s}\leq z_{\rm u}$, $z_{\rm l}$ and $z_{\rm u}$ being the redshift limits of the survey (or the redshift slice used for the calculation).
In this way we are able to sum over the individual corrected inverse accessible volumes for each magnitude bin and obtain an estimate of the galaxy volume number density for each of them. This process allows us to obtain a discrete estimate of the luminosity and the mass functions. However, if these functions are obtained in relatively small redshift intervals ($z_{\rm l}<z<z_{\rm u}$), one has to make sure to remove  absolute magnitude (stellar mass) incomplete bins, in order to avoid to underestimate the number density. Since we do estimate GLF/GSMF in bins of $z$, we make sure that this requirement is satisfied. We do this by using the functions plotted in Figure \ref{fig:Fig6} and calculating the absolute magnitude ($M^{\rm j}_{\rm bright}$ and $M^{\rm j}_{\rm faint}$) and stellar mass ($M_{\rm *, high}$ and $M_{\rm *, low}$) completeness limits at $z_{\rm u}$ and $z_{\rm l}$. Hence we make sure that within the considered redshift slice ($z_{\rm l}<z<z_{\rm u}$) only absolute magnitude/mass bins whose edges fall within these completeness limits are selected for the GLF/GSMF calculation (e.g., \citealp{Loveday-2012}).

Uncertainties on number densities are derived by taking into account both the contribution from shot noise and from uncertainties in photometry and photometric redshifts. The first contribution is taken care of by determining the 84.13 $\%$ confidence Poisson upper and lower limits according to the \citet{Gehrels-1986} recipe in presence of shot noise, so to properly account for the low-counts regime (bright/massive end of the GLF/GSMF). In fact, in first approximation, the GLF (GSMF) must follow a scaled Poisson distribution and the scaling factor within each magnitude bin can be determined, following the approach of \citet{Zhu-2009}, by introducing the effective weight $W_{\rm eff}$

\begin{equation}
\label{eq:eq16}
W_{eff}=\left[ \sum_{s} \frac{1}{(CV_{max})_{s}^{2}}\right] \Bigg / \left[ \sum_{s} \frac{1}{(CV_{max})_{s}}\right]\, ,
\end{equation}

which allows to write the effective number ($N_{\rm eff}$) of galaxies as

\begin{equation}
\label{eq:eq17}
N_{eff}=\left[\sum_{s} \frac{1}{(CV_{max})_{s}}\right] \Bigg / \{W_{eff}\}.
\end{equation}

The second contribution is estimated by our Monte Carlo simulations which allow us to repeat the entire analysis (SED fitting, galaxy properties derivation and derivation of GLF and GSMF) 100 times. This enables us to have number densities distributions for each investigated luminosity (stellar mass) bin on which we can identify $1\sigma$ confidence intervals. The two uncertainty contributions are kept separate due to their asymmetry.
 
 \begin{table*}
%\begin{tiny}
\begin{center}
\caption{{\bf Schechter function best fit values for the GLFs and GSMFs plotted respectively in Figures \ref{fig:Fig7} and \ref{fig:Fig10}}. A double Schechter function is only used for fitting the observed data points obtained in our lowest $z$-bin. For the remaining $z$ bins, a single Schechter function is used with $\alpha$ fixed at a value of $-1.2$.}
\begin{tabular}{llccccc}
 \hline
  \multicolumn{1}{l}{\bf Function} &
  \multicolumn{1}{l}{$\mathbf{z}$ {\bf bin}} &
  \multicolumn{1}{c}{$\mathbf{\alpha_{1}}$} &
  \multicolumn{1}{c}{$\mathbf{\alpha_{2}}$ } &
  \multicolumn{1}{c}{$\mathbf{\phi_{1}^{\ast}}$ } &
  \multicolumn{1}{c}{$\mathbf{\phi_{2}^{\ast}}$ } &
  \multicolumn{1}{c}{$\mathbf{M^{\ast}}$ } \\
  \multicolumn{1}{l}{} &
  \multicolumn{1}{l}{} &
  \multicolumn{1}{c}{} &
  \multicolumn{1}{c}{} &
  \multicolumn{1}{c}{$\mathbf{[10^{-4} {\rm n\ Mpc^{-3}}}]$ } &
  \multicolumn{1}{c}{$\mathbf{[10^{-4} {\rm n\ Mpc^{-3}}}]$ } &
  \multicolumn{1}{c}{[$\mathbf{{\rm mag}}$ or $\mathbf{{\log{(M/M_{\odot})}}}$]} \\
\hline
        & $0<z<$0.2 & $-0.88\pm0.02$ &  $-1.59\pm0.08$ & $39.4\pm0.5$ & $1.0\pm0.5$ & $-21.63\pm0.02$ \\
        & $0.2<z<$0.4 & $-1.2$ (fix.) &  $--$ & $31.5\pm0.2$ & $--$ & $-22.049\pm0.005$ \\
GLF & $0.4<z<$0.6 & $-1.2$ (fix.) &  $--$ & $22.3\pm0.1$ & $--$ & $-22.258\pm0.004$ \\
        & $0.6<z<$0.8 & $-1.2$ (fix.) &  $--$ & $13.50\pm0.04$ & $--$ & $-22.586\pm0.003$ \\
        & $0.8<z<$1.0 & $-1.2$ (fix.) &  $--$ & $8.23\pm0.03$ & $--$ & $-22.880\pm0.003$ \\
\hline
           & $0<z<$0.2 & $-0.92\pm0.02$ &  $-1.68\pm0.06$ & $10.7\pm0.2$ & $0.17\pm0.07$ & $10.72\pm0.01$ \\
           & $0.2<z<$0.4 & $-1.2$ (fix.) &  $--$ & $5.68\pm0.03$ & $--$ & $10.896\pm0.002$ \\
GSMF & $0.4<z<$0.6 & $-1.2$ (fix.) &  $--$ & $4.38\pm0.01$ & $--$ & $10.912\pm0.002$ \\
           & $0.6<z<$0.8 & $-1.2$ (fix.) &  $--$ & $1.94\pm0.01$ & $--$ & $11.129\pm0.002$ \\
           & $0.8<z<$1.0 & $-1.2$ (fix.) &  $--$ & $1.49\pm0.01$ & $--$ & $11.116\pm0.002$ \\
\hline
\end{tabular}
\label{tab:Table2}							    
\end{center}
%  \end{tiny}
\end{table*}

\section{Results}
\label{sec:sec5}
In this section we describe the results obtained for both the GLF and the GSMF. %However, we focus our attention more on the GSMF as it is more suited for studying galaxy mass build up, one of the main aims of this paper. 
We point out that, as a result of our 100 Monte Carlo realisations, when measuring the GLF and GSMF, in each absolute magnitude and stellar mass bin we will refer to the median value of the obtained number density distributions.

\begin{figure*}
\centering
\includegraphics[width=0.4\textwidth]{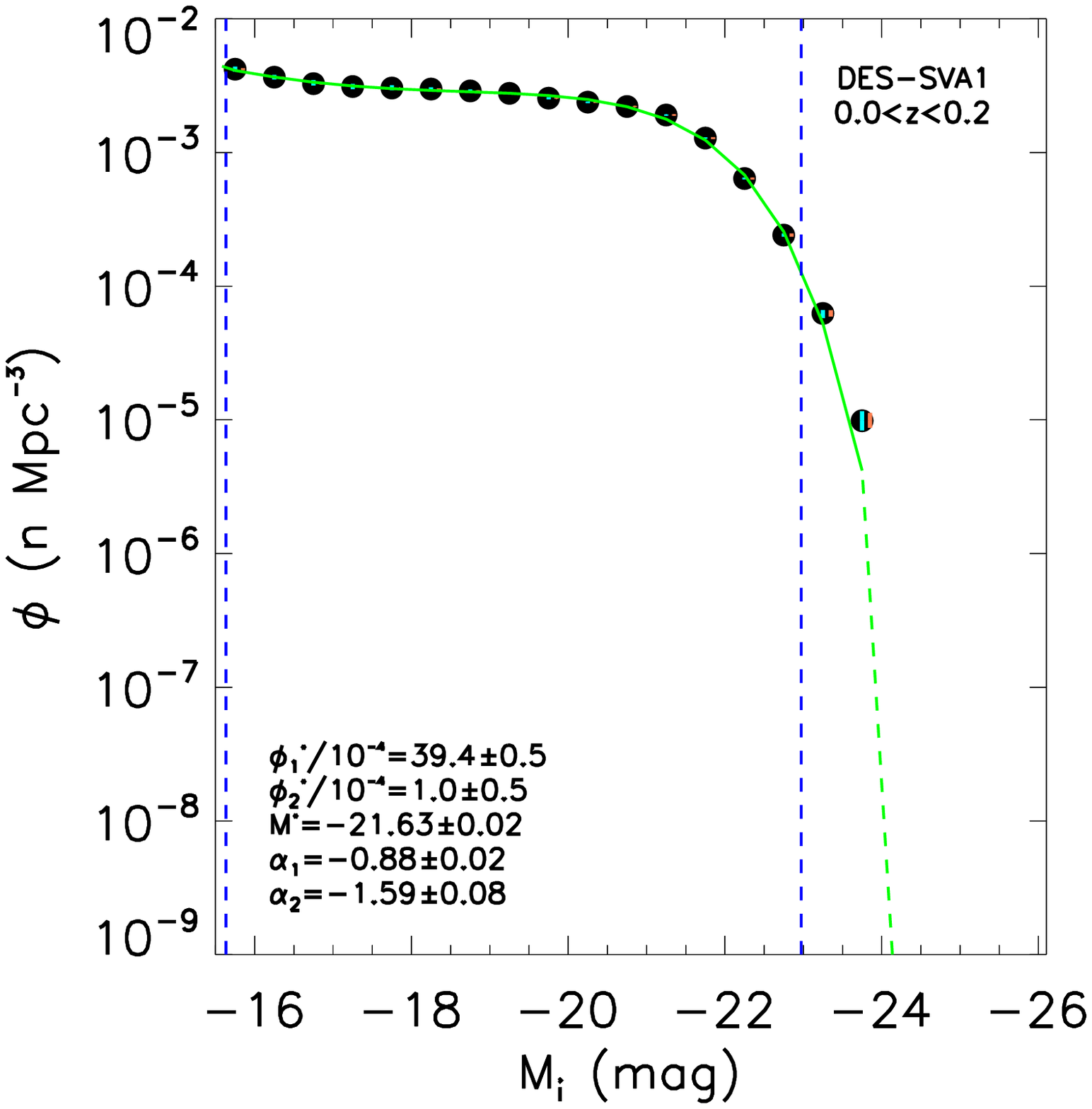}
\includegraphics[width=0.4\textwidth]{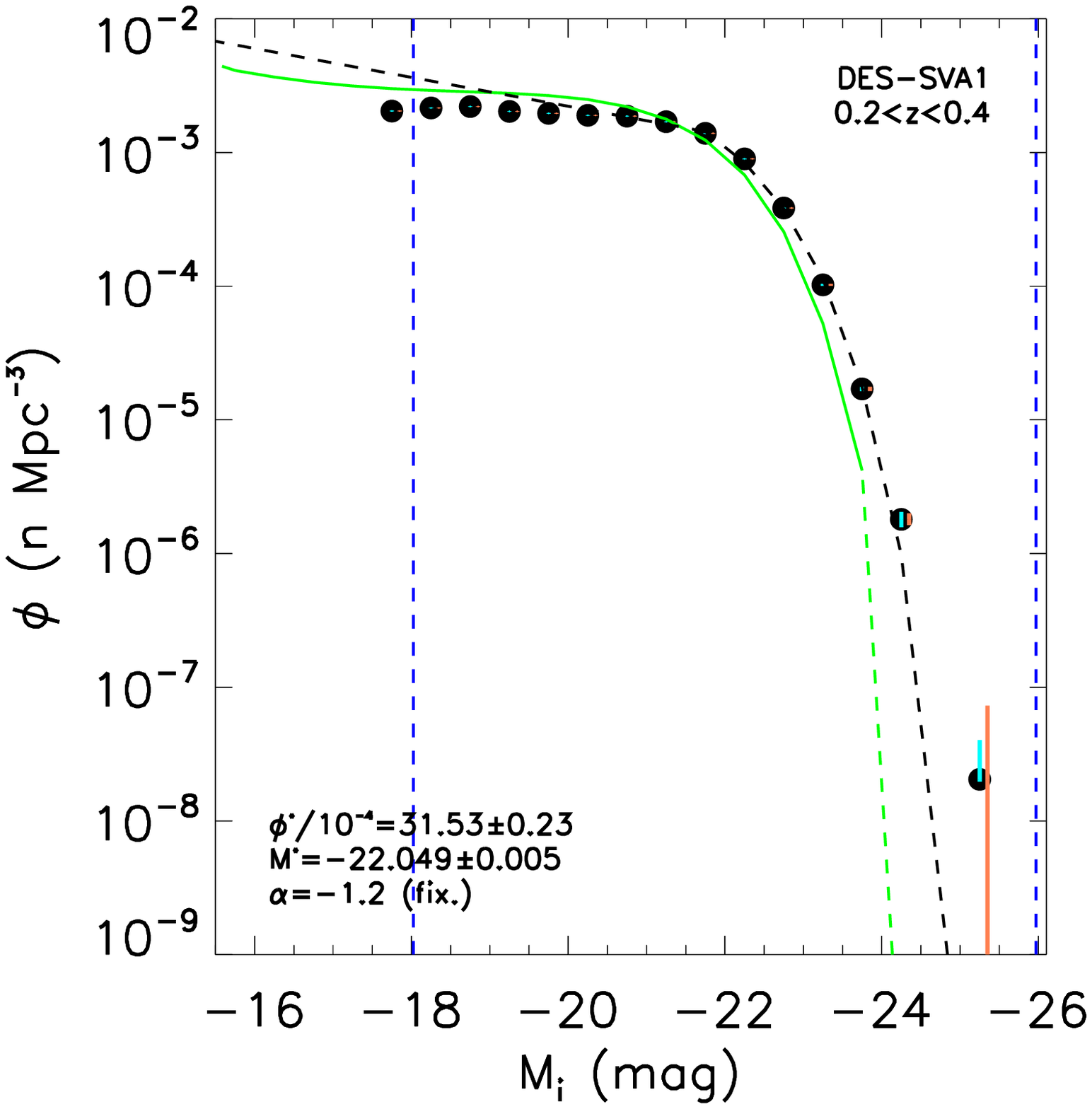}
\includegraphics[width=0.4\textwidth]{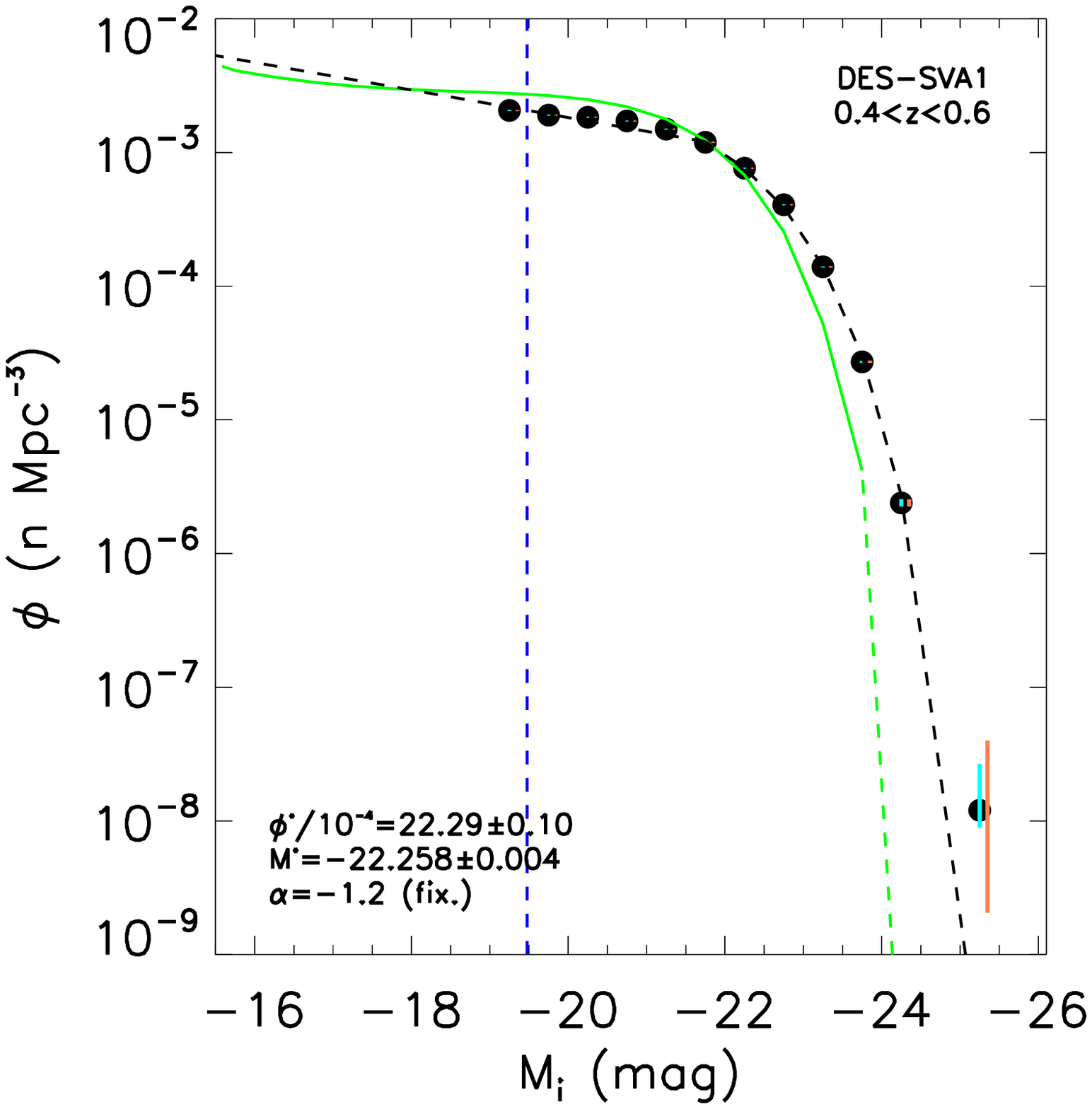}
\includegraphics[width=0.4\textwidth]{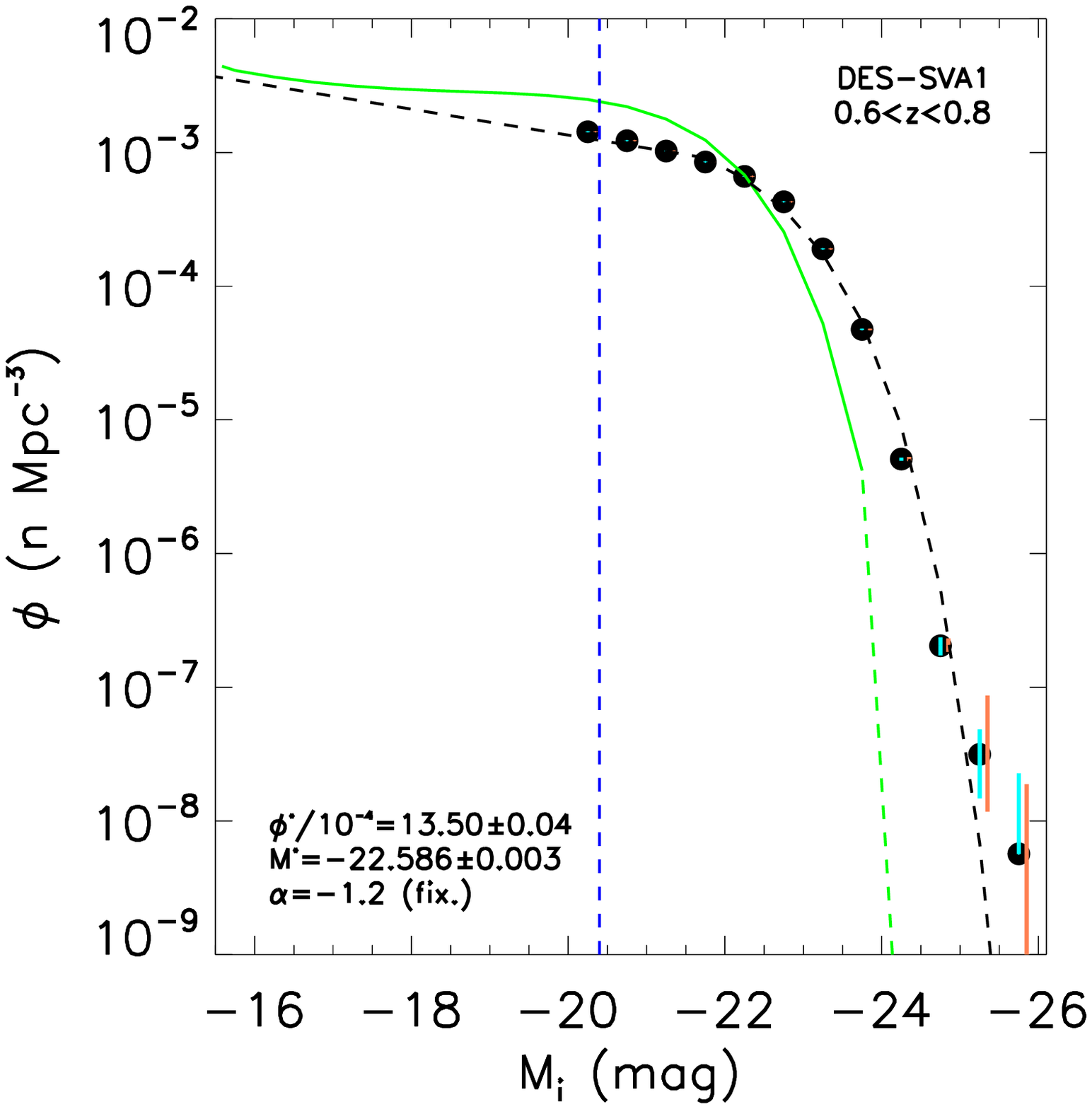}
\includegraphics[width=0.4\textwidth]{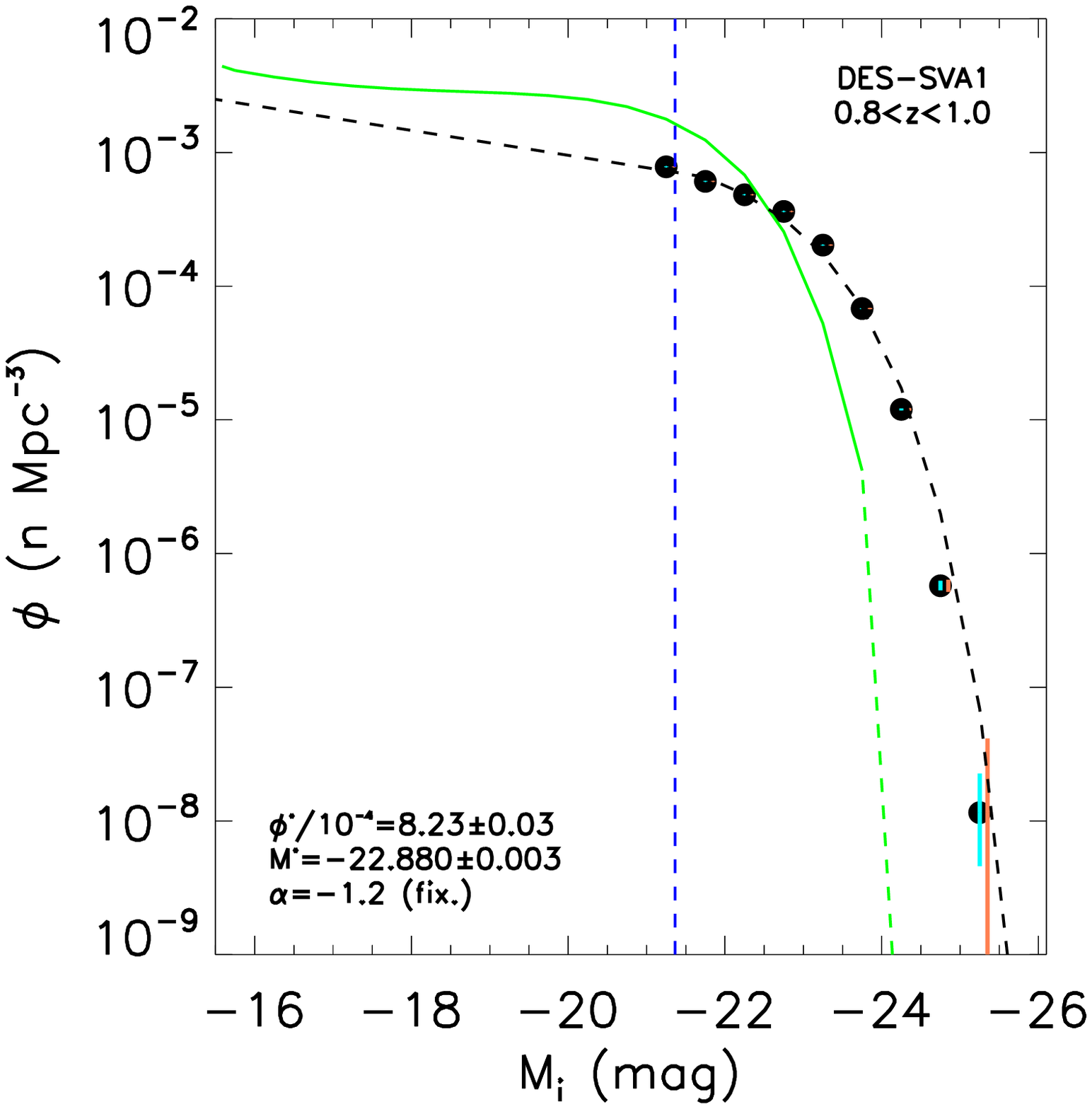}
\caption{Luminosity function in {\it i} band in five redshift bins (see Table \ref{tab:TableA1}). Cyan and coral error bars correspond respectively to Monte-Carlo-simulations (due to uncertainties on photometry and photometric redshifts) and shot-noise uncertainties. The green line stands for the best double-Schechter function fit to the lowest-$z$ data points measured by us. Black dashed lines stand for the best single-Schechter function fits to the $z>0.2$ data points with $-26<M_{\rm i}<-21.4\ {\rm mag}$. In each panel, the best fit parameters are displayed with their uncertainties (see also Table \ref{tab:Table2}). Blue dashed lines represent completeness limits.}
\label{fig:Fig7}
\end{figure*}

\subsection{GLF}
\label{subsec:subsec5.1}

In Figure \ref{fig:Fig7} we show the {\it i}-band GLF in 5 different $\Delta z=0.2$-sized redshift bins in a range between 0 and 1 (see Table \ref{tab:TableA1}).  

As a density dip and upturn at the faint end are evident, a double Schechter function is fitted to the data obtained in the lowest $z$ bin. This fit (green line in the figure) is then also plotted in the higher $z$ bin panels as a reference for highlighting redshift evolution. Such double Schechter function is defined as:

\begin{multline}
%\begin{split}
\phi(M)dM=0.4\ln(10)\left [ \left (\phi_{1}^{\ast} 10^{0.4(\alpha_{1}+1)(M^{\star}-M)} \right ) \right. + \\
 \left. \left (\phi_{2}^{\ast} 10^{0.4(\alpha_{2}+1)(M^{\star}-M)} \right ) \right ] e^{-10^{0.4(M^{\ast}-M)}} dM,
%\end{split}
\label{eq:eq18}
\end{multline}

\noindent where $M^{\ast}$ is the characteristic luminosity, $\alpha_{1}$ and $\alpha_{2}$ are the slopes of the two Schechter functions and $\phi_{1}^{\ast}$ and $\phi_{2}^{\ast}$ are their normalizations. The best fit values obtained are listed in Table \ref{tab:Table2}. 

Focusing on the lowest redshift bin, we notice the presence of a dip at $-20 \lesssim M_{\rm i}\lesssim -18 \ {\rm mag}$ followed by a new steep rise at $M_{\rm i}\gtrsim -18 \ {\rm mag}$, displaying the double-Schechter-like shape seen already at low redshift in the literature (e.g., \citealp{Blanton-2005, Montero-Dorta-2009, Loveday-2012}). This result is reassuring, as we use photometric redshifts, while the majority of the works in the literature are based on spectroscopic redshifts. 

Focusing on the bright end of the GLF, we notice that (as expected) the brightest ($M_{\rm i}\lesssim -23.5\ {\rm mag}$) and rarest galaxies make their entrance in the GLF at higher redshifts ($z>0.2$) only. This is probably due to lack of volume in the lowest redshift bin.

\subsubsection{Comparison with the literature}
\label{subsubsec:subsubsec5.1.1}
In this section we compare our results with those found in studies in the literature carried out on spectroscopic data in the local Universe. This is because we aim at understanding the reliability of our method based on photometric redshifts with respect to those based on spectroscopic ones. In particular, in our comparisons with other measurements of GLF (and of GSMF, see Section \ref{subsec:subsec5.2} ) in the literature, we consider the results obtained for the GAMA survey \citep{Baldry-2012} particularly relevant, as their measurements were carried out on a sky area ($143\ {\rm sq. \ deg}$) almost equivalent to the one used by us ($\sim 155\ {\rm sq. \ deg}$).  

\begin{figure}
\centering
\includegraphics[width=0.48\textwidth]{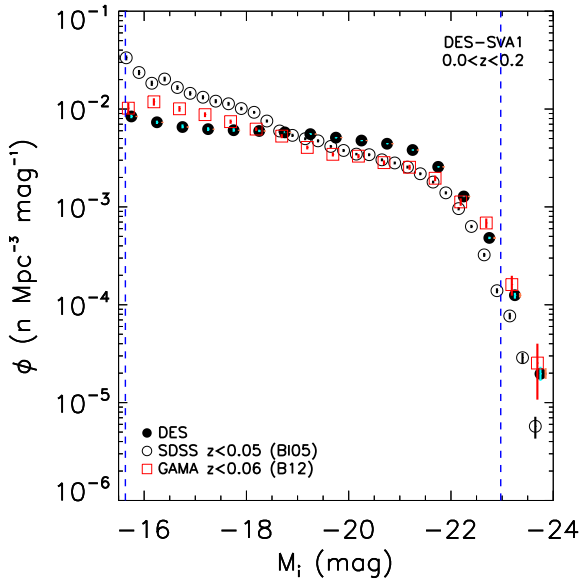}
\caption{Comparison of DES GLF at $z<0.2$ with results from spectroscopic surveys available in the literature. DES (black dots), GAMA (red squares, \citealp{Baldry-2012}) and SDSS (grey circles, \citealp{Blanton-2005}). Cyan and coral (slightly offset for improving visibility) error bars correspond respectively to Monte-Carlo-simulations (due to uncertainties on photometry and photometric redshifts) and shot-noise uncertainties. Dashed blue lines indicate DES completeness limits and error bars are as in Figure \ref{fig:Fig7}.}
\label{fig:Fig8}
\end{figure}

We show such comparison in Figure \ref{fig:Fig8}. In this figure, we compare our $z<0.2$ {\it i}-band GLF with those obtained by \citet{Blanton-2005} using SDSS Data-Release-2 (DR2) data (over an area of $\sim 2221\ {\rm sq. \ deg}$) at $z<0.05$ and by \citet{Baldry-2012} using GAMA data at $z<0.06$. In general we find good agreement with spectroscopic GLFs from the literature, despite some differences that are also present. Our and GAMA GLFs turn out to be similar, despite the different approaches used to estimate galaxy properties and for estimating the actual GLF (differently from us, \citealp{Baldry-2012} use a density-corrected $V_{\rm max}$ method). We do notice, however, that our GLF shifts to higher number densities at $-22.5 \lesssim M_{\rm i}\lesssim-19\ {\rm mag}$ and to lower ones at fainter magnitudes compared to the GAMA one. Both our and GAMA GLFs are similar to the SDSS one at $M_{\rm i}\lesssim-19\ {\rm mag}$. At fainter magnitudes though, they both show lower densities than those measured by \citet{Blanton-2005}. 
\citet{Baldry-2012} associate part of these discrepancies at $M_{\rm i}\gtrsim -18\ {\rm mag}$ (i.e., $L_{\rm i}\lesssim 10^{9}\ {\rm L_{\odot}}$) to the particular way of estimating distances, i.e. whether using distances corrected for large-scale bulk flows using redshifts in the cosmic microwave background frame rather than standard heliocentric-redshift based distances or those referred to the Local Group (LG) frame. \citet{Baldry-2012} did find the discrepancies with the results of \citet{Blanton-2005} to decrease when using the SDSS DR2 catalogue with flow-corrected distances as done with GAMA (see Figure 9b in \citealp{Baldry-2012}), however such discrepancies remained significant. Part of these discrepancies might be due to the method used to correct for surface-brightness incompleteness, which in \citet{Blanton-2005} is based on simulations, while in our case is based on data by relying on deeper DES fields. However, the source of discrepancy between the faint end of Blanton et al.'s GLF and those of DES and GAMA remains unclear to us and it would need to be further investigated.   
We notice that \citet{Baldry-2012} estimated the GAMA GLF at $z<0.1$ also using Loveday et al.'s galaxy sample and method (step-wise maximum likelihood) but k-correcting galaxies at $z=0$ (\citealp{Loveday-2012} k-corrected galaxies to $z=0.1$ as they shifted their {\it i}-band filter to such $z$) and found good agreement with their estimate of the GAMA GLF, which is in turn in good agreement with the DES one.  

\begin{figure*}
\centering
\includegraphics[width=0.33\textwidth]{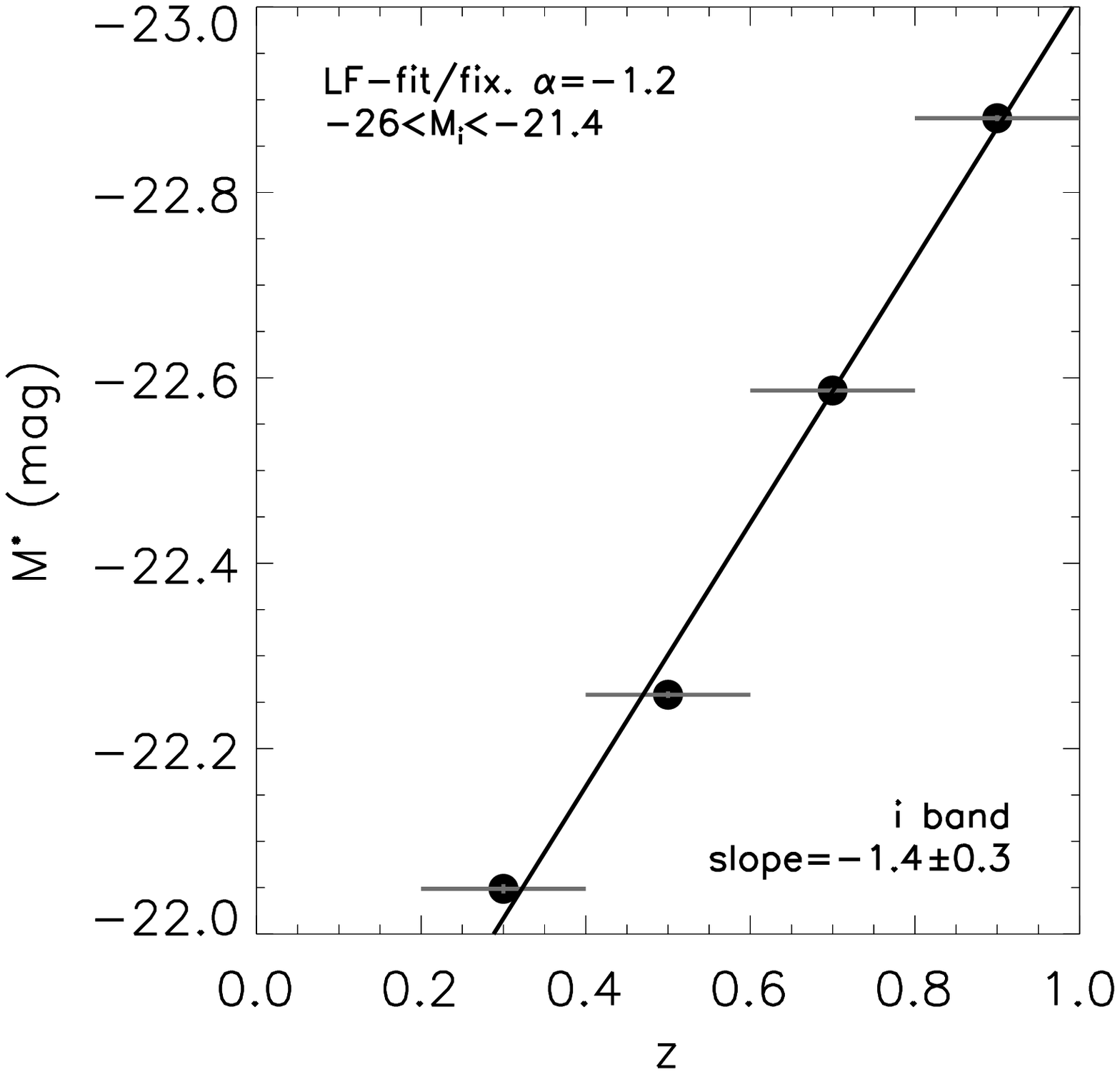}
\includegraphics[width=0.33\textwidth]{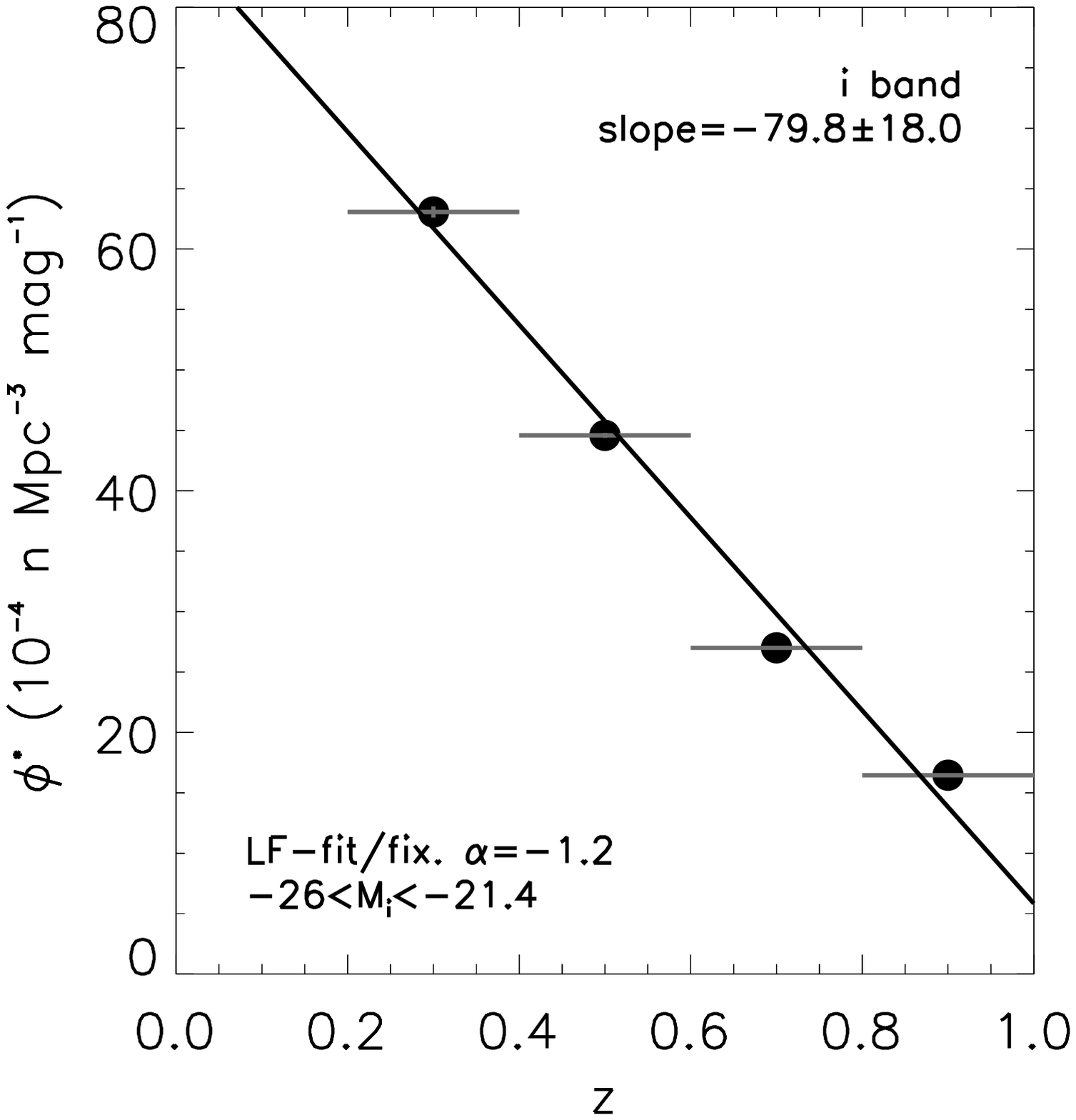}
\includegraphics[width=0.33\textwidth]{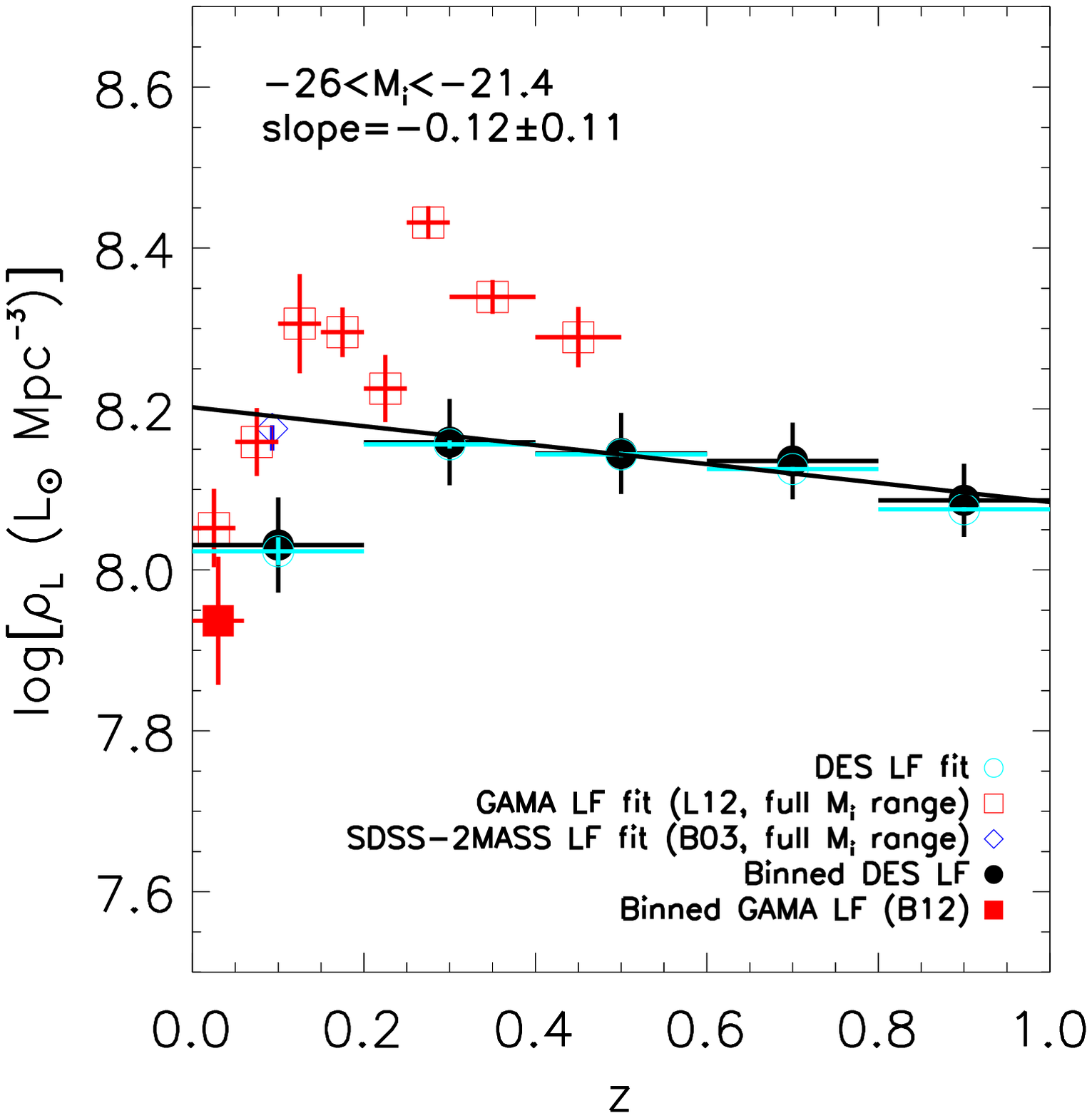}
\caption{Evolution with $z$ of best-fit parameters ($M^{*}$ and $\phi^{*}$) as calculated via the single-Schechter function fit to the $z>0.2$ GLFs and of the luminosity density $\rho_{\rm L}$. {\bf Left-hand panel}: $M^{*}$ vs. $z$. {\bf Centre-panel}:  $\phi^{*}$ vs. $z$. {\bf Right-hand panel}: $\rho_{\rm L}$ vs. $z$. Here $\rho_{\rm L}$ is calculated via two methods, i.e., using the GLF fit (cyan circles) and using the binned GLF (black dots). In addition, the values of $\rho_{\rm L}$ calculated in the lowest $z$ bin are shown for reference. We also add values from the literature, measured from the GLF Schechter fit and/or from the binned function. When possible we adjust the measurement taken from the literature (or re-estimated by us on the binned GLF) to the same luminosity interval ($-26<M_{\rm i}<-21.4$) used in our study. In addition to DES data points we show values from \citet{Loveday-2012} (red open squares, note that these values are referred to the {\it i}-band filter shifted to $z=0.1$), \citet{Baldry-2012} (red filled square) and \citet{Bell-2003} (blue open diamond). In all panels, measurements refer to the luminosity interval $-26<M_{\rm i}<-21.4\ {\rm mag}$ and the black solid line represents the best fit to the data when taking into account uncertainties on both axes.}
\label{fig:Fig9}
\end{figure*}

\subsubsection{Evolution of GLF parameters and of luminosity density}
\label{subsubsec:subsubsec5.1.2}
We identify an absolute magnitude range complete at $\gtrsim 90$ per cent out to $z=1$. We exclude from this exercise the lowest redshift bin (at $z<0.2$) so that we can explore the evolution with cosmic time also of the brightest galaxies ($M_{\rm i}\lesssim -23.5\ {\rm mag}$), which are lacking at $z<0.2$. By doing so, such $\gtrsim 90$ per cent complete luminosity range is identified as $-26<M_{\rm i}<-21.4\ {\rm mag}$ over $0.2<z<1$. As this luminosity range is relatively bright for being able to properly explore the evolution of the faint-end of the GLF, we then fit the GLF measured over this $M_{\rm i}$-range for the four $z$ bins at $z>0.2$ with a single Schechter function and fix the faint-end slope at $\alpha=-1.2$ (fits results are shown in Figure \ref{fig:Fig7}). In this way we are able to homogeneously investigate the evolution of $M^{\ast}$ (left-hand panel of Fig. \ref{fig:Fig9}) and $\phi^{\ast}$ (middle panel of Fig. \ref{fig:Fig9}) with redshift. The chosen value of $\alpha$ is the one best fitting the GLFs of all the explored $z$ bins. However, using any value between $\alpha=-1$ and $\alpha=-1.2$ (a sensible value range for the global population of galaxies, see, e.g., \citealp{Loveday-2012, Stefanon-2013}) does not change the results shown in Figure \ref{fig:Fig9}.  As expected, we find significant evolution (at $>3\sigma$) of both $M^{\ast}$ and $\phi^{\ast}$ with redshift, with $M^{\ast}$ and $\phi^{\ast}$ respectively getting brighter (by $\sim0.9\ {\rm mag}$ from $z=0.2$ to $z=1$) and decreasing (by a factor $\sim4$ from $z=0.2$ to $z=1$) with $z$. 

We also study the evolution of the {\it i}-band luminosity density $\rho_{\rm L}$ with redshift, which we show in the right-hand panel of Figure \ref{fig:Fig9}. We estimate $\rho_{\rm L}$ via two methods: i)  by using the Schechter function fit, i.e. $\rho_{\rm L}=\int_{\rm L_{\rm low}}^{\rm L_{\rm up}} L \phi(L) dL=\phi^{\ast}L^{\ast}[\Gamma(\alpha+2,L_{\rm low}/L^{\ast})-\Gamma(\alpha+2,L_{\rm up}/L^{\ast})]$, where the functions $\Gamma(\alpha+2,L/L^{\ast})$ are upper incomplete gamma functions; ii) by summing over the binned GLF, i.e. $\rho_{\rm L}=\sum^{\rm j=n}_{\rm j=1} L_{\rm j}\phi(L_{\rm j})\Delta L_{\rm j}$, where $j$ is the luminosity bin number. The advantage of method ii) is its independence from the function used for fitting the binned data. However, both methods give consistent results. We find no statistically significant evolution of $\rho_{\rm L}$ between $z=0.2$ and $z=1$. For reference, in the right-hand panel of Figure \ref{fig:Fig9} we also show the value of $\rho_{\rm L}$ obtained for the GLF in our lowest redshift bin $z<0.2$ calculated with both of the methods described [method i) was carried out using the double-Schechter fit reported in Table \ref{tab:Table2}]. However, we remind the reader that the values of $\rho_{\rm L}$ shown at $z<0.2$ are affected by incompleteness due to lack of galaxies with $M_{\rm i}\lesssim-23.5 \ {\rm mag}$.

We also compare our results on $\rho_{\rm L}$ with those available in the literature. In addition to examining in contrast with the density measurements quoted in the studies we are comparing with, when GLFs are available, we also calculate such values ourselves in the same way as done with our data. This is done in order to be as conservative as possible in our comparison, using quantities measured as similarly as possible.  
Note that the same approach will be used in Sections \ref{subsubsec:subsubsec5.2.2} and \ref{subsubsec:subsubsec5.2.3} for GSMF, when studying the evolution of stellar mass ($\rho_{\rm Mstar}$) and number ($\rho_{\rm N}$) spatial densities with cosmic time in comparison with the literature.

In Figure \ref{fig:Fig9} we compare our $\rho_{\rm L}$ values with those derived for SDSS by \citet{Bell-2003} and for GAMA by \citet{Loveday-2012} and by us via \citet{Baldry-2012}'s binned GLF. Note that  the $\rho_{\rm L}$ values quoted by Loveday et al. and Bell et al. were measured over the entire GLF and that Loveday at al.'s measures refer to the {\it i}-band filter shifted to $z=0.1$. However, there is consistency within the uncertainties for the majority of the data points, excluding the two Loveday et al.'s measures at $z\sim 0.275$ and $z\sim 0.35$. Despite this, it is evident that Loveday et al.'s values are systematically higher than ours at comparable redshifts, which could be due to the different luminosity ranges used for estimating $\rho_{\rm L}$. 
Although we investigate luminosity density evolution only at $z>0.2$ and measure no such evolution out to $z=1$, when considering also our $z\sim0.1$ value of $\rho_{\rm L}$ (which is reported as reference but it has to be considered a lower limit as it is partially affected by incompleteness), we do see similar features between \citet{Loveday-2012} 's measurements and ours, i.e., an increase of  $\rho_{\rm L}$ out to $z\sim0.1/0.2$ followed by an average constancy of this quantity. Having said that, Loveday et al.'s values at $z\gtrsim0.2$ do seem quite sparse. The luminosity density in bands other than {\it i} (to our knowledge the evolution of $\rho_{\rm L}$ in this band has been poorly studied) has already been seen to be approximately constant out to $z\sim 1$ in the literature (e.g., see \citealp[and references therein]{Stefanon-2013} for measurements in rest-frame {\it J} and {\it H} obtained using IRAC channels). However, other studies also report an increase of the {\it B}-band $\rho_{\rm L}$ with redshift (e.g., \citealp[and references therein]{Beare-2015}).
As the luminosity density is strictly bound to the dominant stellar population, which can vary at different wavebands, a dependence of the evolution of $\rho_{\rm L}$ with cosmic time on waveband is plausible.

\subsection{GSMF}
\label{subsec:subsec5.2}
\begin{figure*}
\centering
\includegraphics[width=0.4\textwidth]{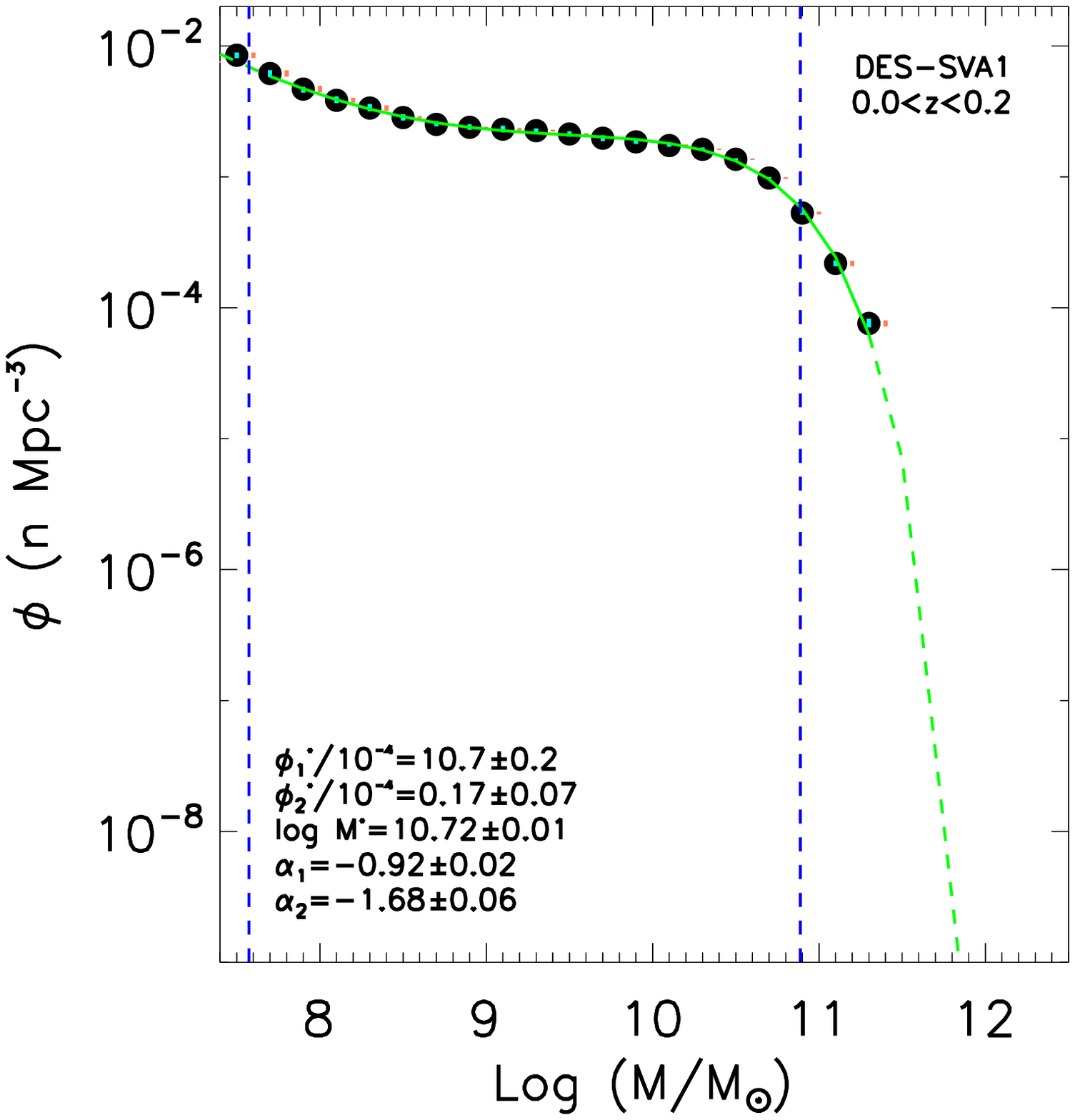}
\includegraphics[width=0.4\textwidth]{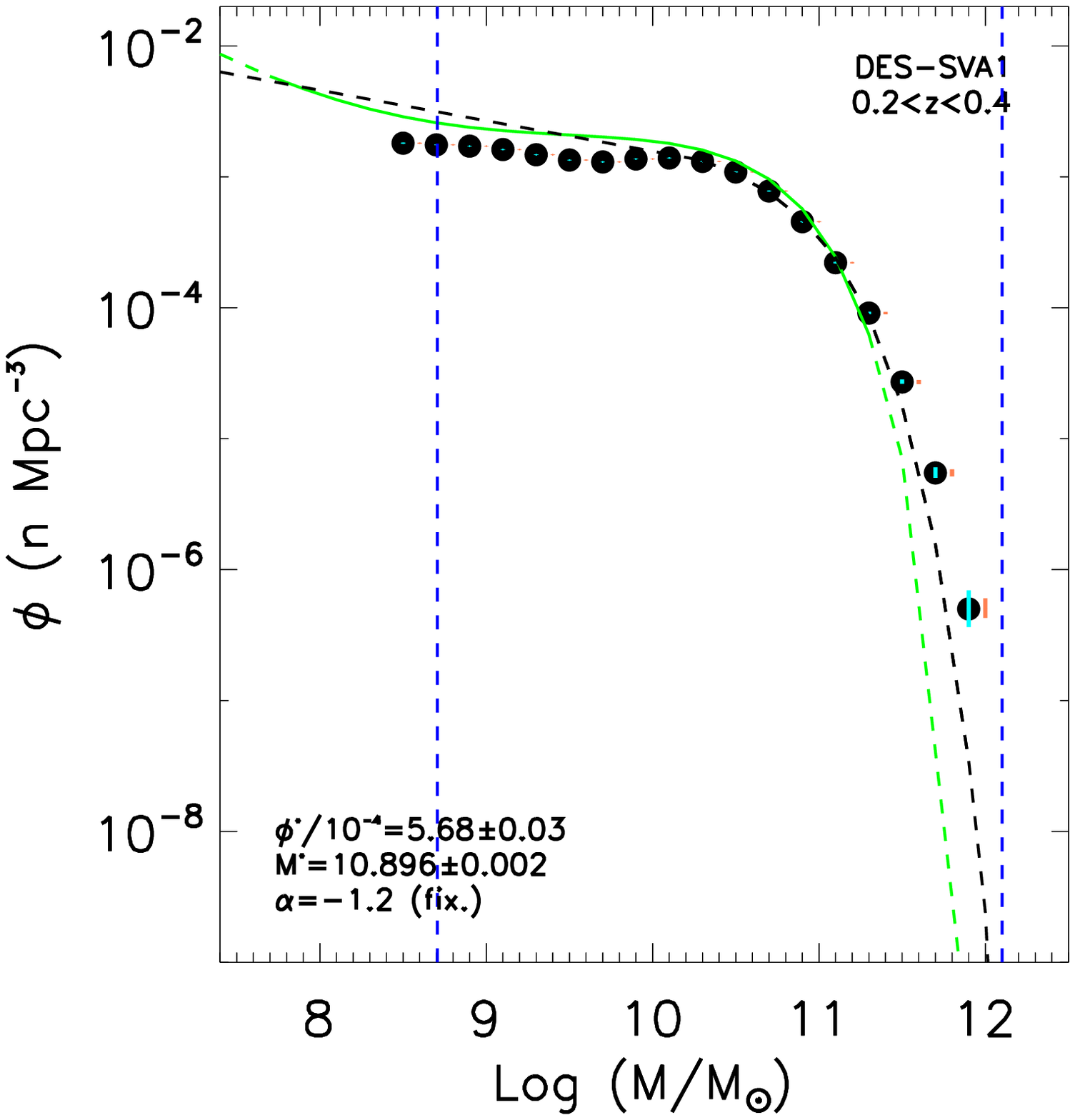}
\includegraphics[width=0.4\textwidth]{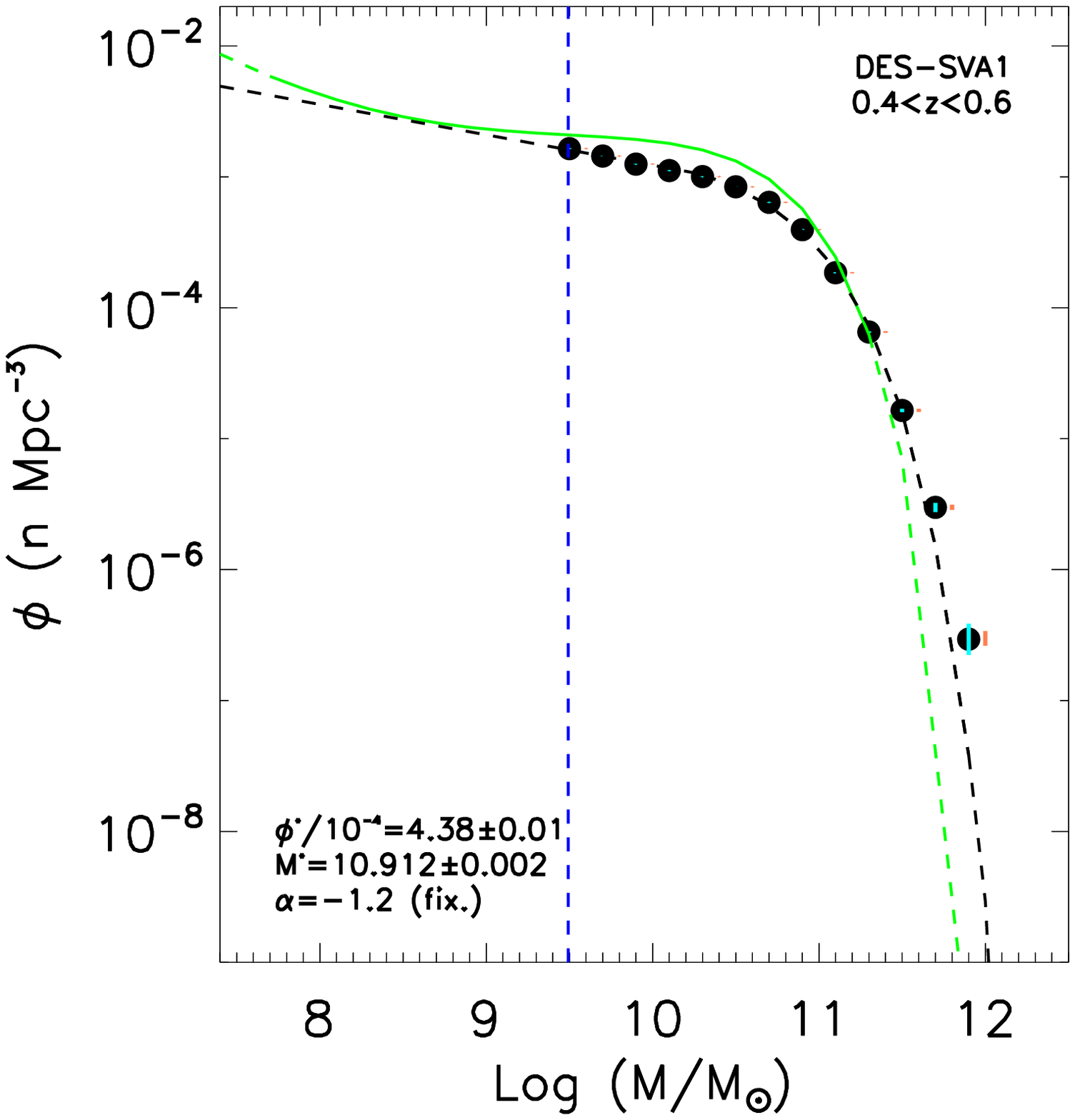}
\includegraphics[width=0.4\textwidth]{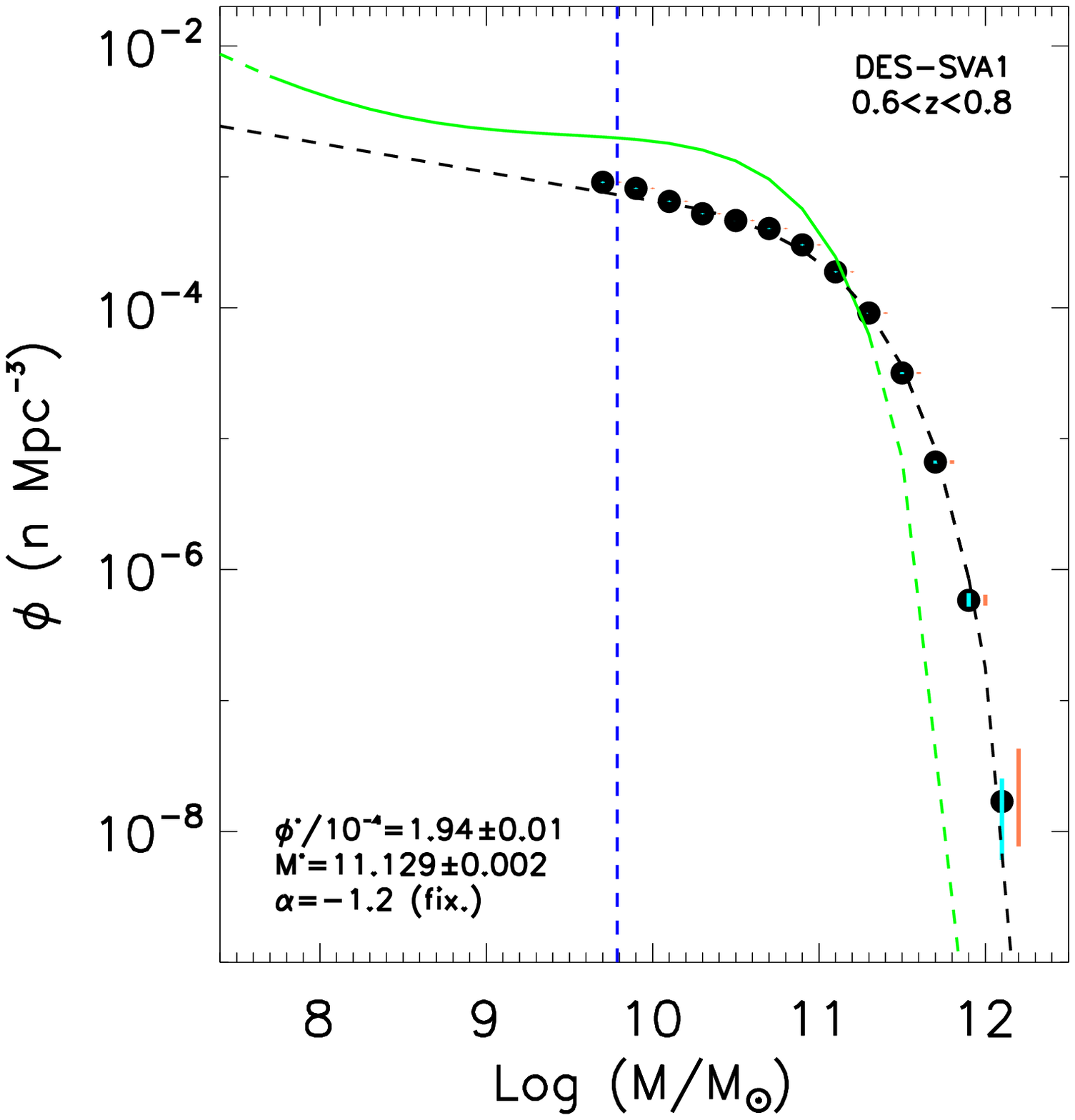}
\includegraphics[width=0.4\textwidth]{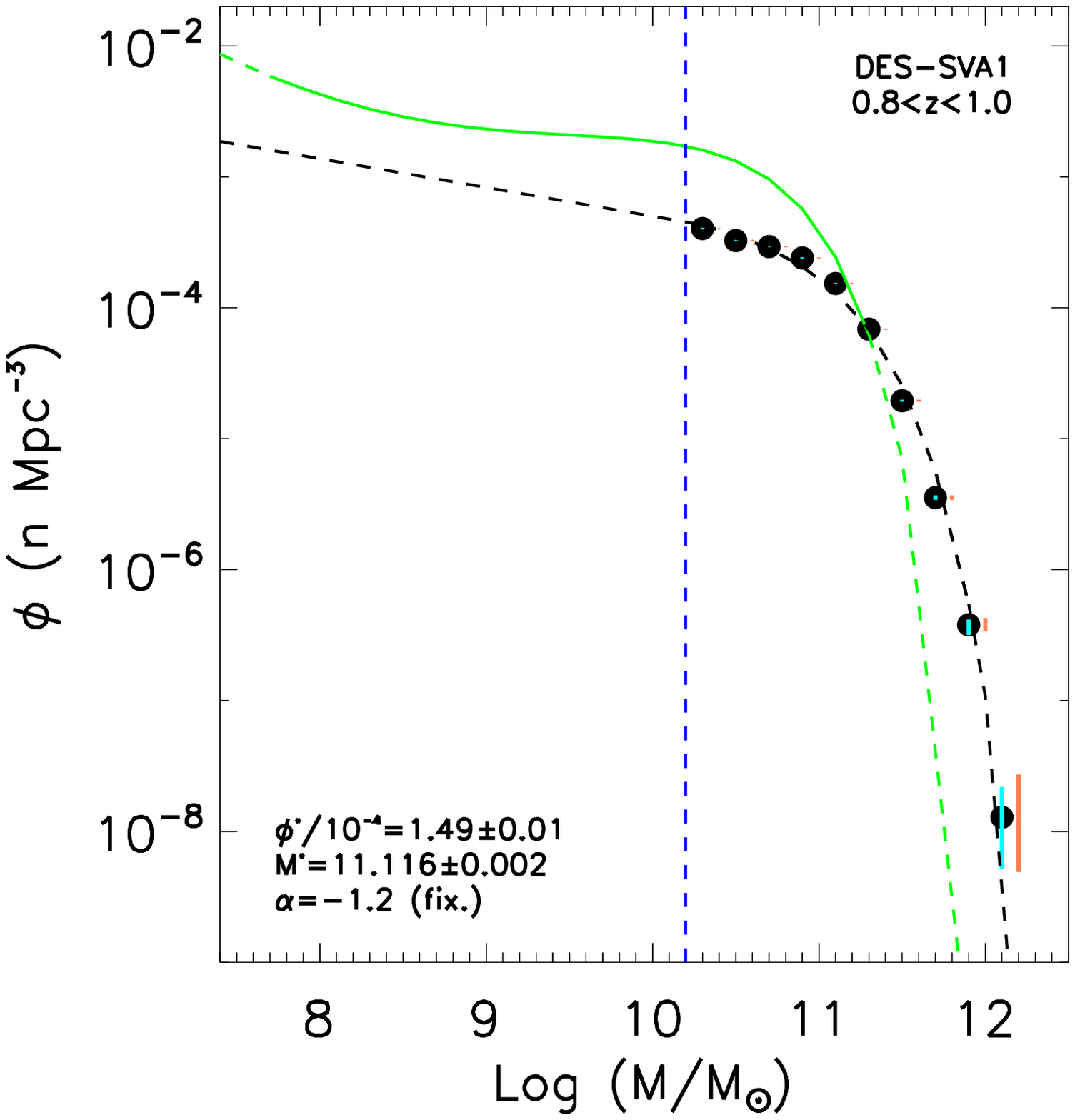}
\caption{Stellar-mass function in five redshift bins (see Table \ref{tab:TableA2}). Cyan and coral error bars correspond respectively to Monte-Carlo-simulations (due to uncertainties on photometry and photometric redshifts) and shot-noise uncertainties. The green line stands fro the best double-Schechter function fit to the lowest-$z$ data points. Black dashed lines stand for the best single-Schechter function fits to the $z>0.2$ data points with $10.2<\log(M/M_{\odot})<12$. In each panel, the best fit parameters are displayed with their uncertainties (see also Table \ref{tab:Table2}). Blue dashed lines represent completeness limits.}
\label{fig:Fig10}
\end{figure*}

We now describe the results of our analysis of the GSMF. As done with the GLF, in Figure \ref{fig:Fig10} we display the median GSMF for the galaxies in our catalogue (see Table \ref{tab:TableA2}), using the same redshift binning scheme and symbols as in Figure \ref{fig:Fig7}. Again, since a density dip and an upturn at the low-mass end are evident in the lowest redshift bin, we fit a double Schechter function to the data points obtained at $z<0.2$. Such function is defined as:

\begin{multline}
%\begin{split}
\phi(M)dM= e^{-M/M^{\ast}} \left [\phi_{1}^{\ast} \left (\frac{M}{M^{\star}} \right )^{\alpha_{1}} + \phi_{2}^{\ast} \left (\frac{M}{M^{\star}} \right )^{\alpha_{2}}  \right ]  \frac{dM}{M^{\ast}},
%\phi(M)dM= e^{-M/M^{\ast}} \left [phi_{1}^{\ast} \left (M/M^{\star} \right ) \right. + \\
 %\left. \left (\phi_{2}^{\ast} 10^{0.4(\alpha_{2}+1)(M^{\star}-M)} \right ) \right ]  dM,
%\end{split}
\label{eq:eq19}
\end{multline}

\noindent whose parameters are the same as those described for equation \ref{eq:eq18}. The best fitting values obtained for the parameters of such fitting are listed in Table \ref{tab:Table2}. 

As for the GLF, a dip is noticeable at $8.5\lesssim \log(M/M_{\odot})\lesssim 10$ in the lowest-$z$-bin GSMF, followed by a steep rise at lower masses, displaying the classic double-Schechter-like shape seen at low $z$ (e.g., \citealp{Baldry-2008, Pozzetti-2010, Baldry-2012, Ilbert-2013, Muzzin-2013}). 
Also here the most massive ($\log(M/M_{\odot})\gtrsim11.3$) and rarest galaxies enter into the GSMF only at $z>0.2$, again probably due to too small a volume to be able to survey these galaxies at lower redshifts.

\subsubsection{Comparison with the literature}
\label{subsubsec:subsubsec5.2.1}
As previously done for the GLF, in this section we compare our results with those found in studies in the literature carried out on spectroscopic data (e.g., \citealp{Bell-2003,Bernardi-2010,Bernardi-2013,Baldry-2012,Moustakas-2013}), in order to assess the reliability of our method based on photometric redshifts with respect to those based on spectroscopic ones. \\
 
\noindent {\it Low Redshift}\\

\begin{figure}
\centering
\includegraphics[width=0.48\textwidth]{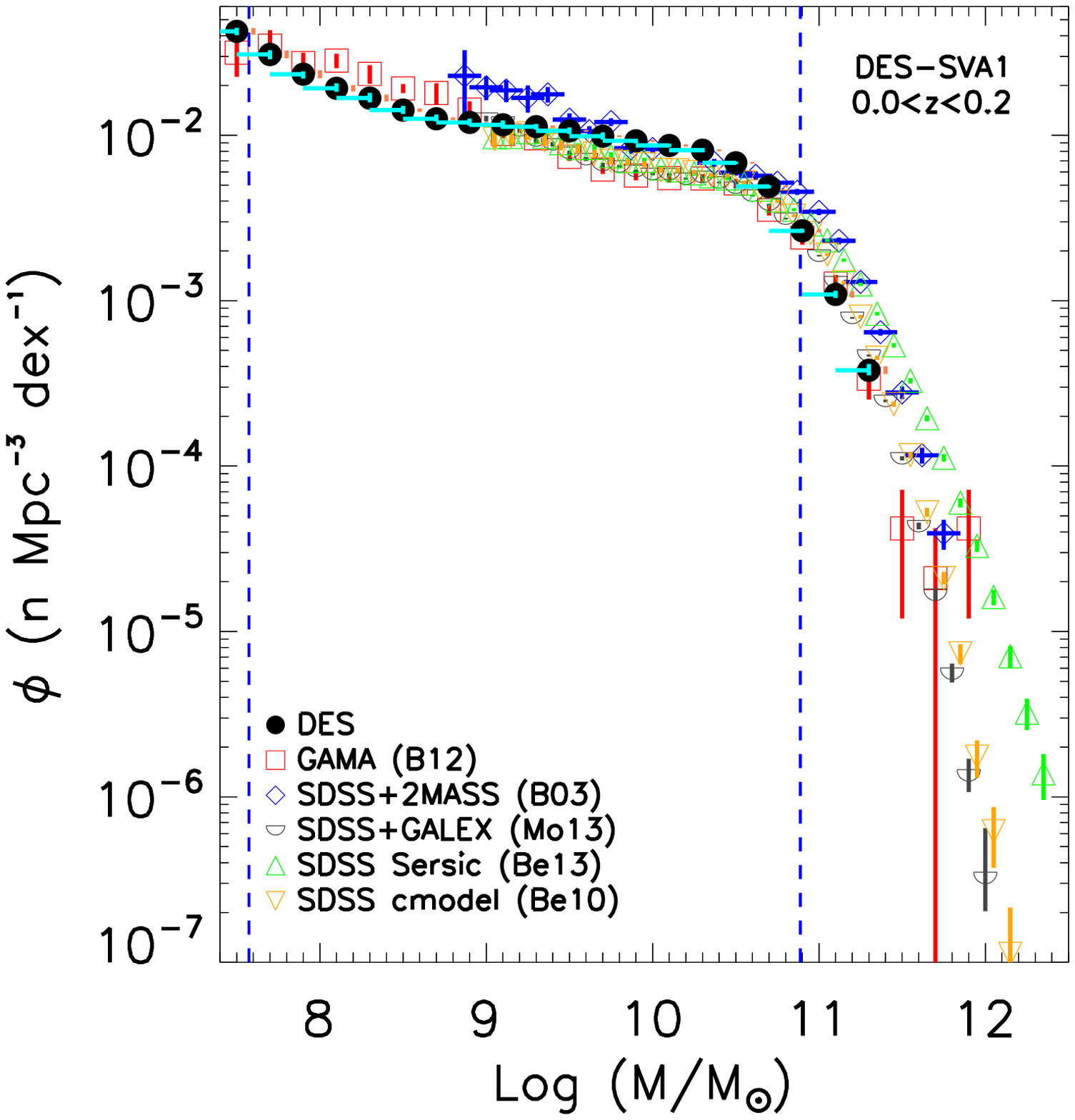}
\caption{Comparison of DES GSMF at $z<0.2$ with results from spectroscopic surveys available in the literature. DES (black dots), GAMA (red squares, \citealp{Baldry-2012}), SDSS+2MASS (blue diamonds, \citealp{Bell-2003}), SDSS+GALEX (grey lower half circles, \citealp{Moustakas-2013}), SDSS with Sersic-profile magnitudes (green upward triangles, \citealp{Bernardi-2013}) and SDSS with {\it cmodel} magnitudes (orange downward triangles, \citealp{Bernardi-2010}). In all panels, dashed blue line indicate DES completeness limits and error bars are as in Figure \ref{fig:Fig10}. Cyan and coral (slightly offset for improving visibility) vertical error bars correspond respectively to Monte-Carlo-simulations (due to uncertainties on photometry and photometric redshifts) and shot-noise uncertainties. Horizontal error bars indicate uncertainties due to IMF variation under the assumption that a systematic mass offset directly translates into a horizontal shift of the data points. We note though, that this is not generally true and that number densities (hence all GSMFs) should be re-calculated using the same IMF.}
\label{fig:Fig11}
\end{figure}

\noindent In Figure \ref{fig:Fig11} we compare the GSMF obtained in our lowest redshift bin with that obtained by \citet{Baldry-2012} at $z<0.06$ for the GAMA survey and with several other spectroscopic surveys, i.e. SDSS/2MASS (\citealp{Bell-2003}, $\sim 410\ {\rm sq. \ deg}$), SDSS (\citealp{Bernardi-2010, Bernardi-2013}, $4681\ {\rm sq. \ deg}$) and SDSS/GALEX (\citealp{Moustakas-2013}, $\sim 2505\ {\rm sq. \ deg}$). As with the GLF, the comparison with GAMA GSMF shows an overall agreement, with DES SGMF showing somewhat larger densities at $\log(M/M_{\odot})\gtrsim 9$ and vice versa at lower masses. Both the GAMA and the DES GSMFs show a double-Schechter-like shape. When comparing individually the double-Schechter function best-fitting parameters we obtain for the DES GSMF (see Table \ref{tab:Table2}) with those obtained by \citet{Baldry-2012} for GAMA, we find statistical consistency at levels between 1 and 3$\sigma$. At $\log(M/M_{\odot})\gtrsim 11$, we estimate our number density values to be affected by stellar mass incompleteness. Furthermore, differently from the GAMA GSMF, we do not find any galaxy with $\log(M/M_{\odot})\gtrsim 11.5$. We do note though, that the high-mass ends of the DES and GAMA GSMFs show good agreement and that the density values at $\log(M/M_{\odot})\gtrsim 11.5$ in the latter are derived on galaxy number counts varying between 1 and 3 galaxies. 

When comparing our GSMF with those of the spectroscopic surveys mentioned above and shown in Figure \ref{fig:Fig11}, we find a general agreement, which is reassuring, given the variation of assumed IMF, stellar mass estimation methods [e.g., SED fitting and from $M_{*}/L$ vs. colour (such as {\it g-r}) relations], ways of estimating galaxy integrated magnitudes (e.g., model magnitudes and Sersic surface-brightness fitting) and the fact that the DES GSMF is based on photometric redshifts rather than spectroscopic ones. However, some discrepancies, especially in specific stellar mass regimes, are also evident. For instance, the SDSS-2MASS GSMF \citep{Bell-2003} shows density values almost always higher than those obtained by the remaining surveys, especially at $\log(M/M_{\odot})\lesssim 9.5$, as also discussed by \citet{Moustakas-2013}.   
In addition, at $\log(M/M_{\odot})\gtrsim 10.8$ our GSMF agrees better with the {\it cmodel}-magnitudes-based GSMF by \citet{Bernardi-2010} than with the Sersic-fit-magnitudes-based one by \citet{Bernardi-2013}. This is expected as we do not use Sersic fitting to estimate our galaxy magnitudes but use magnitudes which are more similar to the SDSS {\it cmodel} magnitudes. In fact, \citet{Bernardi-2013} showed that by using {\it cmodel}-like magnitudes leads to underestimating galaxy total fluxes and stellar masses, especially for massive galaxies with de-Vaucouleurs-like surface brightness profiles. This has the natural consequence of underestimating the spatial densities of such massive galaxies.  
Another important difference is that the DES GSMF does not show the presence of very massive galaxies ($\log(M/M_{\odot})\gtrsim 11.5$). This is probably due to the significantly smaller area probed by us compared to the above-mentioned surveys other than GAMA.\\

\noindent {\it Intermediate/high Redshift}\\

\noindent We now focus on our results obtained in the redshift bins at $z>0.2$. A comparison with studies at  $z>0.2$ is possible as the GSMF is probed out to $z\sim1$ (when only considering spectroscopic surveys) in the literature.

\begin{figure*}
\centering
\includegraphics[width=0.48\textwidth]{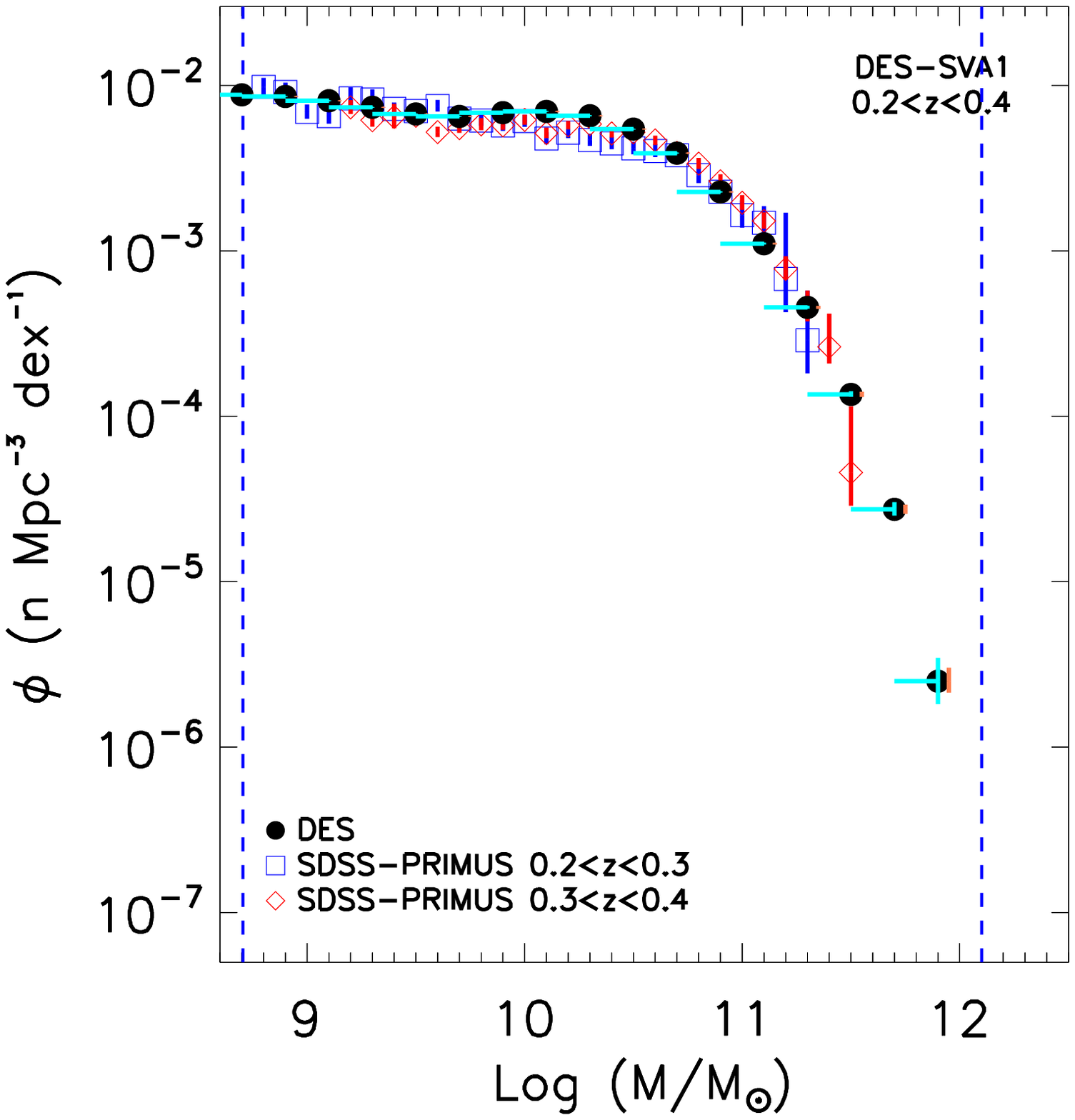}
\includegraphics[width=0.48\textwidth]{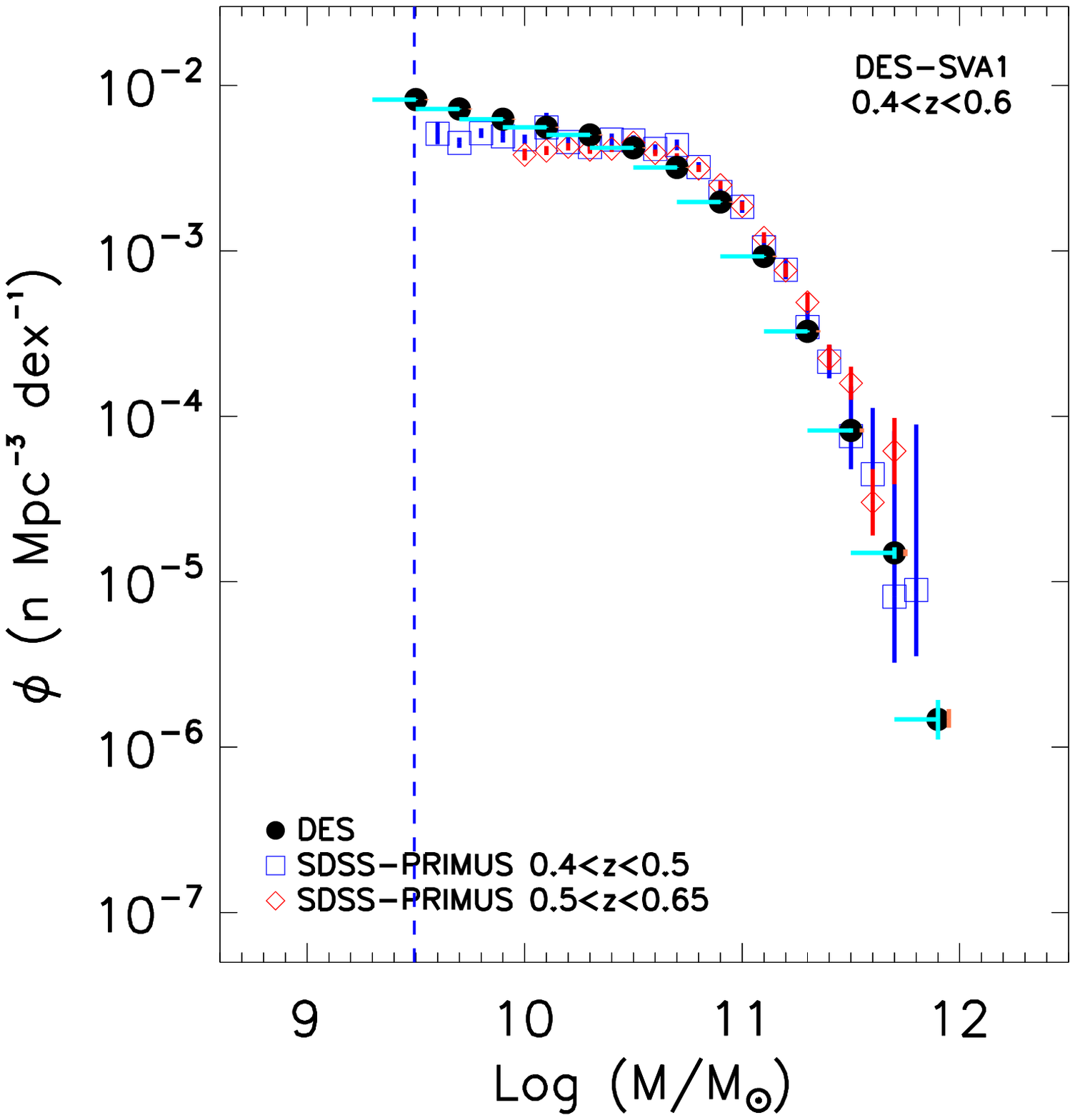}
\includegraphics[width=0.48\textwidth]{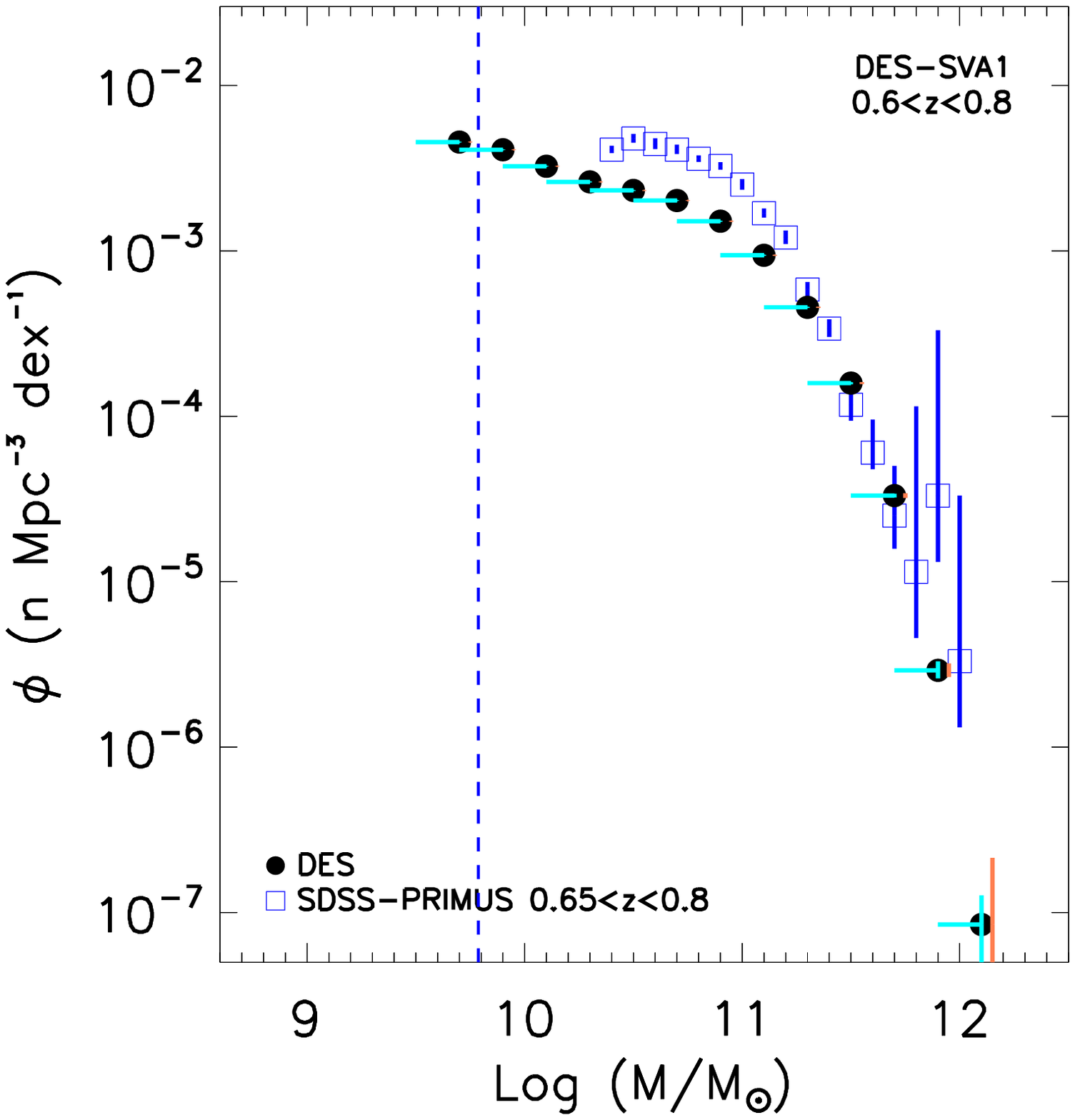}
\includegraphics[width=0.48\textwidth]{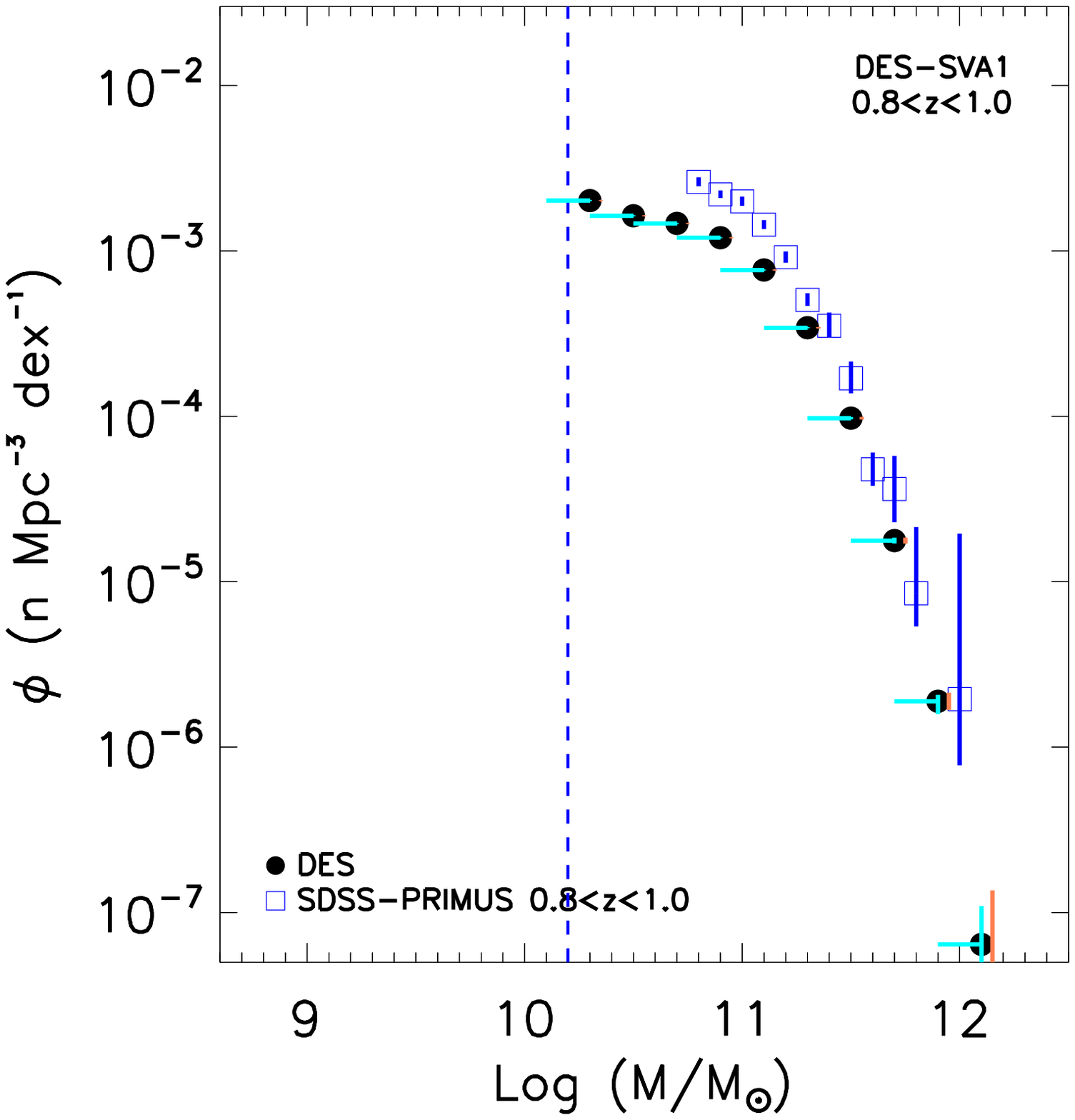}
\caption{Comparison of DES GSMF (black dots) at $0.2<z<0.4$ (upper left panel),at $0.4<z<0.6$ (upper right panel), at $0.6<z<0.8$ (lower left panel), and at $0.8<z<1$ (lower right panel), with those measured by \citet{Moustakas-2013} based on spectroscopic data from PRIMUS. In all panels, dashed blue line indicate DES completeness limits and vertical error bars are as in Figure \ref{fig:Fig10}. Horizontal error bars indicate uncertainties due to IMF variation under the assumption that a systematic mass offset directly translates into a horizontal shift of the data points. We note though, that this is not generally true and that number densities (hence all GSMFs) should be re-calculated using the same IMF.}
\label{fig:Fig12}
\end{figure*}

In Figure \ref{fig:Fig12} we compare the DES GSMF with that obtained by \citet{Moustakas-2013} at $z>0.2$. The latter measured the GSMF at $0.2<z<1$ by using an ${\it i}<23\ {\rm mag}$ flux-limited sample of $\sim 40,000$ galaxies from the PRism MUlti-object Survey (PRIMUS, \citealp{Coil-2011, Cool-2013}), over five fields (COSMOS, XMM-SXDS, XMM-CFHTLS, CDFS and ELAIS-S1) totaling $\sim 5.5\ {\rm sq.\ deg}$ of sky area. The comparison with the results obtained with the PRIMUS galaxy catalogue is particularly useful, as this catalogue was selected using the same band utilised for the DES COMMODORE catalogue and because both these catalogues share the same depth. 

As shown in this figure, DES and the PRIMUS GSMFs at $z<0.6$ are consistent with each other. The agreement between these GSMFs is similar at $z>0.6$, but there are some discrepancies at $\log(M/M_{\odot})\lesssim 11$, where PRIMUS GSMFs show higher number densities. In particular, the latter is affected by incompleteness at $\log(M/M_{\odot})\lesssim 10.5$ (densities at such mass regimes are not publicly available so are not plotted) and seem to lack galaxies with $\log(M/M_{\odot})\gtrsim 11.8$ (this at all redshifts), which are instead present in the DES GSMFs. As the COMMODORE and the PRIMUS galaxy catalogues share the same depth, these differences could be due to cosmic variance (as also discussed by \citealp{Moustakas-2013} themselves). In fact, \citet{Moustakas-2013} showed that despite constructing an area-weighted average of all the 5, $\lesssim1.5 \ {\rm sq.\ deg}$ fields used in their study, the effect of cosmic variance was still significant, especially at $z>0.6$. In fact they found that large-scale overdensities still affect galaxy number densities at $z>0.6$ and at $\log(M/M_{\odot})\gtrsim 10.5$ (see Figures 6-9 in \citealp{Moustakas-2013}), i.e.  exactly over the redshift and stellar mass regimes where we find the largest discrepancies. 

\begin{figure*}
\centering
\includegraphics[width=0.33\textwidth]{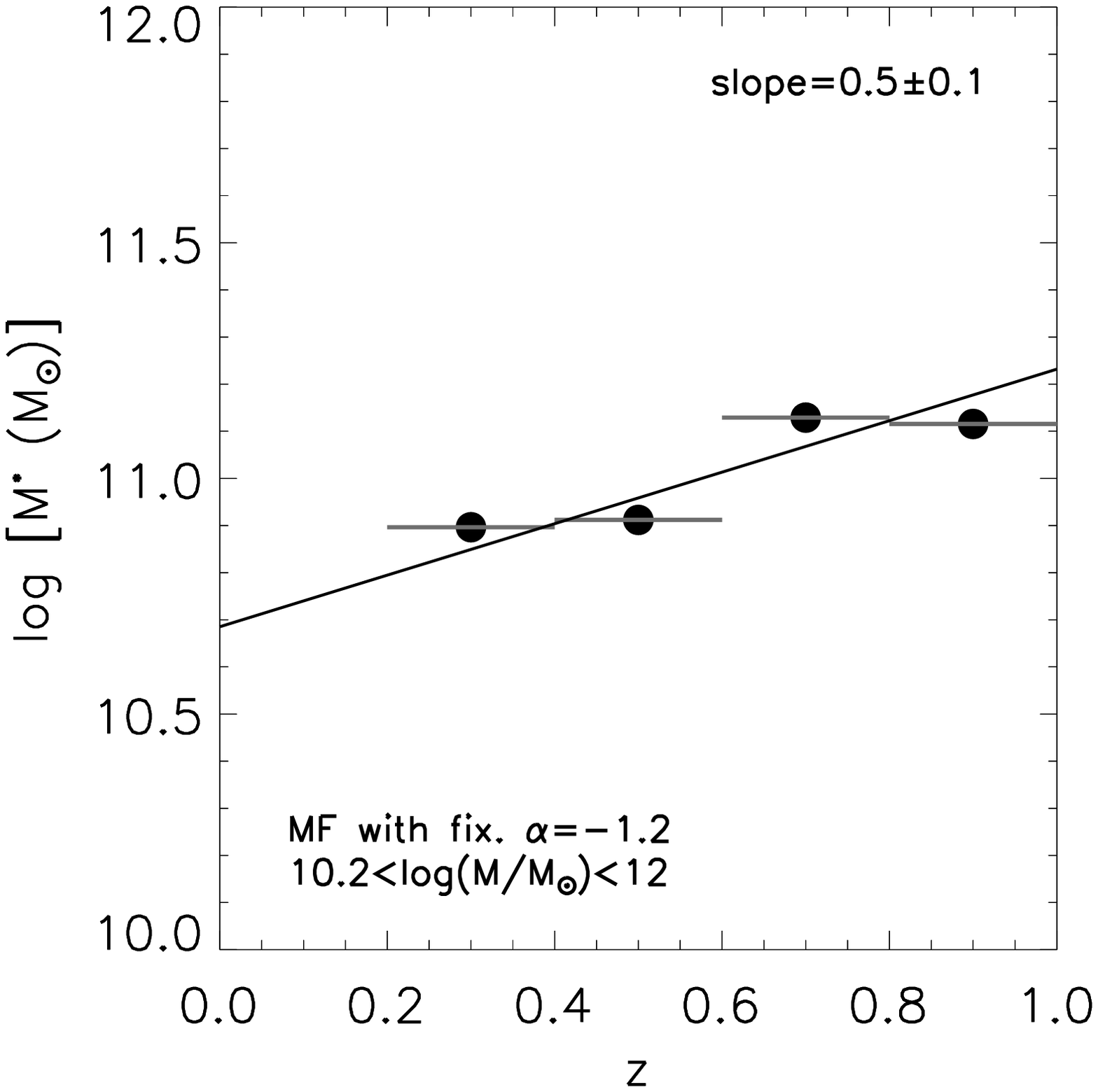}
\includegraphics[width=0.33\textwidth]{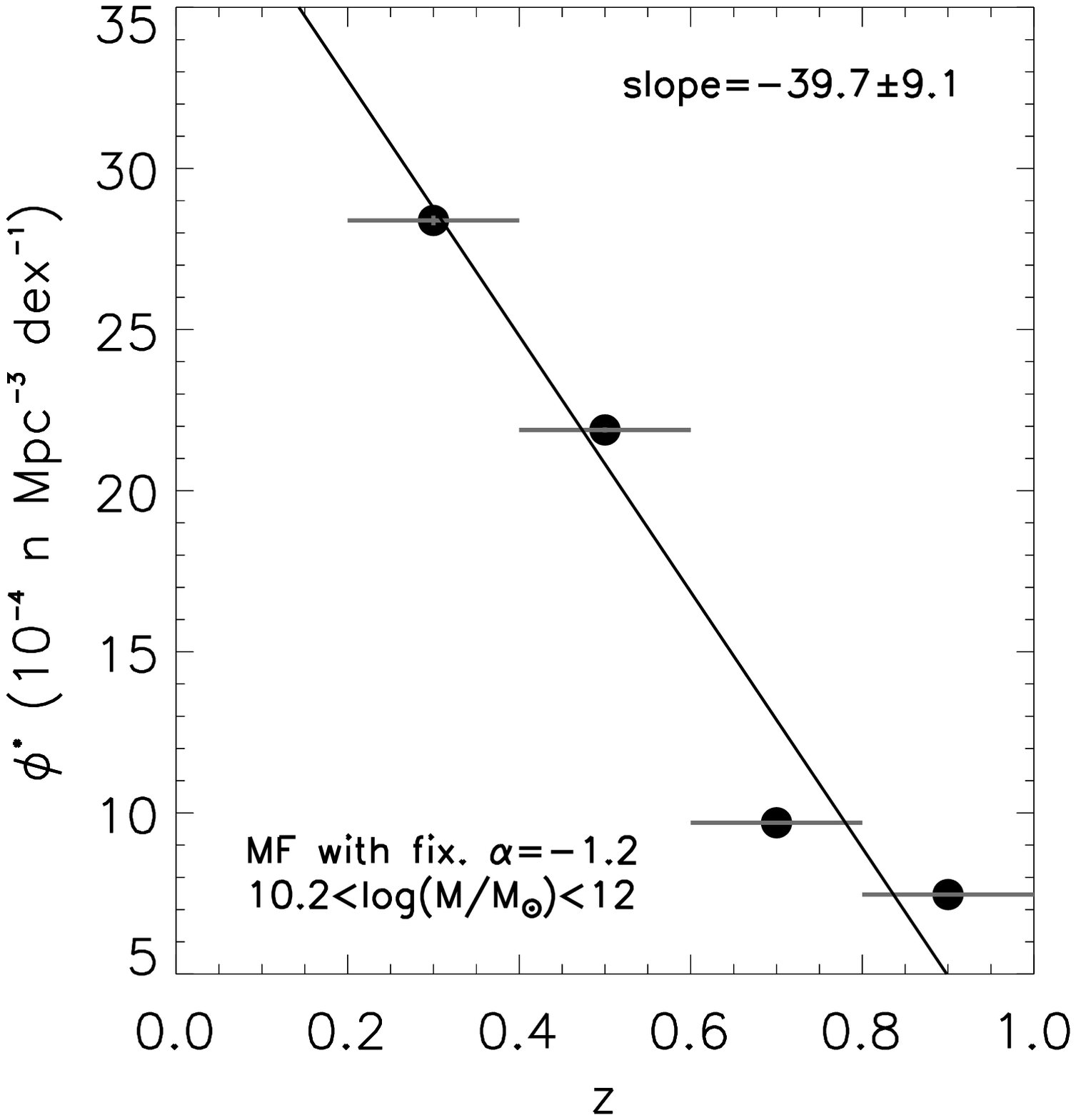}
\includegraphics[width=0.33\textwidth]{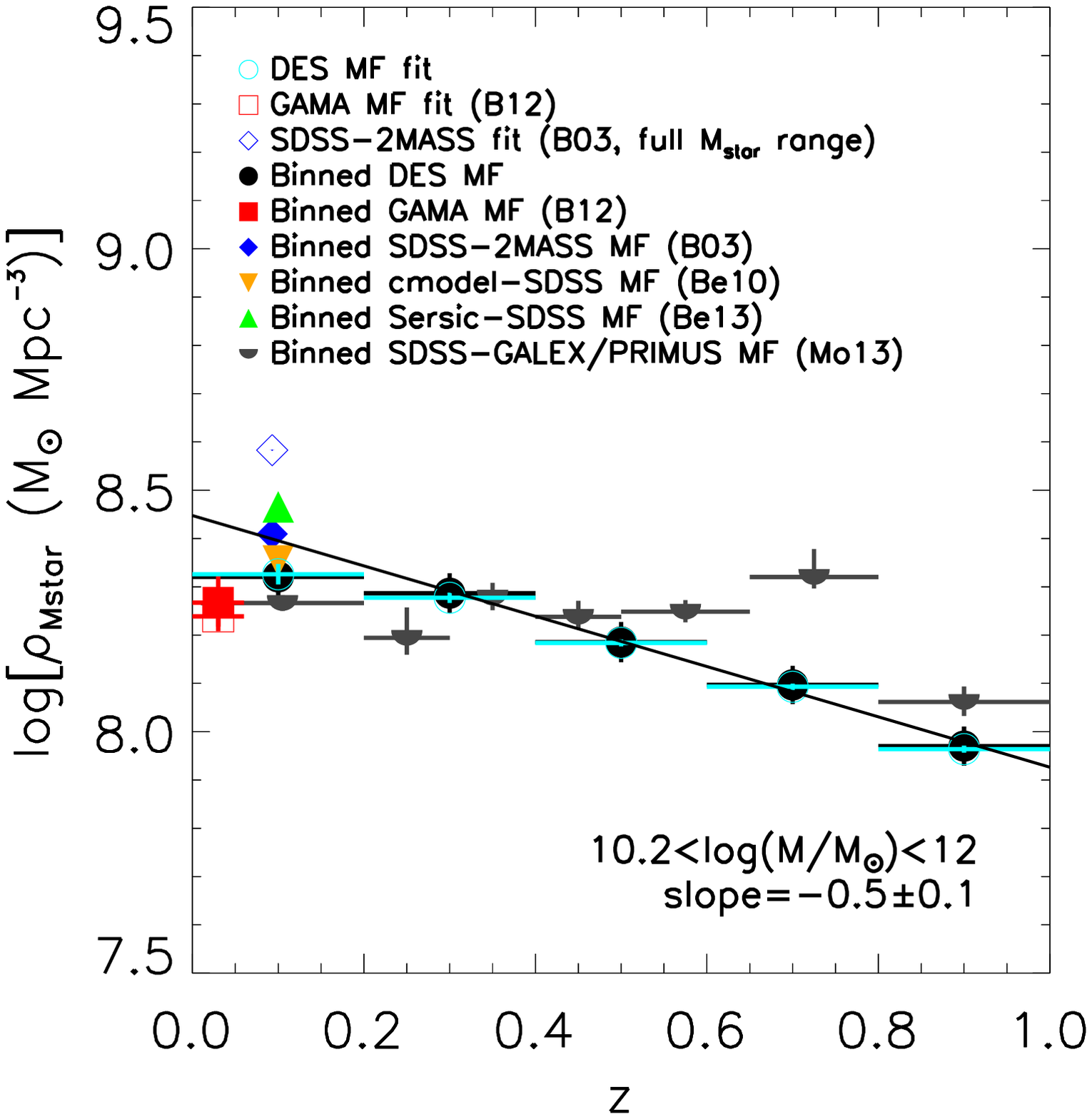}
\caption{Evolution of best-fit parameters ($M^{*}$ and $\phi^{*}$) as calculated via the single-Schechter function fit to the $z>0.2$ GSMFs and of the stellar-mass density $\rho_{\rm Mstar}$ with $z$. {\bf Left-hand panel}: $\log(M^{*})$ vs. $z$. {\bf Centre-panel}:  $\phi^{*}$ vs. $z$. {\bf Right-hand panel}: $\rho_{\rm Mstar}$ vs. $z$. Here, $\rho_{\rm Mstar}$ is calculated via two methods, i.e., using the GSMF fit (cyan circles) and using the binned GSMF (black dots). In addition, the values of $\rho_{\rm Mstar}$ calculated in the lowest $z$ bin are shown for reference. We also add values from the literature, measured from the GSMF Schechter fit and/or from the binned function. When possible we adjust the measurement taken from the literature (or re-estimated by us on the binned GSMF) to the same stellar-mass interval ($10.2<\log(M/M_{\odot})<12$) used in our study. Filled and empty symbols indicate, respectively, measurements carried out using the Schechter fit and via the binned GSMF.  In addition to DES data points (black dots and cyan circles) we show values from \citet{Baldry-2012} (red squares), \citet{Bell-2003} (blue diamonds), \citet{Bernardi-2010} (orange downward triangle), \citet{Bernardi-2013} (green upward triangle) and \citet{Moustakas-2013} (grey lower half circles). In all panels, all measurements refer to the stellar-mass interval $10.2<\log(M/M_{\odot})<12$ and the black solid line represents the best fit to the DES data points at $z>0.2$ when taking into account uncertainties on both axes.}
\label{fig:Fig13}
\end{figure*}

\begin{figure*}
\centering
\includegraphics[width=0.49\textwidth]{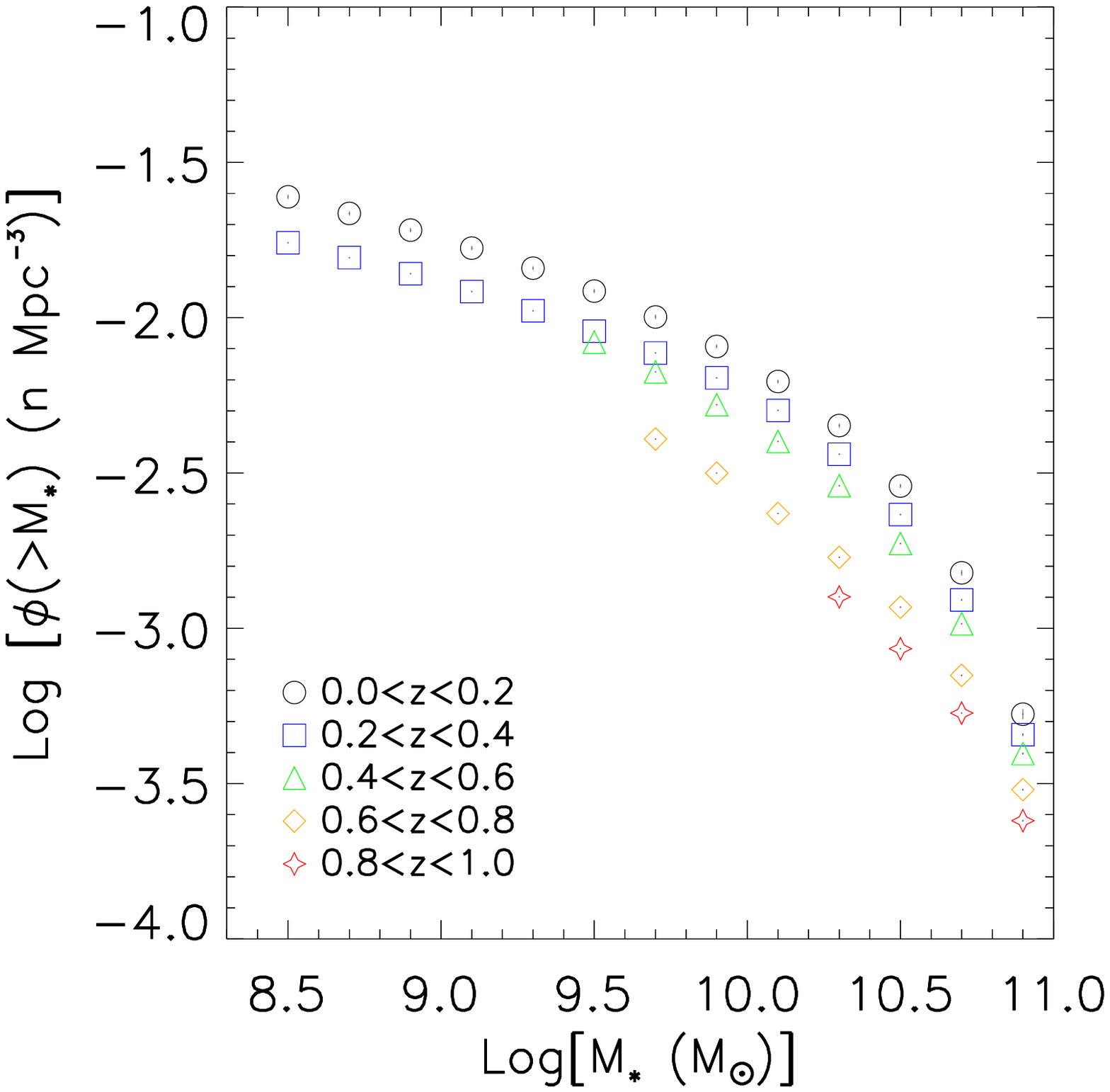}
\includegraphics[width=0.49\textwidth]{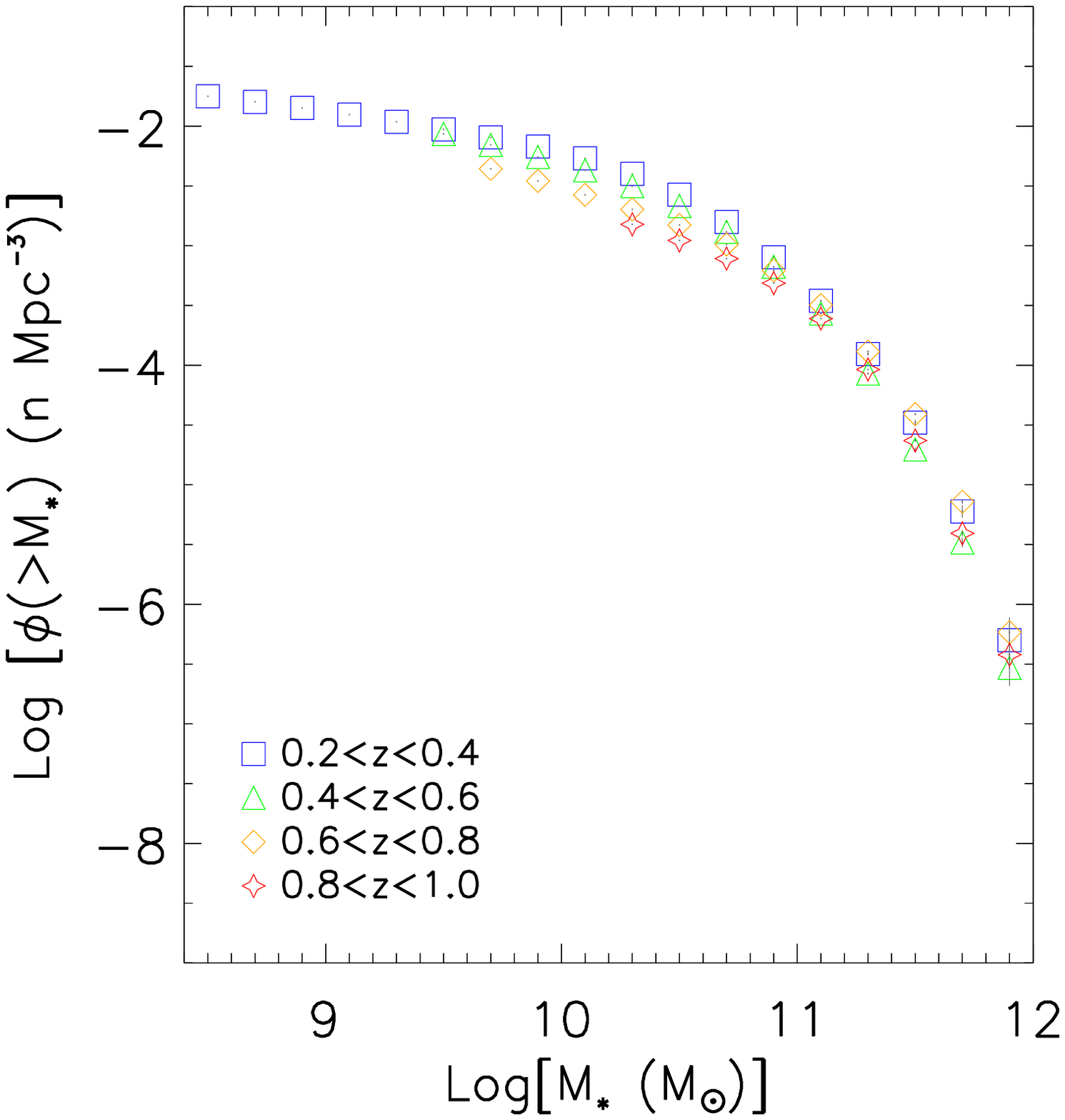}
\caption{Left-hand panel: cumulative GSMFs in all five redshift bins and for $\log(M/M_{\odot})<11$; right-hand panel: cumulative GSMFs at $z>0.2$ and for $\log(M/M_{\odot})<12$.}
\label{fig:Fig14}
\end{figure*}

\subsubsection{Evolution of GSMF parameters and of stellar mass density}
\label{subsubsec:subsubsec5.2.2}
We now repeat for the GSMF the same exercise carried out for the GLF, now studying the evolution of $M^{\ast}$, $\phi^{\ast}$ and $\rho(M_{\rm star})$ with redshift, within a stellar mass range complete at $\gtrsim 90$ per cent level, i.e. $10.2<\log(M/M_{\odot})<12$. We show the results of this analysis in Figures \ref{fig:Fig10} and \ref{fig:Fig13} and find significant evolution at $\gtrsim 3\sigma$ level of all these three quantities. In particular, $\log (M^{\ast})$ (Figure \ref{fig:Fig13}, left-hand panel) increases by $\sim0.2 \ {\rm dex}$ from $z\sim 0.2$ to $z\sim 1$, while both $\phi^{\ast}$ and $\log[\rho(M_{\rm star})]$ (Figure \ref{fig:Fig13}, centre and right-hand panels) decrease with redshift, respectively by a factor $\sim4$ and $\sim 1.05$ from $z=0.2$ to $z=1$. 

Right-hand panel of Figure \ref{fig:Fig13} also shows that there is an overall agreement between our values of $\log[\rho(M_{\rm star})]$ and those based on the literature's studies we compare with \citep{,Bell-2003, Bernardi-2010, Baldry-2012, Moustakas-2013}, especially when we are able to re-calculate the latter in the same way as done for ours. Some more significant discrepancies appear at $z\gtrsim0.7$. \citet{Moustakas-2013}'s measurement at $z\sim0.7$ is significantly higher than ours (but this is known to be the $z$ bin most affected by cosmic variance), while the one at $z\sim0.9$ is partly affected by incompleteness, so despite consistent with ours, it has to be considered as a lower limit. Overall, taking Moustakas et al.'s values alone at $z<0.8$, would suggest constancy of $\rho_{\rm Mstar}$ with redshift. However, as also mentioned for their GSMFs at $z>0.6$ in Figure \ref{fig:Fig12}, this could be due to large-scale structure effects due to the relatively small-area fields ($\lesssim 1.5 \ {\rm sq.\ deg}$ each) over which their measurements were performed.  At low redshift, we notice that the estimate quoted by \citet{Bell-2003} (blue open diamond) is significantly higher. However this is probably due to its derivation from the GSMF fit over the entire spanned stellar mass values. In fact, when we use their binned GSMF and recalculate $\rho_{\rm Mstar}$ in the appropriate $\gtrsim 90$-per-cent complete stellar mass range used for our measurements (see blue full diamond), their result becomes more similar to the rest of the values. Notice also that the value measured on the {\it cmodel}-based GSMF by \citet{Bernardi-2010} is closer to the other values than the one derived from the Sersic-fit-based GSMF by \citet{Bernardi-2013}. This is expected as all the GSMFs studied here were not based on the use of Sersic fitting for measuring integrated magnitudes. Overall, considering our data points and those of the literature (and considering also the possibility that the high-$z$ values of \citealp{Moustakas-2013} are affected by large-scale structure effects) a picture which displays $\rho_{\rm Mstar}$ decreasing with redshift at $0.2<z<1$ seems plausible. Such trend was also seen in several other studies in the literature, e.g. \citet[and references therein]{Marchesini-2009, Muzzin-2013}. However constancy of $\rho_{\rm Mstar}$ with $z$ out to $z=1$ was also reported, for instance, by \citet{Ilbert-2013} and \citet{Moustakas-2013}.      

Our findings of a decreasing stellar mass density and an approximately constant luminosity density with redshift  (respectively right-hand panels of Figures \ref{fig:Fig13} and \ref{fig:Fig9}) can be explained by the fact that while at higher redshift there is less stellar mass available in galaxies, their luminosities do not change substantially because of their stellar populations getting younger and brighter, as one would expect. 

\begin{figure*}
\centering
\includegraphics[width=0.8\textwidth]{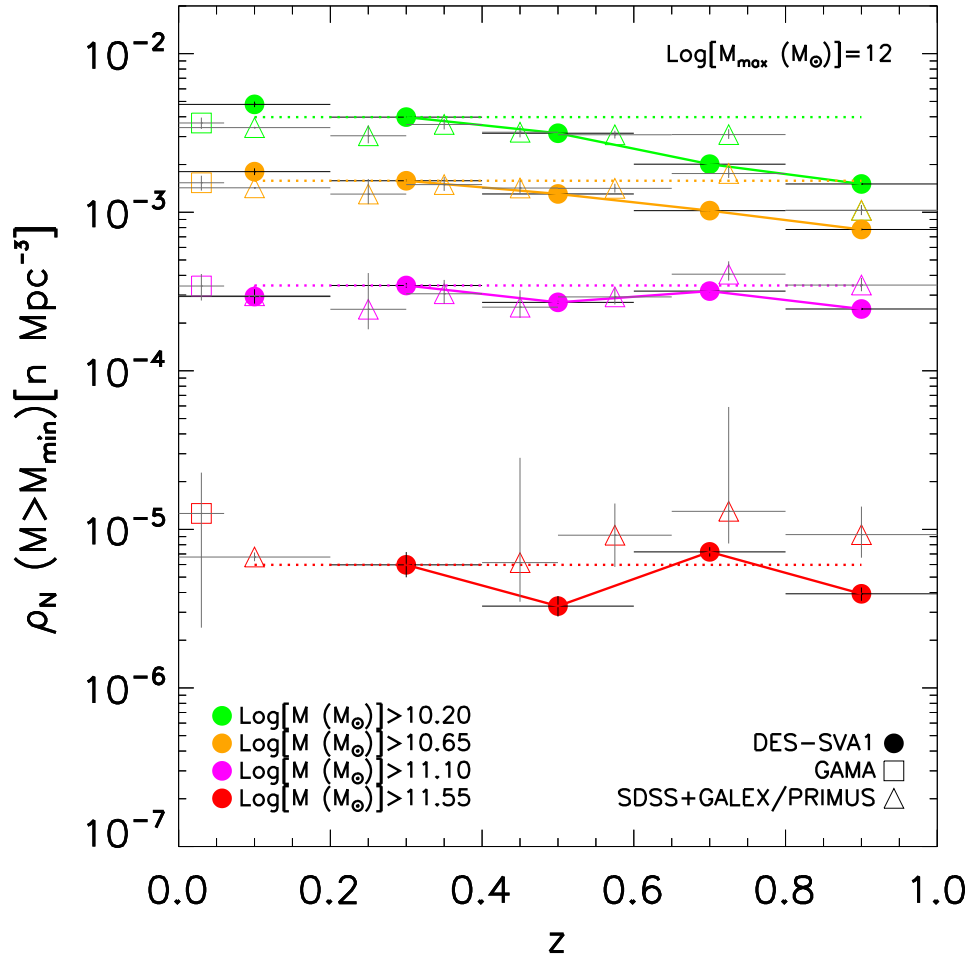}
\caption{Total galaxy number density vs redshift. Densities are calculated within stellar-mass intervals differing for their lower limit. The total stellar-mass range considered is  for $10.2<\log(M/M_{\odot})<12$, unaffected by incompleteness at all redshifts higher than $z=0.2$. Data points in the lowest redshift bins are shown for reference and have to be considered as lower limits (when present) as partly affected by incompleteness of galaxies more massive than  for $\log(M/M_{\odot})\gtrsim11$. We also show values measured for GAMA (squares, \citealp{Baldry-2012}) and for SDSS+GALEX and PRIMUS (triangles, \citealp{Moustakas-2013}).}
\label{fig:Fig15}
\end{figure*}

\subsubsection{Galaxy mass build up}
\label{subsubsec:subsubsec5.2.3}
In order to shed some light on the evolution with cosmic time of galaxy spatial number density as a function of stellar mass, hence on how galaxies build their masses over cosmic time, we analyse and compare among them the cumulative distribution functions obtained in our five redshift bins. 

In Figure \ref{fig:Fig14} we show such cumulative mass functions (derived by summing densities starting from the massive end) in our five redshift bins (left-hand panel) and only in the four higher-$z$ ones (right-hand panel). In the former case, we focus on stellar masses lower than $\log(M/M_{\odot})=11$, as at larger masses our lowest redshift GSMF is affected by stellar mass incompleteness. In the latter case instead, we extend this mass range out to $\log(M/M_{\odot})=12$. In both cases we notice that number densities values of lower-$z$ cumulative functions are always higher than or comparable to (depending on the mass bin considered) those of higher-$z$ ones. Such evolution is indeed mass dependent, as it is more pronounced at low masses, while there is virtually no evolution at the high masses.
Such behaviour can be seen also in Figure \ref{fig:Fig15}, where we plot the total number densities obtained in four different mass ranges as a function of redshift, similarly to what done by \citet{Pozzetti-2010}. We restrict ourselves to $10.2<\log(M/M_{\odot})<12$, a mass range which is $\gtrsim 90$ per cent complete at $0.2<z<1$. As a consequence, the points at $z<0.2$ in Figure \ref{fig:Fig15} have to be considered as lower limits. In particular, the number density of galaxies such that $\log(M/M_{\odot})> 11.55$ is consistent with no evoultion since $z=1$ at $\lesssim 3 \sigma$ level. From Figure \ref{fig:Fig15} the downsizing pattern is clearly visible, in agreement with the results by \citet{Pozzetti-2010}.
 
In this same figure, we also show additional points measured from \citet{Baldry-2012} (for GAMA) and \citet{Moustakas-2013} (for PRIMUS). We do find agreement with the literature, especially for the highest mass bins ($ \log(M/M_{\odot})\gtrsim 11.1$). However, some differences are seen in bins with lower mass limit $\log(M/M_{\odot})<11.1$, where PRIMUS values at $z>0.6$ suggest constancy of spatial number density with redshift rather than a decreasing trend as indicated by our results. This issue, though, can be again the results of large-scale-structure effects. Overall we do find evidence for mass-dependent evolution of galaxy number spatial density as function of redshift, with densities of more massive galaxies remaining about constant at $0<z<1$ or evolving less than densities of less massive galaxies. Similar results were already found in other studies (e.g., \citealp{Pozzetti-2010,Ilbert-2010, Ilbert-2013,Davidzon-2013}), but never over such a large area and in combination with the use of photometric redshifts. However, results claiming a hierarchical-like mass-dependent evolution of galaxy number spatial densities were also found in other studies in the literature (see, for instance, \citealp{Muzzin-2013}). We will further discuss this issue in Section \ref{sec:sec6}.  

\section{Discussion}
\label{sec:sec6}
Studying with precision and consistency the evolution of the GLF and the GSMF over a wide redshift range and over luminosity and stellar mass ranges accessible at all studied redshifts is a difficult task to carry out. This is because, in order to correctly identify or dismiss evolutionary trends able to shed light on galaxy formation (e.g., hierarchical or anti-hierarchical), we need to be able to access large cosmic volumes (i.e., via deep surveys over large sky areas) and measure distances with high precision (ideally via spectroscopy). This is because such approach allows to identify galaxy samples large enough to study with high precision even the extreme parts of the GLF and the GSMF. However, spectroscopy is too time consuming, hence we need to resort to photometric redshifts. In this work we investigate the GLF and the GSMF  out to $z=1$ in comparison with spectroscopic measurements taken from the literature as we aim at validating our analysis methodology, in addition to inferring precious knowledge about galaxy formation and evolution. As our results show general agreement with similar analysis carried out with spectroscopic data over similar or narrower redshift ranges and out to similar or shallower flux depths, we can apply our analysis methodology to the full DES data-set (we currently use only $\sim 3$ per cent of DES final sky coverage), once this becomes available. 

DES {\it i}-band GLF at $z<0.2$ shows the characteristic double-Schechter-function shape seen in several other studies of the GLF (e.g., \citealp{Blanton-2005} and \citealp{Baldry-2012}). Such shape, though, cannot be probed also at higher $z$ as one would need data deeper than the ones currently used. However, by identifying a luminosity range complete at $\gtrsim 90$ per cent level at $0.2<z<1$, we were able to study the evolution of the DES GLF homogeneously within such redshift range. Our results show brightening (of $\sim 0.9\ {\rm mag}$ out to $z=1$) of $M^{*}_{\rm i}$, decrement (by a factor $\sim 4$ out to $z=1$) of $\phi^{*}$ and $\sim$ constancy of the luminosity density $\rho_{\rm L}$ with redshift. Such trends at $0\lesssim z\lesssim 1$ were already seen in several studies of the GLF in the literature, also based on photometric redshifts (e.g, \citealp{Cirasuolo-2010, Marchesini-2012, Stefanon-2013}). However, some other studies also found different trends. For instance, \citet{Ilbert-2005,Prescott-2009,Beare-2015} found an increase of $\rho_{\rm L}$ with redshift out to $z=1$. A direct comparison is particularly difficult for us as we study the {\it i}-band GLF, while all the studies that investigate the evolution of the GLF out to $z\sim1$ use different wavebands. In fact, to our knowledge, ours is the first study of the evolution of the {\it i}-band GLF out to $z>0.5$ ever performed. 

When focusing on the DES GSMF at $z<0.2$, we can make the same considerations done for the local DES GLF, as a double-Schechter-function shape is also seen, in agreement with several studies in the literature (e.g., \citealp{Pozzetti-2010, Baldry-2008, Baldry-2012, Ilbert-2013, Muzzin-2013}). At higher $z$, we repeated the exercise carried out for the GLF, exploring the evolution of the GSMF at $0.2<z<1$ of a $\gtrsim 90$-per-cent stellar-mass complete galaxy sample. Our results showed small increase  (of $\sim 0.2\ {\rm dex}$ out to $z=1$) of $\log M^{*}$ and decrement of both $\phi^{*}$ (by a factor $\sim4$ out to $z=1$) and stellar-mass density $\rho_{\rm Mstar}$ (by a factor $\sim 1.05$ out to $z=1$) with redshift. As the decrement of $\phi^{*}$ is more significant than the increment of $\log (M^{*})$ with $z$, the evolution of $\rho_{\rm Mstar}$ seems to be mainly due to the former. These results are consistent with several studies in the literature (e.g., \citealp{Perez-Gonzalez-2008,Davidzon-2013, Muzzin-2013}), however, some discrepancies can also be found. Note that we now extend the comparison also to studies based on photometric redshifts. For instance, \citet{Ilbert-2013} found constancy of both $\log M^{*}$ and $\rho_{\rm Mstar}$ out to $z\sim1$. Despite this, maybe the most important fact is that the majority of the studies do report results consistent with a mass-dependent evolution of the GSMF of the global galaxy population. Some claimed such dependence to be bottom-up (e.g., \citealp{Muzzin-2013}), while others claimed it to be top-down (e.g., \citealp{Perez-Gonzalez-2008, Pozzetti-2010,Ilbert-2010, Ilbert-2013,Davidzon-2013,Moustakas-2013}). We do find evidence for a top-down galaxy formation scenario. 
In fact we do find that the densities of galaxies with $\log (M/M_{\odot})\gtrsim11.1$ remain constant over the analysed redshift range, while at smaller masses they decrease with redshift and they do it more the lower the masses included in the analysed mass bins (see Figure \ref{fig:Fig15}). This evidence is consistent with the results of \citet{Pozzetti-2010} and of \citet{Moustakas-2013}, the latter when accounting also for the fact that some of their results were shown to be affected by cosmic variance. According to our analysis, most massive galaxies [$\log (M/M_{\odot})\gtrsim11.1$] appear to be already in place by $z=1$, while the mass build-up process is still ongoing from $z=1$ to $z=0$ for less massive galaxies.  

However, despite our findings discussed so far can point at a top-down galaxy formation scenario, we do also have to consider that in our work we have neglected some sources of uncertainties (described previously). Hence our analysis should be repeated taking all uncertainty sources into account and verify the robustness of our results. As an explanatory example of this issue, we note that \citet{Marchesini-2009} found a downsizing-like mass-dependent evolution of the GSMF when considering only random uncertainties and that such mass-dependent evolution was no longer robust when taking into account also systematic uncertainties due to SED-modelling assumptions. In addition, a more thorough study of the impact of systematics (e.g. seeing, airmass, sky brightness) on our measurements should be carried out as well, also using the aid of simulations \citep{Suchyta-2016}. Furthermore the effects of  photo-$z$ pdf uncertainty and estimation methods should also be studied in detail.

Given that the $\gtrsim 90$-per-cent complete luminosity ($-26<M_{\rm i}<-21.4\ {\rm mag}$) and stellar mass ($10.2<\log(M/M_{\odot})<12$) ranges used to explore the evolution of the GLF and GSMF at $0.2<z<1$ do not allow us to study $\alpha$ in a homogeneous way over this redshift range, it is difficult to make reliable considerations on the evolution of the shapes of these functions with $z$. However, despite the fact that we fixed the value of $\alpha$ when fitting the GLFs and GSMFs at $z>0.2$, we have explored the effect of using different values, finding no major changes to the trends with $z$ identified for $M^{*}$,  $\phi^{*}$, $\rho_{\rm L/Mstar}$, suggesting that, within the explored luminosity and stellar mass ranges, the effect of $\alpha$ is small (justifying also a posteriori our choice of fixing its value) and so that the shapes of the GLF and GSMF are not expected to change from $z=1$ to $z=0.2$ for the mass range studied here. 
This would suggest that the physical processes involved in the galaxy stellar-mass build up do not significantly change the shape of the GSMF since $z=1$. Hence one could also infer that the change in the GLF $M^{*}$ may be purely due to stellar evolution, with galaxy stellar populations becoming younger and so brighter at higher $z$.  

\section{ Conclusions}
\label{sec:sec7}
We study the GLF and GSMF of galaxies within the DES COMMODORE catalogue out to $z=1$. This galaxy catalogue was selected within the DES SVA1 data, contains $\sim 4\times\ 10^{6}$ galaxies at $0<z\lesssim1.3$, covers $\sim 155\ {\rm sq.\ deg}$ of sky area (only $\sim 3$ per cent of the final DES footprint) and is characterised by an {\it i}-band depth of $i=23\ {\rm mag}$. Such characteristics are unprecedented for galaxy catalogues and they enable us to study the evolution of GLF and GSMF between $z=1$ and $z=0$ homogeneously with the same statistically-rich data set and free of cosmic variance effects. This is the first time that such a statistically rich and deep galaxy catalogue, covering an area $>20\ {\rm sq.\ deg}$, has ever been used for a comprehensive study of the evolution of GLF and GSMF at $0<z<1$. All the previous studies were either based on deep pencil-beam surveys or large but relatively shallower surveys. Since we utilise photometric redshift pdfs, the aim of this paper is mainly testing our method in comparison with results from similar studies of GLF and GSMF based on spectroscopic redshifts, in addition to studying galaxy formation and evolution since $z=1$.

We investigate the evolution of {\it i}-band GLF and GSMF, their Schechter-fitting parameters and of luminosity, stellar mass and galaxy number densities with cosmic time, attempting to shed some light on how galaxies build up their mass over time (hierarchically or anti-hierarchically?). We first identify the COMMODORE catalogue, then study galaxy completeness as function of surface brightness and apparent magnitude, estimate galaxy physical and detectability-related properties and finally estimate and study GLF and GSMF in five redshift bins. 

Our main results are as follows: 

\begin{itemize}

\item Our low-$z$ estimates of GLF and GSMF show overall agreement with those obtained for GAMA. Our densities are somewhat higher at $M_{\rm i}\lesssim -20\ {\rm mag}$/$\log(M/M_{\odot})\gtrsim9$ then those measured for GAMA and vice versa at fainter magnitudes/lower masses, but the shapes of DES and GAMA functions are similar. They both show a double-Schechter function shape, in agreement with what is found in the literature. When taken individually, all best-fitting parameters of DES and GAMA GSMFs are consistent within $1-3\sigma$, depending on the parameter considered. This result is particularly reassuring as GAMA functions are based on spectroscopic redshifts and are measured over an area very similar to the one used by us. However, some discrepancies are seen when comparing also with other functions in the literature. In particular, DES GLF cannot reproduce the significant excess of faint galaxies seen by \citet{Blanton-2005}. This could be due to the different methods for correcting for surface-brightness incompleteness. The discrepancies seen in the GSMF, are mainly due to lack of area for identifying very massive galaxies, different magnitude estimators and possibly to different ways of estimating stellar masses.\\

\item At higher redshift, there is no available study, to our knowledge, investigating the {\it i}-band GLF. However, we are able to compare DES GSMF with that measured for the PRIMUS survey \citep{Moustakas-2013}, finding very good agreement out to $z=0.6$. At higher redshifts, the effects of cosmic variance in the PRIMUS GSMFs (especially at $z\sim0.7$) are significant and the agreement with the DES one degrades towards low masses. Overall, however, we consider our results promising and the agreement acceptable. \\ 

\item We investigate the evolution of DES GLF and GSMF at $0.2<z<1$ by identifying luminosity and stellar mass ranges complete at $\gtrsim 90$ per cent level. Such ranges, however, do not allow to study the faint/low-mass end slope $\alpha$, which we keep fixed at $\alpha=-1.2$.  
For the GLF, we find $M^{*}$ brightening and $\phi^{*}$ decreasing with $z$. Such effects approximately balance each other, producing a constant luminosity density $\rho_{\rm L}$ with $z$ ($\rho_{\rm L}\propto (1+z)^{-0.12\pm0.11}$). With regards to the GSMF,  $M^{*}$ is found to slightly increase with $z$, while $\phi^{*}$ to significantly decrease with $z$, resulting in a mildly decreasing stellar-mass density $\rho_{\rm Mstar}$ with $z$ ($\rho_{\rm Mstar}\propto (1+z)^{-0.5\pm0.1}$). These results suggest that, within the $\gtrsim 90$-per-cent complete stellar-mass range investigated, the physical processes regulating galaxy mass build-up do not modify the shape of the GSMF between $z=1$ and $z=0.2$. Finally, by investigating the galaxy number density $\rho_{\rm N}$ in different mass intervals, we do find evidence for mass-dependent evolution of the DES GSMF, following a downsizing pattern. In particular, the number densities of galaxies with $\log (M/M_{\odot}>11.1)$ are found to be about constant from $z\sim1$ to $z\sim0.2$, while when including smaller masses such densities decrease with $z$ and they do it more the smaller the masses. Our measurements are directly compared with those from PRIMUS and, despite some variation of the latter due to cosmic variance, are found to be in agreement.\\

\item Given our results and how our measurements compare with those based on spectroscopic data, we consider our method for studying GLF and GSMF based on photometric redshift satisfactory. We have only used $\sim 3$ per cent of the final DES footprint and we plan to build on this study to take full advantage of DES full area when the full data set becomes available. In addition, we also plan to test our analysis on simulations tailored on DES and also to compare our observation-based results with those we obtained by analysing simulated galaxies in the same way. This will help us not only with assessing the robustness of our analysis method, but also with assessing how well theoretical models can reproduce observed GLFs and GSMFs.

\end{itemize}

\section*{Acknowledgments}
This paper has gone through internal review by the DES collaboration.
 
DC thanks Maurizio Paolillo, Ivan Baldry, Guido Walter Pettinari, Lado Samushia, Joanne D. Cohn, Douglas L. Tucker, Chiara Spiniello and Maurizio Salaris for valuable scientific conversations and Gary Burton and Edd Edmondson respectively for providing help with code parallelisation and with IT-related matters.

DC particularly thanks the following people for their outstanding moral support and the help provided throughout the past few years, without which this work would have not been possible: Angela Del Gaudio, Spartaco Capozzi, Alessandro Capozzi, Marcello Capozzi, Claudio Gino Capozzi, Margherita Lanzi, Davide Bianchi, Nora Siklodi, Matthew Withers, Hedda Gressel, Rob Baker, Robert Fairall, Milan Kreuschitz Markovi\v{c}, Dida Markovi\v{c}, Susana Sampaio Dias, Olivia Umurerwa Rutazibwa, James Dennis, Miguel Marraco Eibar, Jan Guthrie, Miguel Fern\'{a}ndez L\'{o}pez, Alex Panayides, Karen Claeys, Ambra Sottile, Tiziana La Piana, Emma Vittorio, Claudio Nicol\`{o}, Francesco Pace, H\'{e}ctor Gil Mar\'{i}n, Claire Le Cras, Xan Morice-Atkinson, David Wilkinson, Jojo Pratt, Leonidas Christodoulou, Guido Walter Pettinari, Lucas Lombriser, Arna Karick, Spencer Craig, Tesla Jeltema, William Wester, Joshua Frieman, Rae Bull, Maurizio Paolillo, Ivan Baldry, Chris Collins, Giuseppe Longo, Alkistis Pourtsidou, Ben Bose, Alicia Bueno Belloso, Jennifer Pollack, Rossana Ruggeri, Deike Striez, Monica Rizzo, Magdalena Moszy\'{n}ska, the staff from ``WorkPlaceWellness'' and the ``Talking Change'' NHS trust in Portsmouth.

Funding for the DES Projects has been provided by the U.S. Department of Energy, the U.S. National Science Foundation, the Ministry of Science and Education of Spain, 
the Science and Technology Facilities Council of the United Kingdom, the Higher Education Funding Council for England, the National Center for Supercomputing 
Applications at the University of Illinois at Urbana-Champaign, the Kavli Institute of Cosmological Physics at the University of Chicago, 
the Center for Cosmology and Astro-Particle Physics at the Ohio State University,
the Mitchell Institute for Fundamental Physics and Astronomy at Texas A\&M University, Financiadora de Estudos e Projetos, 
Funda{\c c}{\~a}o Carlos Chagas Filho de Amparo {\`a} Pesquisa do Estado do Rio de Janeiro, Conselho Nacional de Desenvolvimento Cient{\'i}fico e Tecnol{\'o}gico and 
the Minist{\'e}rio da Ci{\^e}ncia, Tecnologia e Inova{\c c}{\~a}o, the Deutsche Forschungsgemeinschaft and the Collaborating Institutions in the Dark Energy Survey. 

The Collaborating Institutions are Argonne National Laboratory, the University of California at Santa Cruz, the University of Cambridge, Centro de Investigaciones Energ{\'e}ticas, 
Medioambientales y Tecnol{\'o}gicas-Madrid, the University of Chicago, University College London, the DES-Brazil Consortium, the University of Edinburgh, 
the Eidgen{\"o}ssische Technische Hochschule (ETH) Z{\"u}rich, 
Fermi National Accelerator Laboratory, the University of Illinois at Urbana-Champaign, the Institut de Ci{\`e}ncies de l'Espai (IEEC/CSIC), 
the Institut de F{\'i}sica d'Altes Energies, Lawrence Berkeley National Laboratory, the Ludwig-Maximilians Universit{\"a}t M{\"u}nchen and the associated Excellence Cluster Universe, 
the University of Michigan, the National Optical Astronomy Observatory, the University of Nottingham, The Ohio State University, the University of Pennsylvania, the University of Portsmouth, 
SLAC National Accelerator Laboratory, Stanford University, the University of Sussex, Texas A\&M University, and the OzDES Membership Consortium.

The DES data management system is supported by the National Science Foundation under Grant Number AST-1138766.
The DES participants from Spanish institutions are partially supported by MINECO under grants AYA2015-71825, ESP2015-88861, FPA2015-68048, SEV-2012-0234, SEV-2012-0249, and MDM-2015-0509, some of which include ERDF funds from the European Union. IFAE is partially funded by the CERCA program of the Generalitat de Catalunya.

We are grateful for the extraordinary contributions of our CTIO colleagues and the DECam Construction, Commissioning and Science Verification
teams in achieving the excellent instrument and telescope conditions that have made this work possible.  The success of this project also 
relies critically on the expertise and dedication of the DES Data Management group.

The STILTS and TOPCAT softwares by \citet{Taylor-2006} were significantly used in the making of this work.

\bibliographystyle{mn2e}

\bibliography{Reference}

\section*{Affiliations}
$^{1}$Institute of Cosmology and Gravitation, University of Portsmouth, Portsmouth, PO1 3FX, UK\\
$^{2}$SEPnet, South East Physics Network, (www.sepnet.ac.uk)\\
$^{3}$Kavli Institute for Particle Astrophysics \& Cosmology, P. O. Box 2450, Stanford University, Stanford, CA 94305, USA\\
$^{4}$SLAC National Accelerator Laboratory, Menlo Park, CA 94025, USA\\
$^{5}$Centro de Investigaciones Energeticas, Medioambientales y Tecnologicas (CIEMAT), Madrid, Spain \\
$^{6}$Wisconsin IceCube Particle Astrophysics Center (WIPAC), Madison, WI 53703, USA\\
$^{7}$Department of Physics, University of WisconsinMadison, Madison, WI 53706, USA\\
$^{8}$Department of Astronomy, University of Illinois, 1002 W. Green Street, Urbana, IL 61801, USA\\
$^{9}$National Center for Supercomputing Applications, 1205 West Clark St., Urbana, IL 61801, USA\\
$^{10}$Fermi National Accelerator Laboratory, P. O. Box 500, Batavia, IL 60510, USA\\
$^{11}$Aix Marseille Universit\'{e}, CNRS, LAM (Laboratoire D'Astrophysique de Marseille) UMR 7326, 13388, Marseille, France\\
$^{12}$ESA/ESTEC, Noordwijk, The Netherlands\\
$^{13}$Observat\'{o}rio Nacional Rua Gal. Jos\'{e} Cristino 77, Rio de Janeiro, RJ - 20921-400, Brazil\\
$^{14}$Laborat\'{o}rio Interinstitucional de e-Astronomia-LIneA, Rua Gal. Jos\'{e} Cristino 77, Rio de Janeiro, RJ - 20921-400, Brazil\\
$^{15}$CNRS, UMR 7095, Institut d'Astrophysique de Paris, F-75014, Paris, France\\
$^{16}$Department of Physics \& Astronomy, University College London, Gower Street, London, WC1E 6BT, UK\\
$^{17}$Sorbonne Universit\'{e}s, UPMC Univ Paris 06, UMR 7095, Institut d'Astrophysique de Paris, F-75014, Paris, France\\
$^{18}$Department of Physics, Stanford University, 382 Via Pueblo Mall, Stanford, CA 94305, USA\\
$^{19}$Kavli Institute for Cosmology, University of Cambridge, Madingley Road, Cambridge CB3 0HA, UK\\
$^{20}$Institute of Astronomy, University of Cambridge, Madingley Road, Cambridge CB3 0HA, UK\\
$^{21}$George P. and Cynthia Woods Mitchell Institute for Fundamental Physics and Astronomy, Department of Physics and Astronomy, Texas A\&M University, College Station, TX 77843, USA\\
$^{22}$Department of Physics and Electronics, Rhodes University, PO Box 94, Grahamstown, 6140, South Africa\\
$^{23}$Institut de F\'{\i}sica d'Altes Energies (IFAE), The Barcelona Institute of Science and Technology, Campus UAB, 08193 Bellaterra (Barcelona) Spain\\
$^{24}$Department of Physics and Astronomy, University of Pennsylvania, Philadelphia, PA 19104, USA\\
$^{25}$Department of Physics, IIT Hyderabad, Kandi, Telangana 502285, India\\
$^{26}$Department of Astronomy, University of Michigan, Ann Arbor, MI 48109, USA\\
$^{27}$Department of Physics, University of Michigan, Ann Arbor, MI 48109, USA\\
$^{28}$Institut de Ci\`encies de l'Espai, IEEC-CSIC, Campus UAB, Carrer de Can Magrans, s/n,  08193 Bellaterra, Barcelona, Spain\\
$^{29}$Kavli Institute for Cosmological Physics, University of Chicago, Chicago, IL 60637, USA\\
$^{30}$Instituto de Fisica Teorica UAM/CSIC, Universidad Autonoma de Madrid, 28049 Madrid, Spain\\
$^{31}$Department of Physics, ETH Zurich, Wolfgang-Pauli-Strasse 16, CH-8093 Zurich, Switzerland\\
$^{32}$Astronomy Department, University of Washington, Box 351580, Seattle, WA 98195, USA\\
$^{33}$Cerro Tololo Inter-American Observatory, National Optical Astronomy Observatory, Casilla 603, La Serena, Chile\\
$^{34}$Santa Cruz Institute for Particle Physics, Santa Cruz, CA 95064, USA\\
$^{35}$Australian Astronomical Observatory, North Ryde, NSW 2113, Australia\\
$^{36}$Argonne National Laboratory, 9700 South Cass Avenue, Lemont, IL 60439, USA\\
$^{37}$Departamento de F\'isica Matem\'atica, Instituto de F\'isica, Universidade de S\~ao Paulo, CP 66318, S\~ao Paulo, SP, 05314-970, Brazil\\
$^{38}$Center for Cosmology and Astro-Particle Physics, The Ohio State University, Columbus, OH 43210, USA\\
$^{39}$Department of Astronomy, The Ohio State University, Columbus, OH 43210, USA\\
$^{40}$Instituci\'o Catalana de Recerca i Estudis Avan\c{c}ats, E-08010 Barcelona, Spain\\
$^{41}$Jet Propulsion Laboratory, California Institute of Technology, 4800 Oak Grove Dr., Pasadena, CA 91109, USA\\
$^{42}$Department of Physics and Astronomy, Pevensey Building, University of Sussex, Brighton, BN1 9QH, UK\\
$^{43}$School of Physics and Astronomy, University of Southampton,  Southampton, SO17 1BJ, UK\\
$^{44}$Instituto de F\'isica Gleb Wataghin, Universidade Estadual de Campinas, 13083-859, Campinas, SP, Brazil\\
$^{45}$Computer Science and Mathematics Division, Oak Ridge National Laboratory, Oak Ridge, TN 37831\\
%$^{45}$\\
%$^{46}$\\
%$^{47}$\\
%Space Sciences Laboratory and Theoretical Astrophysics Center, University of California, Berkley CA 94720

\appendix

\section{Tables}

\begin{table*}
%\begin{tiny}
\begin{center}
\caption{Binned {\it i}-band luminosity function in the five redshift bins studied. The full table is available in electronic format online.}
\begin{tabular}{cccccc}
\hline
  \multicolumn{1}{c}{$\mathbf{M_{\rm i}}$} &
  \multicolumn{1}{c}{$\mathbf{\phi}$} &
  \multicolumn{1}{c}{$\mathbf{\sigma_{\rm shot}^{\rm low}}$} &
  \multicolumn{1}{c}{$\mathbf{\sigma_{\rm shot}^{\rm up}}$} &
  \multicolumn{1}{c}{$\mathbf{\sigma_{\rm phot}^{\rm low}}$} &
  \multicolumn{1}{c}{$\mathbf{\sigma_{\rm phot}^{\rm up}}$} \\
  \multicolumn{1}{c}{({\rm mag})}&
  \multicolumn{1}{c}{$({\rm n\ Mpc^{-3}/10^{-4}})$} &
  \multicolumn{1}{c}{$({\rm n\ Mpc^{-3}/10^{-4}})$} &
  \multicolumn{1}{c}{$({\rm n\ Mpc^{-3}/10^{-4}})$} &
  \multicolumn{1}{c}{$({\rm n\ Mpc^{-3}/10^{-4}})$} &
  \multicolumn{1}{c}{$({\rm n\ Mpc^{-3}/10^{-4}})$} \\
  \multicolumn{1}{c}{1}&
  \multicolumn{1}{c}{2}&
  \multicolumn{1}{c}{3}&
  \multicolumn{1}{c}{4}&
  \multicolumn{1}{c}{5}&
  \multicolumn{1}{c}{6} \\
\hline
  \multicolumn{6}{c}{$0<z<0.2$} \\
\hline
\hline   
-23.75   &   0.10   &     0.01     &        0.01     &		0.01      & 		 0.02 \\
-23.25   &   0.63   &     0.03     &        0.03     &		0.05      & 		 0.04 \\
-22.75   &   2.41   &     0.06     &        0.06     &		0.06      & 		 0.07 \\
-22.25   &   6.4     &      0.1	     &	       0.1	 &		0.09	     &		 0.1   \\ 
-21.75   &   12.8   &     0.2	     &	       0.2	 &		0.1	     &		 0.1   \\
%-21.25   &   19.1   &     0.2	     &	       0.2	 &		0.2	     &		 0.2   \\ 
%-20.75   &   22.1   &     0.2	     &	       0.2	 &		0.2	     &		 0.3   \\
%-20.25   &   23.8   &     0.3	     &	       0.3	 &		0.4	     &		 0.3   \\ 
%-19.75   &   25.5   &     0.3	     &	       0.3	 &		0.5	     &		 0.4   \\ 
%-19.25   &   27.7   &     0.3	     &	       0.3	 &		0.5	     &		 0.6   \\
.....          &   .....      &     .....	     &	       .....	 &		.....	     &		 .....    \\
\hline
  \multicolumn{6}{c}{$0.2<z<0.4$} \\
\hline
\hline   
-25.25  &    0.000204  &     0.0002  &    0.0005	&     0.000008	& 	 0.0002  \\
-24.25  &    0.018         &     0.002    &    0.002	&     0.002		& 	 0.002    \\
-23.75  &    0.170         &     0.006    &    0.006	&     0.006		& 	 0.006    \\
-23.25  &    1.02           &     0.01       &    0.01	&     0.02		& 	 0.02       \\
-22.75  &    3.84           &     0.03       &    0.03	&     0.03		& 	 0.03       \\
.....        &   .....                &     .....	      &	       .....	&      .....	         &          .....        \\
\hline
\hline
\end{tabular}
\label{tab:TableA1}                               
\end{center}
% \end{tiny}
\end{table*}

\begin{table*}
%\begin{tiny}
\begin{center}
\caption{Binned stellar mass function in the five redshift bins studied. The full table is available in electronic format online.}
\begin{tabular}{cccccc}
\hline
  \multicolumn{1}{c}{$\mathbf{Log (M/M_{\rm \odot})}$} &
  \multicolumn{1}{c}{$\mathbf{\phi}$} &
  \multicolumn{1}{c}{$\mathbf{\sigma_{\rm shot}^{\rm low}}$} &
  \multicolumn{1}{c}{$\mathbf{\sigma_{\rm shot}^{\rm up}}$} &
  \multicolumn{1}{c}{$\mathbf{\sigma_{\rm phot}^{\rm low}}$} &
  \multicolumn{1}{c}{$\mathbf{\sigma_{\rm phot}^{\rm up}}$} \\
  \multicolumn{1}{c}{}&
  \multicolumn{1}{c}{$({\rm n\ Mpc^{-3}/10^{-4}})$} &
  \multicolumn{1}{c}{$({\rm n\ Mpc^{-3}/10^{-4}})$} &
  \multicolumn{1}{c}{$({\rm n\ Mpc^{-3}/10^{-4}})$} &
  \multicolumn{1}{c}{$({\rm n\ Mpc^{-3}/10^{-4}})$} &
  \multicolumn{1}{c}{$({\rm n\ Mpc^{-3}/10^{-4}})$} \\
  \multicolumn{1}{c}{1}&
  \multicolumn{1}{c}{2}&
  \multicolumn{1}{c}{3}&
  \multicolumn{1}{c}{4}&
  \multicolumn{1}{c}{5}&
  \multicolumn{1}{c}{6} \\
\hline
  \multicolumn{6}{c}{$0<z<0.2$} \\
\hline
\hline   
6.9  &	 83     &   2    &  3     &   4     &  4     \\
7.1  &	 74     &   2    &  3     &   4     &  3     \\ 
7.3  & 	 76     &   3    &  4     &   3     &  5     \\
7.5  & 	 85     &   4    &  4     &   4     &  4     \\ 
7.7  & 	 62     &   3    &  3     &   4     &  4     \\
.... &   ....   &   .... &  ....  &   ....  &  ....  \\ 
\hline
  \multicolumn{6}{c}{$0.2<z<0.4$} \\
\hline
\hline   
8.5  &   18.1	&   0.1  &  0.1   &   0.2   &  0.2   \\
8.7  &   17.56	&   0.09 &  0.09  &   0.07  &  0.09  \\
8.9  &   17.17	&   0.07 &  0.07  &   0.07  &  0.06  \\
9.1  &   16.17	&   0.06 &  0.06  &   0.06  &  0.06  \\
9.3  &   14.76	&   0.06 &  0.06  &   0.05  &  0.06  \\
.... &   ....   &   .... &  ....  &   ....  &  ....  \\ 
\hline
\hline
\end{tabular}
\label{tab:TableA2}                               
\end{center}
% \end{tiny}
\end{table*}

\clearpage

\subsection*{Supporting Information}
\noindent Additional Supporting Information can be found in the online version of this article:\\

\noindent {\bf Table \ref{tab:TableA1}:} The binned GLFs obtained for the full galaxy sample in the five $z$ bins studied in this paper and plotted in Figure \ref{fig:Fig7}. $X=M_{\rm i}$ ({\rm mag}) and $Y=\phi(M_{\rm i})\ ({\rm n\ Mpc^{-3}})/10^{-4}$. 
 
\noindent {\bf Table \ref{tab:TableA2}:} The binned GSMFs obtained for the full galaxy sample in the five $z$ bins studied in this paper and plotted in Figure \ref{fig:Fig7}. $X=\log (M/M_{\rm \odot}$) and $Y=\phi[\log (M/M_{\rm \odot})]\ ({\rm n\ Mpc^{-3}})/10^{-4}$.

\end{document}